\documentclass[preprint]{elsarticle} 

\makeatletter
	\def\ps@pprintTitle{%
 	\let\@oddhead\@empty
	\let\@evenhead\@empty
	\def\@oddfoot{\centerline{\thepage}}%
	\let\@evenfoot\@oddfoot}
\makeatother

\usepackage{etoolbox}
\patchcmd{\MaketitleBox}{\footnotesize\itshape\elsaddress\par\vskip36pt}{\footnotesize\itshape\elsaddress\par\parbox[b][36pt]{\linewidth}{\vfill\hfill\textnormal{\today}\hfill\null\vfill}}{}{}%
\patchcmd{\pprintMaketitle}{\footnotesize\itshape\elsaddress\par\vskip36pt}{\footnotesize\itshape\elsaddress\par\parbox[b][36pt]{\linewidth}{\vfill\hfill\textnormal{\today}\hfill\null\vfill}}{}{}%

\usepackage[colorlinks=true,%
            breaklinks=true,%
            linkcolor=blue,
            urlcolor=blue,%
            citecolor=blue,%
            pdftitle={Title of paper}, 
            pdfkeywords={Keywords},	
            pdfauthor={Authors},		
            bookmarksopen=false,
            pdfpagemode=UseNone]{hyperref}

\usepackage[margin=0.9in]{geometry}
\usepackage[T1]{fontenc}
\usepackage[english]{babel}
\usepackage{icomma}
\usepackage[latin1]{inputenc} 
\usepackage{subfiles}
\usepackage{tabularray}
\biboptions{numbers,sort&compress}
\usepackage{subcaption}
\usepackage[justification=centering]{subcaption}

\usepackage{amsmath}
\usepackage{pifont}

\newcommand{\xmark}{\text{\ding{55}}}
\usepackage{mathtools}
\usepackage{siunitx}
\usepackage{bbm}
\usepackage{amsmath}
\usepackage{amsfonts}
\usepackage{amsthm}
\usepackage{amssymb}
\usepackage{array}
\usepackage{algorithm} 
\usepackage{algorithmicx}
\usepackage{algpseudocode}
\usepackage{subcaption}
\usepackage{siunitx}
\usepackage{amsmath}
\usepackage{mathrsfs}
\usepackage{xpatch}

\usepackage{color}
\usepackage{url}
\usepackage{cleveref}
\usepackage{graphicx}
\usepackage{breakcites}
\usepackage{soul} 
\usepackage{enumitem}
\usepackage[title,titletoc,toc]{appendix}

\newtheoremstyle{mytheoremstyle}{5pt}{5pt}{\itshape}{}{\bfseries}{.}{.5em}{} 
\theoremstyle{mytheoremstyle}

\newtheoremstyle{myremarkstyle}{3pt}{3pt}{\itshape}{}{\bfseries}{.}{.5em}{} 
\theoremstyle{myremarkstyle}

\usepackage{pifont}

{\noindent {\textbf{Proof}:} }%
{\hfill $\Box$ \\ }

\newcommand{\bit}{\begin{itemize}}
\newcommand{\eit}{\end{itemize}}
\newcommand{\ben}{\begin{enumerate}}
\newcommand{\een}{\end{enumerate}}

\newcommand{\Hquad}{\hspace{0.5em}}

\begin{document}
\begin{frontmatter}
    \title{Effects of Artificial Collisions, Filtering, and Nonlocal Closure Approaches on Hermite-based Vlasov-Poisson Simulations}
    \author[1,2]{Opal Issan\corref{cor1}}\ead{oissan@ucsd.edu}
    \author[2]{Oleksandr Chapurin}
    \author[2]{Oleksandr Koshkarov}
    \author[2]{Gian Luca Delzanno}
    
    \cortext[cor1]{Corresponding author}
    \address[1]{Department of Mechanical and Aerospace Engineering, University of California San Diego, La Jolla, CA, USA}
    \address[2]{T-5 Applied Mathematics and Plasma Physics Group, Los Alamos National Laboratory, Los Alamos, NM, USA}
    
    \begin{abstract}
    Kinetic simulations of collisionless plasmas are computationally challenging due to phase space mixing and filamentation, resulting in fine-scale velocity structures. 
    This study compares three methods developed to reduce artifacts related to limited velocity resolution in Hermite-based Vlasov-Poisson simulations: artificial collisions, filtering, and nonlocal closure approaches. 
    We evaluate each method's performance in approximating the linear kinetic response function and suppressing recurrence in linear and nonlinear regimes. 
    Numerical simulations of Landau damping demonstrate that artificial collisions, particularly higher orders of the Lenard-Bernstein collisional operator, most effectively recover the correct damping rate across a range of wavenumbers.
    Moreover, Hou-Li filtering and nonlocal closures underdamp high wavenumber modes in linear simulations, and the Lenard-Bernstein collisional operator overdamps low wavenumber modes in both linear and nonlinear simulations. 
    This study demonstrates that hypercollisions offer a robust approach to kinetic simulations, accurately capturing collisionless dynamics with limited velocity resolution.
    \end{abstract}	
\end{frontmatter}

\section{Introduction}\label{sec:introduction}
Kinetic simulations of collisionless plasmas are important for many applications, e.g. astrophysics, geophysics, and laboratory fusion devices. Solving the collisionless Vlasov equation numerically is challenging due to its high dimensionality, Hamiltonian structure, and significant spatial and temporal scale separation. 
Furthermore, kinetic solvers struggle to accurately capture phase-space mixing, in which the distribution function develops increasingly fine structures in velocity space, known as \textit{filamentation}~\cite{grant_1967_hermite}. As shown by the solution $f(x,v,t) = f(x - vt, v, 0)$ to the free-streaming equation $\partial_t f + v \partial_x f = 0$, these structures cannot be resolved after a finite time determined by the resolution of the velocity space. Filamentation can also cause \textit{recurrence} in Eulerian solvers (grid or spectral), where the solution is artificially periodic in time~\cite{canosa_1974_recurrence, cheng_1976_recurrence}.

A large class of kinetic spectral solvers use Hermite functions in velocity space, weighted by a Gaussian distribution~\cite{schumer_1998_sw_aw, schouri_1976_hermite, grad_1949_sw, joyce_1971_hermite, armstrong_1967_hermite, grant_1967_hermite, black_2013_spectra, issan_2024_antisymmetry, chapurin_2024_hybrid,  pagliantini_2023_adaptive, pagliantini_2023_rk, camporeale_2006, delzanno_2015_jcp, koshkarov_2021, roytershetyn_2018_sps, celebre_2023_hermite, pezzi_2019_hermite, pezzi_2016_recurrence, bourdiec_2006_hermite, loureiro_2016_hermite, camporeale_2016}. Hermite-based spectral expansions are advantageous for accurately approximating near-Maxwellian distributions with only a few basis functions. 
However, the spectral method exhibits recurrence phenomena when filamentation develops beyond the scales that the finite Hermite resolution can resolve.
An insightful mechanical analogy from \citet{hammett_1993_closure} describes the Hermite moment system as a semi-infinite chain of springs and masses. In this analogy, an initial perturbation of the lowest Hermite mode corresponds to displacing the first mass in the chain. Truncating the Hermite system is akin to placing a wall at the end of a finite spring-mass chain. The wall then reflects the perturbation energy, resulting in an inverse cascade from the last mass back to the first that artificially reconstructs the initial perturbation. 
The recurrence period in Hermite simulations scales with $\sim \sqrt{N_{v}} / k$~\cite{cheng_1976_recurrence}, where $N_{v}$ is the number of Hermite moments and $k$ is the spatial wavenumber of the initial perturbation. In nature, filamentation exhibits a cutoff scale at which diffusive scattering effects become significant. To adequately resolve such scales, a very high resolution in velocity is necessary; for instance, the slow solar wind requires $N_{v} \sim 1,800$~\cite{pezzi_2019_hermite} and tokamak edge simulations require $N_{v} \sim 180$~\cite{loureiro_2016_hermite}, which becomes computationally prohibitive for long-time or three-dimensional simulations. Nevertheless, a low-resolution Hermite spectral representation is analogous to assuming that the particle distribution function lacks fine-scale structures in velocity space, which can lead to numerical instabilities in the case of filamentation. This phenomenon is common in other spectral methods, known as \textit{spectral blocking}~\cite{boyd_2001_spectral}, where noise accumulates in high-order spectral modes near the truncation limit. 

The most common closure for the Hermite spectral approximation is closure by truncation, which assumes that the last spectral moment is zero. However, the Landau root~\cite{landau_46} does not emerge as a discrete eigenvalue with the closure by truncation, regardless of velocity resolution.~\citet{grant_1967_hermite} were the first to show that adding artificial collisions to the finite system with the closure by truncation can restore the correct Landau damping~\cite{ng_1999_spectra, ng_2004_spectra, pezzi_2016_recurrence}. Previous work by~\cite{grant_1967_hermite, black_2013_spectra} use the Lenard Bernstein (LB)~\cite{lenard_bernstein_1958_collisions} collisional operator. Similarly, recent work by~\cite{camporeale_2016, delzanno_2015_jcp, koshkarov_2021, chapurin_2024_hybrid, pagliantini_2023_adaptive, pagliantini_2023_rk, roytershetyn_2018_sps} use a hypercollisional operator, which damps more strongly the higher-order Hermite moments in comparison to the LB operator.
Filtering techniques, such as Klimas~\cite{klimas_1987_filter} and Hou-Li~\cite{hou_li_2007_filter}, are also used to suppress recurrence and filamentation artifacts~\cite{klimas_2018_recurrence, cai_2018_suppression, di_2019_filtered, filbet_2022_conservative, cheng_1976_recurrence}. 
Another method for incorporating kinetic physics into moment solvers is with a linear nonlocal closure, often referred to as~\textit{Landau-fluid closures}~\cite{hammet_1990_closure, hammet_1992_closure}. This approach matches the closure coefficients to the asymptotic limits of the linear kinetic response function. Landau-fluid closures were initially formulated for fluid models incorporating three and four moments and were later generalized to models involving higher moments in Hermite space~\cite{smith_1997_closure}.
Therefore, it becomes essential to incorporate a modified term or closure that introduces dissipation into the discrete system to prevent filamentation issues while ensuring that the chosen approach accurately captures the system's average dynamics.

In this paper, we provide a comprehensive comparison of artificial collisions, filtering, and nonlocal closure approaches. We examine the discrete system approximation of the linear kinetic response function and evaluate each method's ability to recover the Landau damping dispersion relation. Additionally, we assess each method's suppression of recurrence and filamentation artifacts by simulating a Langmuir wave's linear and nonlinear Landau damping. 
This paper builds on previous work by~\citet{pezzi_2016_recurrence}, which shows that the LB collisional operator significantly modifies the collisionless dynamics. Specifically, at the same collisional rate that recovers the Landau root, nonlinear waves damp due to LB collisions instead of reaching saturation. 
We replicate these findings and assess whether similar issues arise with hypercollisions, filtering, and nonlocal closure approaches. 
The analysis shows that hypercollisions, particularly higher orders of the LB collision operator, most effectively recover the correct damping rate across a range of wavenumbers, particularly in limited velocity resolution.

This paper is organized as follows. Section~\ref{sec:section-2} briefly reviews the one-dimensional collisionless Vlasov-Poisson equations, linear theory, and the velocity discretization of the particle distribution function using the asymmetrically weighted Hermite spectral expansion. Section~\ref{sec:section-3} describes the three methods to mitigate recurrence: (1) artificial collisions, (2) filtering, and (3) nonlocal closures, along with their ability to approximate linear kinetic dynamics. We numerically test these methods on the linear and nonlinear Landau damping benchmark problem in section~\ref{sec:section-4}, and the concluding remarks are given in section~\ref{sec:section-5}.

\section{Vlasov-Poisson Equations: Hermite Spectral Discretization in Velocity}\label{sec:section-2}
We present the one-dimensional collisionless Vlasov-Poisson equations in section~\ref{sec:vlasov_poisson}. Section~\ref{sec:linear_theory} derives the well-known linear kinetic response function. We introduce the asymmetrically weighted Hermite expansion and its approximation of the response function in section~\ref{sec:aw_hermite_expansion}.

\subsection{Vlasov-Poisson Equations}\label{sec:vlasov_poisson}
We consider the one-dimensional Vlasov-Poisson equations, which model the interaction of collisionless charged particles with a self-consistent electric field. The plasma is composed of electrons and immobile background ions. The one-dimensional (normalized) Vlasov-Poisson equations are
\begin{align}
    \left(\frac{\partial}{\partial t} + v \frac{\partial}{\partial x}+ \frac{\partial \phi(x, t)}{\partial x}\frac{\partial}{\partial v}\right) f(x, v, t)  &= 0, \label{vlasov-continuum}\\
    -\frac{\partial^{2} \phi(x, t)}{\partial x^{2}} &= 1 - \int_{\mathbb{R}} f(x,v, t) \mathrm{d}v, \label{poisson-continuum}
\end{align}
where $f(x, v, t)$ is the electron distribution function, $\phi(x, t)$ is the electrostatic potential given by $E(x, t) = - \partial \phi(x, t)/ \partial x$ such that $E(x, t)$ is the electric field. We consider an unbounded velocity coordinate $v \in \mathbb{R}$, a periodic spatial coordinate $x \in [0, \ell]$, where $\ell$ is the length of the spatial domain, and time $t \geq 0$. All quantities in the Vlasov-Poisson equations~\eqref{vlasov-continuum}--\eqref{poisson-continuum} are normalized as follows:
\begin{equation*}
   t \coloneqq t_{d} \omega_{pe}, \qquad x \coloneqq \frac{x_{d}}{\lambda_{D}}, \qquad v \coloneqq \frac{v_{d}}{v_{te}}, \qquad f \coloneqq f_{d}\frac{v_{te}}{n_{e}},  \qquad \phi \coloneqq \phi_{d} \frac{e \lambda_{D}^2}{T_{e}}, 
\end{equation*}
where the subscript `$d$' indicates the dimensional quantities, $e$ is the positive elementary charge, $\omega_{pe} \coloneqq \sqrt{4 \pi e^2 n_{e}/m_{e}}$ is the electron plasma frequency, $m_{e}$ is the electron mass, $n_{e}$ is the reference electron density, $v_{te} \coloneqq \sqrt{T_{e}/m_{e}}$ is the electron thermal velocity, $T_{e}$ is a reference electron temperature, $\lambda_{D} \coloneqq v_{te}/\omega_{pe} = \sqrt{T_{e}/4\pi e^{2}n_{e}}$ is the electron Debye length.

\subsection{Linear Response Function}\label{sec:linear_theory}
The electron distribution function and electrostatic potential can be expressed as the sum of equilibrium and small-amplitude perturbation components
\begin{alignat*}{3}
    f(x, v, t) &= f_{0}(v) + \tilde{f}(x, v, t),\qquad &&\mathrm{s.t.} \qquad \tilde{f} \ll f_{0}, \\
    \phi(x, t) &= \phi_{0}(x) + \tilde{\phi}(x, t), \qquad &&\mathrm{s.t.} \qquad \phi_{0}(x)=0.
\end{alignat*}
We are interested in the asymptotic solutions at large times, in which it is sufficient to assume that the first-order perturbed quantities vary as 
\begin{equation}\label{normal-modes}
    \tilde{f}(x, v, t) = \exp(-i\omega t + ikx)\hat{f}(v) \qquad \mathrm{and} \qquad \tilde{\phi}(x, t) = \exp(-i\omega t + ikx) \hat{\phi}, 
\end{equation}
where the wavenumber $k$ is real and the frequency $\omega = \omega_{r} + i \gamma$ is complex; see~\cite{gary_1993_theory, hunana_2019_pade, fitzpatrick_2015_plasma} for more details; then $\gamma \in \mathbb{R}$ is the damping/growth rate and $\omega_{r} \in \mathbb{R}$ is the oscillation frequency. We substitute Eq.~\eqref{normal-modes} into the linearized Vlasov equation~\eqref{vlasov-continuum}, resulting in 
\begin{equation}\label{linearized-vlasov}
    -i\omega \hat{f}(v) + ikv \hat{f}(v) + ik\hat{\phi}\frac{\mathrm{d}f_{0}}{\mathrm{d}v} = 0 \qquad \Rightarrow \qquad \hat{f}(v) = k\hat{\phi} \frac{\mathrm{d} f_{0}/\mathrm{d} v}{\omega-kv}.
\end{equation}
For a Maxwellian equilibrium, that is, $f_{0}(v) = \exp\left(-v^2 / 2\right)/ \sqrt{2 \pi}$, the perturbed density is described via the following linear kinetic response function $R(\xi)$:
\begin{equation}\label{response-function-definition}
    \hat{n} \coloneqq \int_{\mathbb{R}} \hat{f}(v) \mathrm{d} v  = k\hat{\phi}\int_{\mathbb{R}} \frac{\mathrm{d} f_{0}/\mathrm{d} v}{\omega - kv} \mathrm{d} v = \hat{\phi} R(\xi), \qquad \mathrm{s.t.} \qquad R(\xi) \coloneqq 1 + \xi Z(\xi) \qquad \mathrm{and} \qquad \xi \coloneqq \frac{\omega}{\sqrt{2} |k|},
\end{equation}
where the plasma dispersion function is $Z(\xi) = \pi^{-1/2} \int_{\mathbb{R}} \exp(-s^2)/(s-\xi) \mathrm{d} s$ for $\mathrm{Im}(\xi)>0$ and is analytically continued for $\mathrm{Im}(\xi) \leq 0$. 
There is a subtle distinction between defining $\xi \coloneqq \omega/(\sqrt{2}k)$ and $\xi \coloneqq \omega/(\sqrt{2}|k|)$. We adopt the latter definition because it allows the use of the plasma dispersion function $Z(\xi)$ as originally defined by \citet{fried_conte_1961}. In contrast, the former definition requires a redefinition of $Z(\xi)$; see \citet[\S 2.2]{hunana_2019_pade}.
The asymptotic expansion of the response function is given by~\cite[\S 3]{hunana_2019_pade}:
\begin{alignat}{3}
R(\xi) &= 1 + i \sqrt{\pi} \xi -2\xi^2 -i\sqrt{\pi} \xi^3 + \frac{4}{3} \xi^4 + i \frac{\sqrt{\pi}}{2} \xi^5 - \frac{8}{15} \xi^6 -i\frac{\sqrt{\pi}}{6} \xi^7+\frac{16}{105}\xi^8 + \ldots \qquad &&\mathrm{for} \qquad |\xi|\ll 1, \label{analytic-response-asymptotics-low}\\
R(\xi) &= i \sigma \sqrt{\pi} \xi \exp(-\xi^2) -\frac{1}{2\xi^2} -\frac{3}{4\xi^4} -\frac{15}{8\xi^6} -\frac{105}{16\xi^8} -\frac{945}{32\xi^{10}} -\frac{10395}{64\xi^{12}} - \frac{135135}{128\xi^{14}} + \ldots \qquad &&\mathrm{for} \qquad |\xi| \gg 1,\label{analytic-response-asymptotics-high}
\end{alignat}
where 
\begin{equation*}
    \sigma = \begin{cases*}
        0 & \text{if } $\text{Im}(\xi) >0$,\\
        1 & \text{if } $\text{Im}(\xi) =0$,\\
        2 & \text{if } $\text{Im}(\xi) <0$.
    \end{cases*}
\end{equation*}
The response function captures key microscopic dynamics of the linear Vlasov-Poisson system, such as wave-particle resonance. 

\subsection{Asymmetrically Weighted Hermite Expansion}\label{sec:aw_hermite_expansion}
We discretize the electron distribution function in velocity space via an asymmetrically weighted (AW) Hermite spectral expansion, where the weight function does not equal unity. Hermite-based velocity discretizations inherently separate different scales (analogous to other orthogonal bases, e.g. Fourier), offering insight into the propagation of free energy (anisotropy or inhomogeneity in the initial distribution function) across scales in both linear and nonlinear processes~\cite{parker_2015_hermite}. Additionally, the AW Hermite expansion, in particular, conserves mass, momentum, and energy (if coupled with a conserving spatial and temporal integrator). The AW Hermite expansion is given by 
\begin{equation}\label{aw-expansion}
    \hat{f}(v) \approx \sum_{n=0}^{N_{v}-1} \hat{C}_{n} \psi_{n}(v) \qquad \mathrm{and} \qquad f_{0}(v) = \frac{1}{\sqrt{2}} \psi_{0}(v), 
\end{equation}
where 
\begin{align}
    \psi_{n}(v) &\coloneqq (\pi 2^n n!)^{-\frac{1}{2}} \mathcal{H}_{n}\left(\frac{v}{\sqrt{2}}\right) \exp{\left(-\frac{v^2}{2}\right)}, \label{hermite-basis-function}\\
     \psi^{n}(v) &\coloneqq (2^n n!)^{-\frac{1}{2}} \mathcal{H}_{n}\left(\frac{v}{\sqrt{2}}\right), \nonumber\\
    \mathcal{H}_{n}(v) &\coloneqq (-1)^n \exp{\left(v^2\right)} \frac{\mathrm{d}^{n}}{\mathrm{d}v^n} \exp{\left(-v^2\right)}.\nonumber
\end{align}
The limit relation of the AW Hermite basis function~\cite[\S 22.15]{abramowitz_1964_math} is given by
\begin{alignat}{3}
    \lim_{n\to\infty} \psi_{n}(v) \propto \begin{cases}
        \cos(\sqrt{n} v)\exp\left(-\frac{v^2}{2}\right) \qquad &\text{if n is even}, \label{scale-hermite}\\
        \sin(\sqrt{n-1}v) \exp\left(-\frac{v^2}{2}\right)\qquad &\text{if n is odd},
    \end{cases}
\end{alignat}
such that the shortest wave-length $\lambda_{v}$ in velocity space is resolved by the last Hermite basis of $N_{v}-1$, i.e $\lambda_{v} \approx 2\pi/\sqrt{N_{v}-1}$. 
The orthogonality and recursive properties of the AW Hermite basis~\cite[\S 22.2--22.8]{abramowitz_1964_math} are 
\begin{align}
    \int_{\mathbb{R}} \psi_{n}(v)\psi^{m}(v) \mathrm{d} v &= \sqrt{2} \delta_{n, m},\label{orthogonality-AW}\\
    \frac{\mathrm{d}\psi_{n}}{\mathrm{d}v}  &= - \sqrt{n+1}\psi_{n+1}(v) \label{recursive-AW-1},\\
    v\psi_{n}(v) &= \sqrt{n+1} \psi_{n+1}(v) + \sqrt{n} \psi_{n-1}(v) \label{recursive-AW-2}, 
\end{align}
where $\delta_{n,m}$ is the Kronecker delta function.
Inserting the spectral expansion in Eq.~\eqref{aw-expansion} in the linearized Vlasov equation~\eqref{linearized-vlasov} and employing the orthogonality in Eq.~\eqref{orthogonality-AW} and recursive properties in Eqns.~\eqref{recursive-AW-1}--\eqref{recursive-AW-2} results in 
\begin{equation}\label{linear-C-equation}
     -i \omega \hat{C}_{n} + \underbrace{ik \left( \sqrt{n+1}\hat{C}_{n+1} + \sqrt{n}\hat{C}_{n-1}\right)}_{\text{advection}} - \underbrace{\frac{ik \hat{\phi}}{\sqrt{2}} \delta_{n, 1}}_{\text{acceleration}}= 0,
\end{equation}
with the convention of $\hat{C}_{<0} = 0$ and the closure by truncation $\hat{C}_{N_{v}} = 0$, such that the closure term introduces an error only in the advection term $v\partial_{x} f$. We later consider other closure ideas in section~\ref{sec:nonlocal_closure}. Then, in vector form, Eq.~\eqref{linear-C-equation} becomes
\begin{equation}\label{linear-C-equation-vector-form}
    \omega \bar{C} = k \bar{\bar{{A}}}\bar{C} + k\hat{\phi}\bar{B} \qquad \Rightarrow \qquad \left(\xi \mathbb{I}_{N_{v}} - \frac{k}{\sqrt{2}|k|} \bar{\bar{A}}\right) \bar{C} = \frac{k\hat{\phi}}{\sqrt{2} |k|} \bar{B},
\end{equation}
where $\bar{C} \coloneqq [\hat{C}_{0}, \hat{C}_{1}, \hat{C}_{2}, \ldots, \hat{C}_{N_{v}-1}]^{\top} \in \mathbb{R}^{N_{v}}, \bar{B} \coloneqq [0 \Hquad -\frac{1}{\sqrt{2}} \Hquad 0 \ldots 0]^\top \in \mathbb{R}^{N_{v}}, \mathbb{I}_{N_{v}} \in \mathbb{R}^{N_{v} \times N_{v}}$ is the identity matrix, the variable $\xi$ is defined in Eq.~\eqref{response-function-definition}, and 
\begin{equation}\label{advection-operator}
    \bar{\bar{A}} \coloneqq \begin{bmatrix}
    0& \sqrt{1} & 0 & \ldots & & & \\
    \sqrt{1}& 0& \sqrt{2} & 0 &\ldots && \\
     0 & \sqrt{2} & 0 & \sqrt{3} & \\
     & & & \ddots & &   \\
     & \ldots & 0& \sqrt{N_{v}-2} & 0 & \sqrt{N_{v}-1} \\
     & & \ldots &0 & \sqrt{N_{v}-1}& 0 
    \end{bmatrix} \in \mathbb{R}^{N_{v} \times N_{v}}.
\end{equation}
Therefore, inserting Eq.~\eqref{linear-C-equation-vector-form} in the response function definition in Eq.~\eqref{response-function-definition} results in the following Hermite approximation to the response function 
\begin{align}
    R^{\mathrm{aw}}_{N_{v}}(\xi) &\coloneqq \frac{\hat{n}}{\hat{\phi}} = \frac{\sqrt{2}\hat{C}_{0}}{\hat{\phi}} = \frac{\sqrt{2}}{\hat{\phi}} [1\Hquad 0 \Hquad 0 \ldots 0 ] \left(\xi \mathbb{I}_{N_{v}} -\frac{k}{\sqrt{2}|k|}\bar{\bar{A}}\right)^{-1}\left(\frac{k\hat{\phi}}{\sqrt{2}|k|}\bar{B}\right) \label{approximate-response-function}\\
    &=\frac{-k}{\sqrt{2}|k|} [1\Hquad 0 \Hquad 0 \ldots 0 ]\left(\xi \mathbb{I}_{N_{v}} -\frac{k}{\sqrt{2}|k|}\bar{\bar{A}}\right)^{-1}\left[ 0 \Hquad 1 \Hquad 0 \ldots 0 \right]^{\top}. \nonumber
\end{align}
We compute the response function symbolically, which limits the computation to a modest velocity resolution. For example, some analytical results are as follows
\begin{align*}
    R^{\mathrm{aw}}_{3}(\xi) &= \frac{-1}{2 \xi^{2} - 3}, \\
    R^{\mathrm{aw}}_{4}(\xi) &= \frac{3 - 2 \xi^{2}}{4 \xi^{4} - 12 \xi^{2} + 3}, \\
    R^{\mathrm{aw}}_{5}(\xi) &= \frac{7 - 2 \xi^{2}}{4 \xi^{4} - 20 \xi^{2} + 15}, \\
    R^{\mathrm{aw}}_{6}(\xi) &= \frac{- 4 \xi^{4} + 24 \xi^{2} - 15}{8 \xi^{6} - 60 \xi^{4} + 90 \xi^{2} - 15}.
\end{align*}
The asymptotic results for $|\xi| \ll 1$ (Taylor/Maclaurin series) are 
\begin{align*}
    R^{\mathrm{aw}}_{3}(\xi) &= \frac{1}{3} + \frac{2 \xi^2}{9} + \frac{4 \xi^4}{27} + \mathcal{O}\left(\xi^6\right),\\
    R^{\mathrm{aw}}_{4}(\xi) &= \boxed{1} + \frac{10 \xi^2}{3} + 12 \xi^4 + \mathcal{O}\left(\xi^6\right),\\
    R^{\mathrm{aw}}_{5}(\xi) &= \frac{7}{15} + \frac{22 \xi^2}{45} + \frac{356 \xi^4}{675} + \mathcal{O}\left(\xi^6\right),\\
    R^{\mathrm{aw}}_{6}(\xi) &= \boxed{1} + \frac{22 \xi^2}{5} + \frac{68 \xi^4}{3} + \mathcal{O}\left(\xi^6\right),
\end{align*}
and for $|\xi| \gg 1$ (Laurent series) are
\begin{align*}
    R^{\mathrm{aw}}_{3}(\xi) &= {\boxed{-\frac{1}{2 \xi^2} - \frac{3}{4 \xi^4}}} - \frac{9}{8 \xi^6} + \mathcal{O}\left(\frac{1}{\xi^{8}}\right),\\
    R^{\mathrm{aw}}_{4}(\xi) &= {\boxed{-\frac{1}{2 \xi^2} -\frac{3}{4 \xi^4} -\frac{15}{8 \xi^6}}} - \frac{81}{16 \xi^8} + \mathcal{O}\left(\frac{1}{\xi^{10}}\right),\\
    R^{\mathrm{aw}}_{5}(\xi) &= \boxed{-\frac{1}{2 \xi^2} - \frac{3}{4 \xi^4} -\frac{15}{8 \xi^6} - \frac{105}{16 \xi^8}}- \frac{825}{32 \xi^{10}}+ \mathcal{O}\left(\frac{1}{\xi^{12}}\right),\\
    R^{\mathrm{aw}}_{6}(\xi) &=  {\boxed{-\frac{1}{2 \xi^2} - \frac{3}{4 \xi^4} - \frac{15}{8 \xi^6} - \frac{105}{16 \xi^8} - \frac{945}{32 \xi^{10}} }} -\frac{9675}{64 \xi^{12}} + \mathcal{O}\left(\frac{1}{\xi^{14}}\right).
\end{align*}
The terms in the box above match the analytic response function coefficients at the respective asymptotic limit; see Eqns.~\eqref{analytic-response-asymptotics-low}--\eqref{analytic-response-asymptotics-high}. 

\begin{figure}
    \centering
    \includegraphics[width=0.6\linewidth]{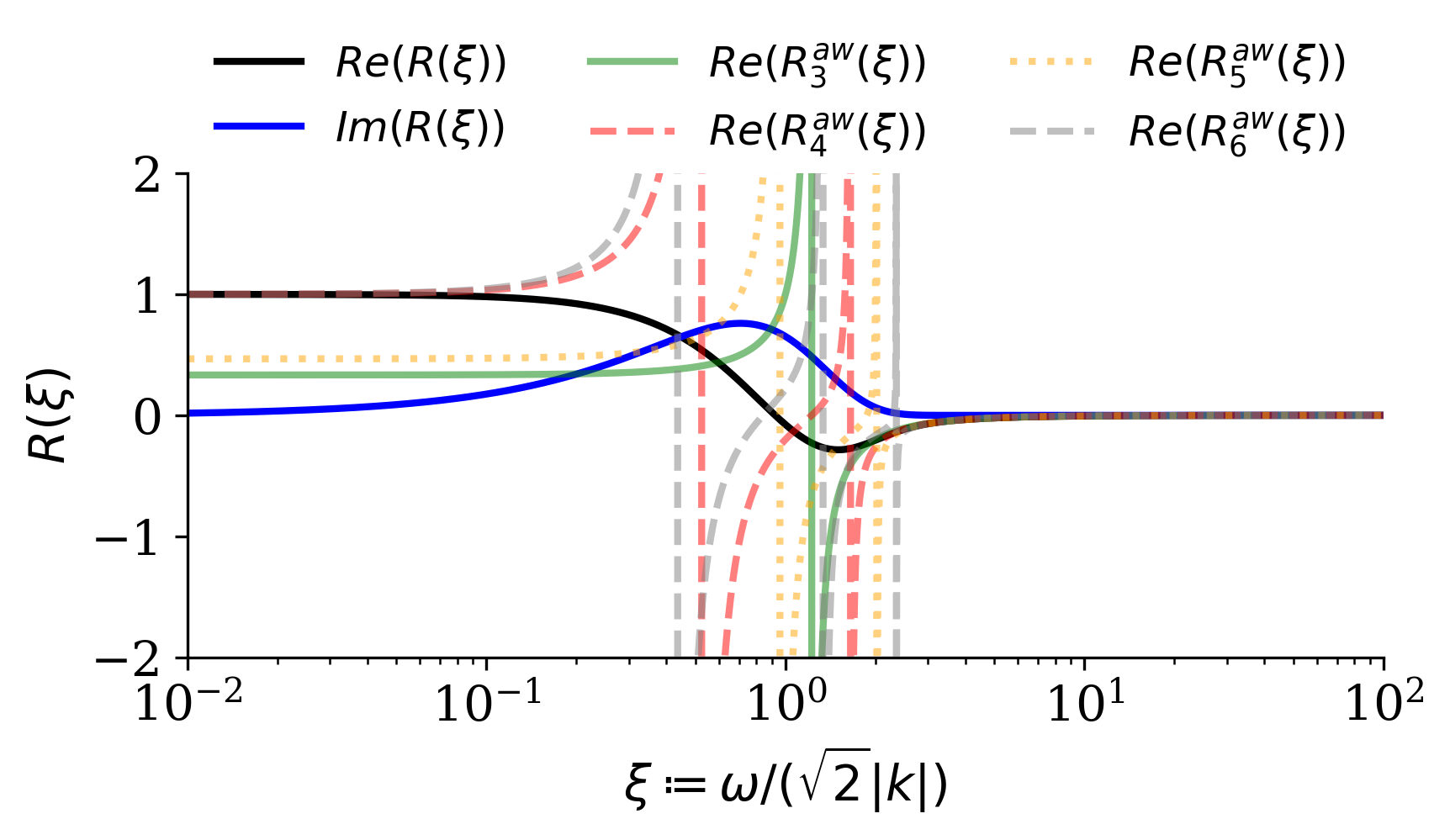}
    \caption{The collisionless Hermite response function approximation with closure by truncation and $N_{v}=3, 4, 5, 6$. As expected, the approximations are most accurate in the high-phase velocity region $|\xi| \gg 1$ as there are weak kinetic effects in this region. In the small phase velocity region $|\xi| \ll 1$, the even $N_{v}$ approximations are more accurate than the odd $N_{v}$ approximations (up to the zeroth order approximation).}
    \label{fig:response-function-AW-Hermite}
\end{figure}

The Hermite response function approximations have the correct asymptotic behavior in the fluid limit $|\xi| \gg 1$, i.e. when the phase velocity is much larger than the thermal velocity. In this region, increasing the number of moments improves the order of accuracy of the asymptotic expansion. However, in the adiabatic limit of $|\xi| \ll 1$, the accuracy of the approximation does not improve as we increase the number of Hermite moments~$N_{v}$. Additionally, models with even $N_{v}$ are more accurate in the $|\xi| \ll 1$ limit in comparison to models with odd $N_{v}$ (at least to the zeroth order approximation)~\cite{gillot_2021_bt}. The mathematical reasoning is that the eigenvalues of the skew-symmetric advection matrix $ik \bar{\bar{A}}$ defined in Eq.~\eqref{advection-operator} are purely imaginary in the even-dimension case, whereas, in the odd-dimension case, there is an additional unpaired zero eigenvalue. Figure~\ref{fig:response-function-AW-Hermite} shows the Hermite response function approximation for models with three to six moments. The Hermite approximations have purely real poles in the $|\xi| \sim 1$ region that cause the response function to grow unbounded. Furthermore, the kinetic response function $R(\xi)$ has both real and imaginary components for purely real $\xi$, whereas the approximation $R_{N_{v}}^{\mathrm{aw}}(\xi)$ is purely real for purely real $\xi$. A purely real response function indicates that the approximate discrete system cannot sustain a phase shift between the density and electrostatic potential perturbations. 

\section{Linear Spectral Deformation via Artificial Collisions, Filtering, and Nonlocal Closures}\label{sec:section-3}
This section outlines three approaches for modifying the linearized spectral coefficient evolution equation~\eqref{linear-C-equation} to accurately capture wave-particle resonance effects. These approaches comprise artificial collisions (section~\ref{sec:artificial_collisions}), filtering (section~\ref{sec:filtering}), and nonlocal closure (section~\ref{sec:nonlocal_closure}).
We compare the methods' ability to approximate the analytic linear response function in section~\ref{sec:response-function-comparison} and the dispersion relation in section~\ref{sec:eigenvalue-analysis}.

\subsection{Artificial Collisions}\label{sec:artificial_collisions}
Previous work by~\cite{grant_1967_hermite, camporeale_2016, delzanno_2015_jcp, koshkarov_2021, vencels_2015_sps, parker_2016_thesis, gajewski_1977_artificial_collisions, black_2013_spectra, pueschel_2010_recurrence, loureiro_2016_hermite} include artificial collisions to introduce velocity-space dissipation, which allows the solvers to recover the Landau root~\cite{landau_46} as a discrete eigenvalue of the discretized Vlasov-Poisson system. Hypercollisionality is similar in principle to hyperviscosity methods commonly used in spectral discretizations of hydrodynamic turbulence~\cite{passot_1988_hyperviscosity, boyd_2001_spectral}, as both introduce higher-order dissipation to manage small-scale features and prevent unphysical recurrence or energy accumulation. \citet{joyce_1971_hermite} introduce a hypercollisional operator based on powers of the LB~\cite{lenard_bernstein_1958_collisions} collisional operator, i.e.
\begin{equation}\label{hypercollisions}
    \mathcal{C}_{\mathrm{hyper}}(f) \coloneqq \nu \mathcal{D}^{2\alpha-1}\tilde{\mathcal{D}}^{2\alpha-1} f \qquad \mathrm{with} \qquad \mathcal{D}f \coloneqq \frac{\partial}{\partial v}f, \qquad \tilde{\mathcal{D}}f\coloneqq \left(\frac{\partial }{\partial v} + v\right)f, \qquad \mathcal{C}_{\mathrm{hyper}}(f_{0}) = 0,
\end{equation}
where $\nu \in \mathbb{R}_{+}$ is the normalized collisional frequency and $\alpha \in \mathbb{N}_{\geq 1}$ controls the order of dissipation. 
The artificial collision operator is added to the right-hand side of Eq.~\eqref{vlasov-continuum}. If $\alpha \geq 2$, the hypercollisional operator conserves mass, momentum, and energy~\cite{funaro_2021_hyper}.
The Hermite basis function $\psi_{n}(v)$ defined in Eq.~\eqref{hermite-basis-function} is an eigenfunction of the collisional operator with eigenvalue $-\nu n!/(n-2\alpha+1)!$. 
The operator in Eq.~\eqref{hypercollisions} recovers the LB~\cite{lenard_bernstein_1958_collisions} collisional operator when $\alpha=1$. Similarly, ~\citet{camporeale_2016} adopted a hypercollisional operator that corresponds to the operator in Eq.~\eqref{hypercollisions} with $\alpha=2$ (up to a normalizing factor), which was adopted in various studies, e.g.~\cite{delzanno_2015_jcp, koshkarov_2021, chapurin_2024_hybrid, pagliantini_2023_adaptive, pagliantini_2023_rk, roytershetyn_2018_sps}.
Following the recursive Hermite properties in Eq.~\eqref{recursive-AW-1}--\eqref{recursive-AW-2}, see~\citet{funaro_2021_hyper}, the normalized artificial collisions operator in Hermite space acts as
\begin{equation}\label{hypercollisions-normalized}
    \mathcal{C}_{\mathrm{hyper}}(\hat{C}_{n})  = -\nu \underbrace{\frac{n!}{(n-2\alpha+1)!}
    \underbrace{\frac{(N_{v}-2\alpha)!}{(N_{v}-1)!}}_{\text{normalizing factor}}}_{\text{damping rate } \propto n^{2\alpha-1}}\hat{C}_{n}
\end{equation}
such that the last Hermite moment $\hat{C}_{N_{v}-1}$ damping rate is $\nu \in \mathbb{R}_{+}$. The hypercollisional operator should always be applied in a convergence sense, ensuring that it does not substantially alter the collisionless physics of interest. An important feature of the operator is that the damping rate increases with larger $n$ Hermite coefficients, scaling as $n^{2\alpha - 1}$.

\subsection{Filtering}\label{sec:filtering}
\begin{figure}
    \centering
    \begin{subfigure}{0.49\textwidth}
        \caption{Damping rate $N_{v}=20$}
        \includegraphics[width=\textwidth]{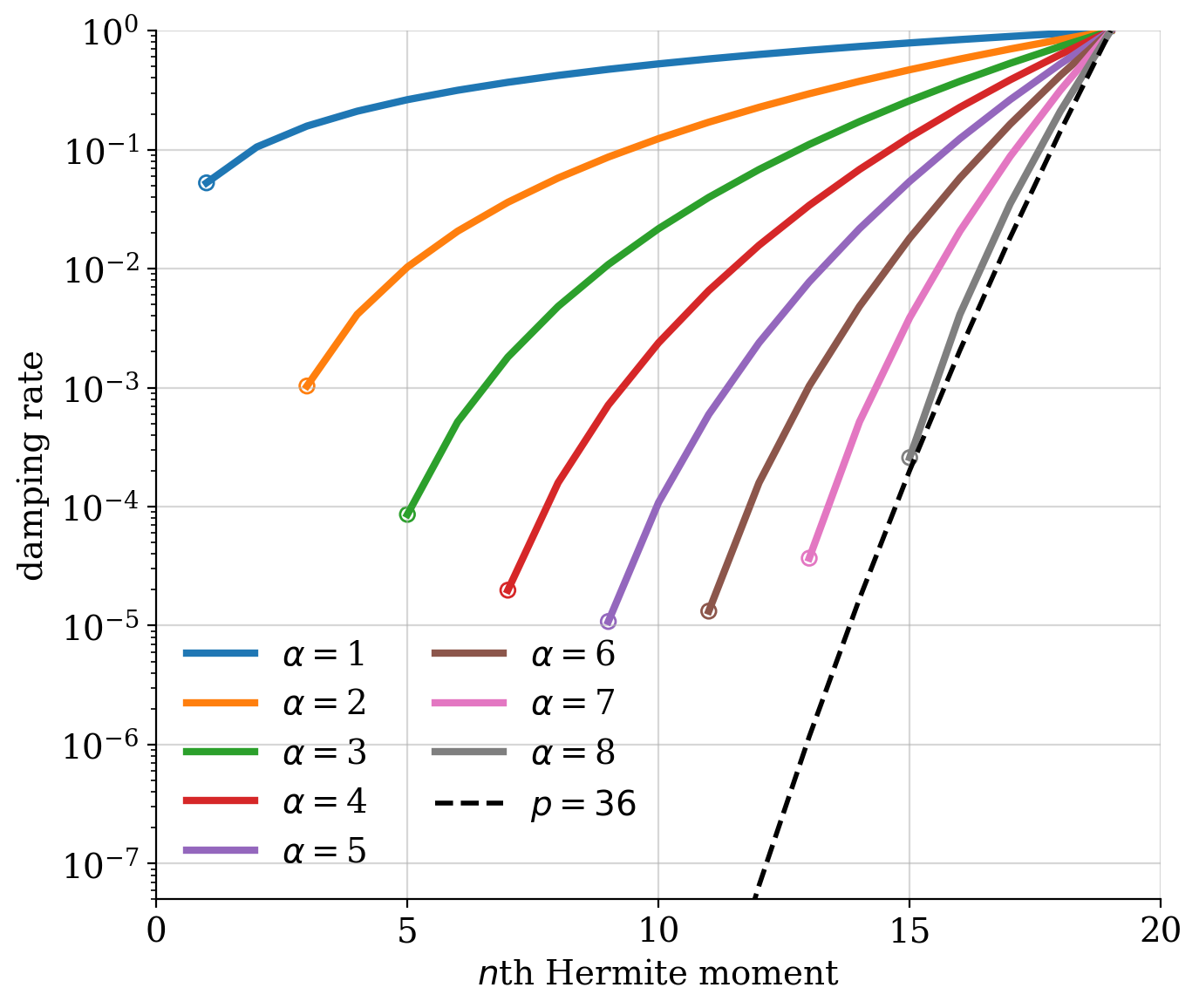}
    \end{subfigure}
    \begin{subfigure}{0.49\textwidth}
        \caption{Damping rate $N_{v}=10^3$}
        \includegraphics[width=\textwidth]{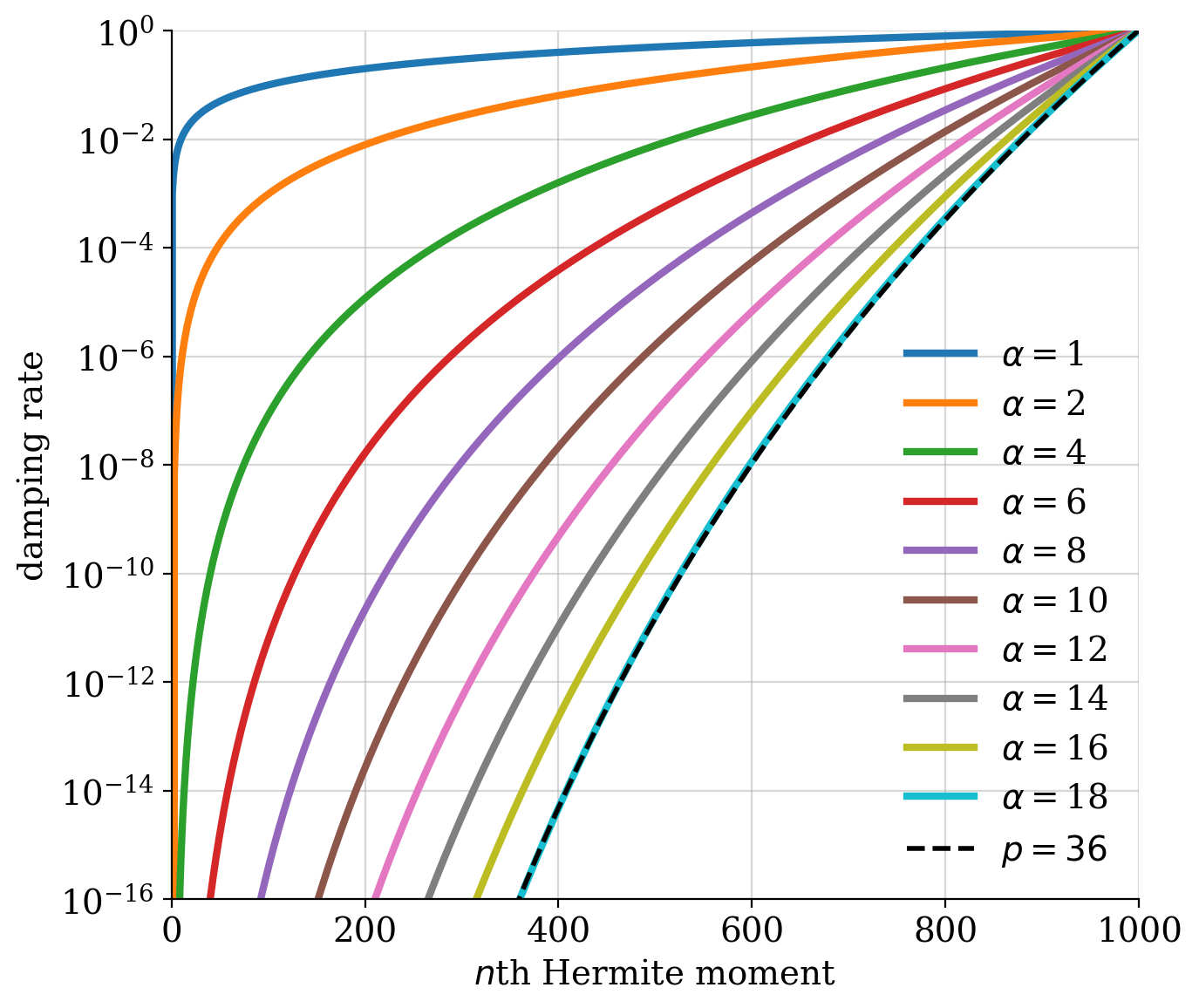}
    \end{subfigure}
    \caption{Damping rates for artificial collisions at various values of $\alpha$ (solid lines), as described in Eq.~\eqref{hypercollisions-normalized}, compared to the filtering approach with $p=36$ (dashed line), as outlined in Eq.~\eqref{filtering-normalized}. For $N_{v} = 20$, corresponding to subfigure~(a), artificial collisions with $\alpha = 8$ produce a damping rate comparable to filtering. Similarly, for $N_v = 10^3$, corresponding to subfigure~(b), artificial collisions with $\alpha = 18$ yield a damping rate similar to filtering. }
    \label{fig:damping_rate}
\end{figure}

The low-pass exponential filter introduced in~\citet{cai_2018_suppression} for the Hermite system, based on \citet{hou_li_2007_filter} filter, modifies the Hermite moments as follows
\begin{equation*}
    \hat{C}_{n} \rightarrow \hat{C}_{n} \sigma\left(\frac{n}{N_{v}-1}\right) \qquad \mathrm{and} \qquad \sigma\left(\frac{n}{N_{v}-1}\right) \coloneqq \exp \left(- \chi {\left(\frac{n}{N_{v}-1}\right)}^{p}\right).
\end{equation*}
This filter corresponds to computing the solution to the modified equation, where the following term is added to the right-hand side of the Vlasov equation~\eqref{vlasov-continuum}:
\begin{equation*}
    \mathcal{C}_{\mathrm{HouLi}}(f) = \chi \frac{(-1)^{p+1}}{\Delta t (N_{v}-1)^{p}} \hat{\mathcal{D}}^{p} f, \qquad \mathrm{with} \qquad \hat{\mathcal{D}}f \coloneqq \frac{\partial}{\partial v} \left[\exp\left(-\frac{v^2}{2}\right) \frac{\partial}{\partial v} \left( \exp \left( \frac{v^2}{2} \right) f\right)\right], \qquad \mathcal{C}_{\mathrm{HouLi}}(f_{0}) = 0,
\end{equation*}
where $\Delta t \in \mathbb{R}_{>0}$ is the time step, and  $\chi \in \mathbb{R}$ and $p \in \mathbb{N}_{>0}$ are tunable parameters of the exponential filter. Following the recursive Hermite properties in Eq.~\eqref{recursive-AW-1}--\eqref{recursive-AW-2}, the filtering operator in Hermite space acts as
\begin{equation}\label{filtering-normalized}
    \mathcal{C}_{\mathrm{HouLi}}(\hat{C}_{n})  = -\frac{\chi}{\Delta t} \underbrace{\left(\frac{n}{N_{v} - 1}\right)^{p}}_{\text{damping rate } \propto n^{p}}\hat{C}_{n}.
\end{equation}
\citet{hou_li_2007_filter} use \( p = 36 \) for spatial Fourier smoothing in hydrodynamic simulations, a parameter later adopted for AW Hermite-based solvers (see, e.g., \cite{parker_2015_hermite, parker_2016_thesis, cai_2018_suppression, di_2019_filtered, bessemoulin_2022_dg}). In this paper, we adopt the same scaling $p=36$.
We note that filtering and artificial collisions are fundamentally equivalent methods in the sense that a combination of parameters can be found so that both methods yield very similar damping rates. This is demonstrated in Fig.~\ref{fig:damping_rate}. 
Specifically, for $N_{v}=20$, artificial collisions with $\alpha=8$ have a damping rate comparable to that of filtering. When $N_{v}=10^3$, artificial collisions with $\alpha=18$ show a damping rate similar to that of filtering. 
In \ref{sec:Appendix-A}, we discuss another commonly used filter developed by~\citet{klimas_1987_filter}. The Klimas filter does not correspond to a true velocity space dissipation and hence the Landau root is not a discrete mode of the Klimas filtered AW Hermite discretization with closure by truncation.

\subsection{Hermite Hammett-Perkins Nonlocal Linear Closures}\label{sec:nonlocal_closure}
~\citet{hammet_1990_closure} developed a nonlocal linear closure that accounts for collisionless Landau damping effects in a fluid model. This closure approximates the highest moment in the fluid hierarchy by a linear combination of lower-order moments in Fourier space. 
The transformation between Hermite moments and fluid moments is nonlinear and invertible, such that a linear fluid moment closure corresponds to a nonlinear Hermite moment closure, and \textit{vice versa}~\cite[\S 3]{smith_1997_closure}. The transformation from fluid moments to Hermite moments is 
\begin{alignat*}{3}
    \hat{n} &= \sqrt{2} \hat{C}_{0} \qquad &&\Longrightarrow \qquad \hat{C}_{0} = \frac{\hat{n}}{\sqrt{2}}\\
    \hat{u} &= \frac{\hat{C}_{1}}{\hat{C}_{0}} \qquad &&\Longrightarrow \qquad \hat{C}_{1} = \frac{ \hat{u} \hat{n}}{\sqrt{2}} \\
    \hat{p} &= 2\hat{C}_{2} + \sqrt{2}\hat{C}_{0} - \frac{\sqrt{2}\hat{C}_{1}^2}{\hat{C}_{0}}\qquad &&\Longrightarrow \qquad \hat{C}_{2} = \frac{\hat{p} - \hat{n}+ \hat{u}^2 \hat{n}}{2} \\
     \hat{q} &= 2\sqrt{3} \hat{C}_{3} - \frac{6\hat{C}_{2} \hat{C}_{1}}{\hat{C}_{0}} + \frac{2\sqrt{2}\hat{C}_{1}^3}{\hat{C}_{0}^2} \qquad &&\Longrightarrow \qquad \hat{C}_{3} =\frac{\hat{q} + 3\hat{u}  \left(\hat{p} - \hat{n} \right)  + \hat{u}^3 \hat{n} }{2\sqrt{3}} 
\end{alignat*}
where the first four fluid moments are density $\hat{n}$, bulk velocity $\hat{u}$, pressure $\hat{p}$, and heat flux $\hat{q}$:
\begin{equation*}
    \hat{n} \coloneqq \int_{\mathbb{R}} \hat{f}(v) \mathrm{d} v, \qquad
    \hat{n}\hat{u} \coloneqq \int_{\mathbb{R}}v \hat{f}(v) \mathrm{d} v, \qquad
    \hat{p} \coloneqq \int_{\mathbb{R}} (v-\hat{u})^2 \hat{f}(v) \mathrm{d} v, \qquad
    \hat{q} \coloneqq \int_{\mathbb{R}} (v-\hat{u})^3 \hat{f}(v) \mathrm{d} v.
\end{equation*}
\citet{hammet_1990_closure} derived a closure for three- and four-moment fluid models, e.g. the three-moment closure relates the heat flux to the pressure and density as follows:
\begin{equation*}
    \hat{q} = -i \frac{2\sqrt{2}}{\sqrt{\pi}} \frac{k}{|k|} \left(\hat{p} - \hat{n}\right).
\end{equation*}
In Hermite space, this leads to a nonlinear closure expressed as $\hat{C}_{3} = \mathcal{F}(\hat{C}_{2}, \hat{C}_{1}, \hat{C}_{0})$. By neglecting quadratic and higher-order perturbations, the first-order approximation yields the following closure
\begin{equation*}
    \hat{C}_{3} = -i\frac{2\sqrt{2}}{\sqrt{3\pi}} \frac{k}{|k|} \hat{C}_{2}.
\end{equation*}

Following the work by~\citet{smith_1997_closure, hammett_1993_closure}, we generalize the mathematical approach to higher $N_{v}$ dimensions for a nonlocal linear closure of the form 
\begin{equation}\label{nonlocal-closure-form}
    \hat{C}_{N_{v}} = \sum_{n=N_{v} - N_{m}}^{N_{v}-1} i \mu_{n} \frac{k}{|k|} \hat{C}_{n}.
\end{equation}
The number of lower order Hermite moments used to close the equations, which corresponds to the number of adjustable parameters in the linear closure, is denoted by $N_{m} \in \mathbb{N}_{>0}$.
The linear closure coefficients $\mu_{n} \in \mathbb{C}$ are chosen by matching the approximate response function to the analytic response function at the asymptotic limit of $|\xi| \ll 1$, see Eq.~\eqref{analytic-response-asymptotics-low}. The linear closure retains the same order of accuracy at the fluid limit $\xi \gg 1$ as the closure by truncation. Since we solve for the approximate response function symbolically, computational limitations require keeping $N_{m}$ relatively small. To the best of our knowledge, this study is the first to interpret the nonlocal closure approach as a method to mitigate recurrence and suppress filamentation artifacts in kinetic simulations. The concept of nonlocal Hermite closure is similar to the approach proposed by \citet{eliasson_2001_outflow}, which introduced absorbing velocity boundary conditions in Fourier velocity discretizations to reduce recurrence effects. 

The closure is nonlocal in space, since for example the term $ik/|k| g(k)$ in configuration space becomes $\int_{0}^{\ell}[g(x + y) - g(x- y)]/y \mathrm{d} y$, where $g(x)$ is an arbitrary function and $x \in [0, \ell]$. For practical computational purposes, local (e.g. finite-difference) numerical solvers can approximate the nonlocal closure as a sum of Lorentzians in Fourier space which solves a modified Helmholtz equation in configuration space $x$, see~\citet{dimits_2014} for more details.

\subsection{Linear Response Function Approximation}\label{sec:response-function-comparison}
The procedure to derive the approximate response function with artificial collisions, filtering, or a nonlocal closure approach follows the same formula as Eq.~\eqref{approximate-response-function}. The only difference is in the advection matrix $\bar{\bar{A}} \in \mathbb{C}^{N_{v} \times N_{v}}$ which is defined in Eq.~\eqref{advection-operator}. For artificial collisions and filtering, the diagonal elements of the advection matrix are adjusted to match the corresponding dissipation operator. For the nonlocal closure, modifications are made to the entries in the last row of the advection matrix. 

\begin{figure}
    \centering
    \begin{subfigure}{0.45\textwidth}
        \caption{Convergence rate comparison}
        \label{fig:convergence-rate-comparison}
        \includegraphics[width=\textwidth]{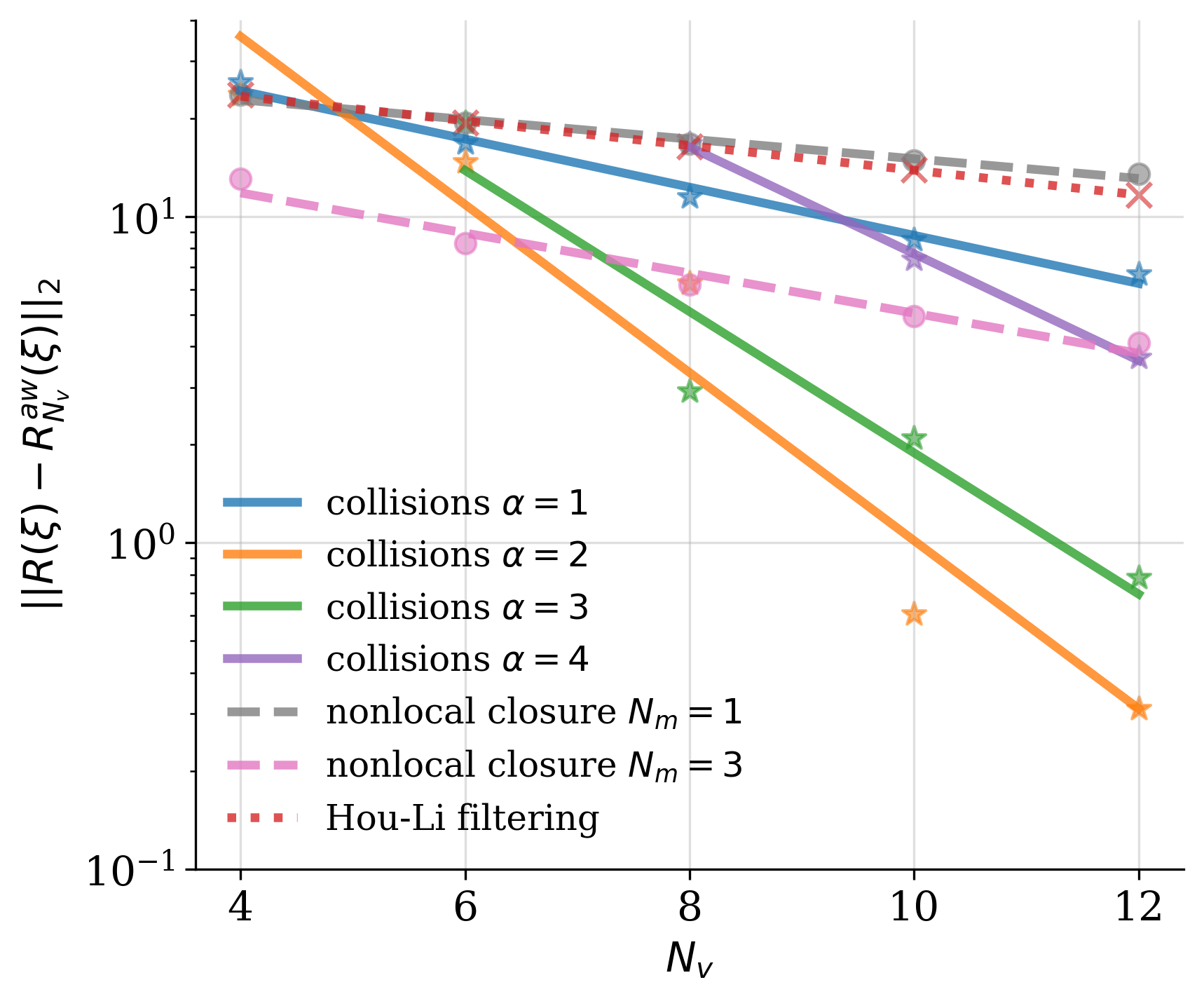}
    \end{subfigure}
    \begin{subfigure}{0.45\textwidth}
        \caption{Absolute error for $N_{v}=12$}
        \label{fig:response-function-absolute-error-Nv-12}
        \includegraphics[width=\textwidth]{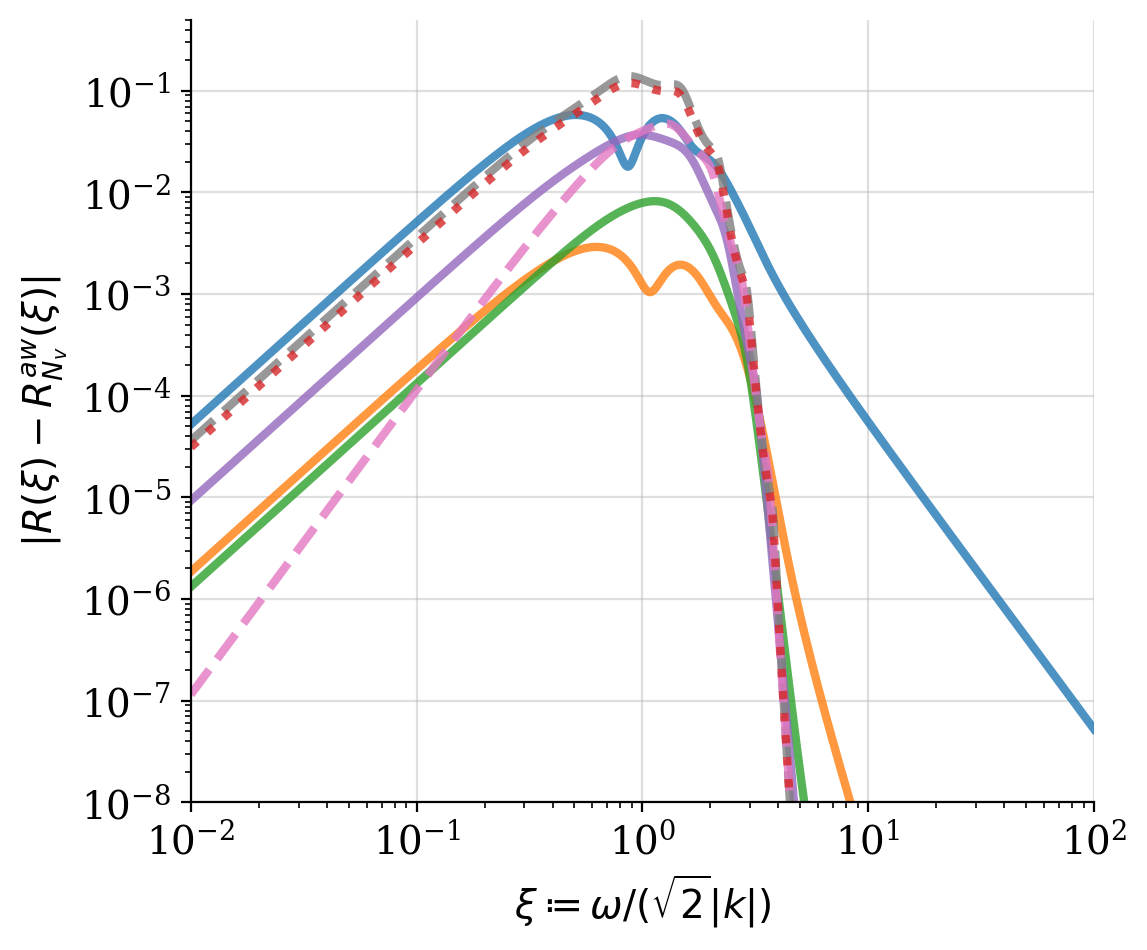}
    \end{subfigure}
    \caption{The approximate linear kinetic response function showing (a) convergence rate and (b) absolute error for $N_{v}=12$. Results are compared across the artificial collisions (solid lines), nonlocal closure (dashed lines), and filtering (dotted line) approaches. Artificial collisions with $\alpha=2$ (as in \citet{camporeale_2016}) and $\alpha=3$ exhibit the highest accuracy and fastest convergence toward the analytic response function.}
\end{figure}

The convergence rates of the different methods are shown 
in Figure~\ref{fig:convergence-rate-comparison}. The absolute error of the response function is evaluated using the $L_{2}$ norm on a log-uniform grid of $\xi \in [10^{-2}, 10^{2}]$ with $10^{5}$ nodes (more nodes do not change the error significantly). The absolute error is plotted with the optimal tunable parameter of the method, such that the approximate response function matches the analytic response at the asymptotic limit of $|\xi| \ll 1$. We indicate the optimal parameters for each method in Table~\ref{tab:optimal_parameters}.  For artificial collisions, the tunable parameter is the artificial collisional rate $\nu \in \mathbb{R}_{+}$ in Eq.~\eqref{hypercollisions-normalized}. For the nonlocal closure with $N_{m}=1$, the tunable parameter is the coefficient in the linear relation $\mu_{n} \in \mathbb{C}$ in Eq.~\eqref{nonlocal-closure-form}. For filtering, the tunable parameter is $\chi/\Delta t \in \mathbb{R}_{+}$ in Eq.~\eqref{filtering-normalized}. There is not a clear trend in optimal parameter variation as a function of the velocity resolution $N_{v}$. Interestingly, the optimal closure coefficients with $N_{m}=3$, show that $\mu_{N_{v}-2} = 0$ for all considered even $N_{v}$, such that a nonlocal closure with $N_{m}=2$ and $N_{m}=1$ are equivalent. The approximate response function for both artificial collisions and filtering depends on the wavenumber $k$, i.e. $R^{aw}_{N_{v}}(\xi, k)$. We present the results with $k=1$, and additional tests (not shown here) indicate that varying $k$ does not significantly influence the convergence or accuracy of the methods. 
The most accurate methods are hypercollisions with $\alpha=2$ and $\alpha=3$. The hypercollisional operator with $\alpha=2$ converges about four times faster than that with $\alpha=1$ (LB collisions). Accuracy decreases when $\alpha=4$, indicating that $\alpha=2$ and $\alpha=3$ represent the optimal range for the order of artificial collisions.
Additionally, the nonlocal closure method with $N_{m}=3$ converges approximately twice as fast as with $N_{m}=1$. As shown by~\citet{smith_1997_closure}, the convergence rate of the nonlocal closure improves as $N_{m}$ increases. However, matching the closure coefficient becomes increasingly difficult as $N_{m}$ and $N_{v}$ increase since calculating the approximate response function requires a symbolic matrix inversion, see Eq.~\eqref{approximate-response-function}, with $N_{m}$ nonlinear solves to match to the asymptotic limit. 
The filtering approximation converges slowly, similar to the nonlocal closure method with $N_{m}=1$. 

Figure~\ref{fig:response-function-absolute-error-Nv-12} shows the absolute error of the approximate response function for $N_{v}=12$. In all cases, the largest error occurs in the wave-particle resonance region $\xi \sim 1$. However, these approximations show a significant improvement compared to closure by truncation without dissipation, as shown in Figure~\ref{fig:response-function-AW-Hermite}. In all approximations, the response function poles are complex, resulting in damping.
For artificial collisions, the results with $\alpha=2$ and $\alpha=3$ are most accurate at the wave-particle resonance region $\xi \sim 1$.
For nonlocal closures, as expected, increasing $N_{m}$, i.e., the number of coefficients matched in the limit of $|\xi| \ll 1$, improves the accuracy of the response function. 
Lastly, the filtering approximation error is similar to the nonlocal closure method with $N_{m}=1$.
Overall, given the constraints on the nonlocal closure method parameter $N_m$ imposed by symbolic computation limitations, hypercollisions with $\alpha=2$ or $\alpha=3$ converge the fastest and are the most accurate approaches for approximating the response function under limited velocity resolution.

\noindent
\begin{table}
\caption{The optimal tunable parameter of each method based on matching the asymptotic limit $\xi \ll 1$ (adiabatic limit). The three methods are artificial collisions in Eq.~\eqref{hypercollisions-normalized}, nonlocal closure in Eq.~\eqref{nonlocal-closure-form}, and filtering in Eq.~\eqref{filtering-normalized}. }
\centering
\SetTblrInner{rowsep=1pt}
\begin{tblr}{c |c |c | c| c| c| c | c }
$N_{v}$ & \shortstack{collisions \\ $\alpha=1$ \\ $\nu$ } & \shortstack{collisions\\ $\alpha=2$ \\ $\nu$}  & \shortstack{collisions \\ $\alpha=3$ \\ $\nu$} & \shortstack{collisions \\ $\alpha=4$ \\ $\nu$} & \shortstack{nonlocal closure \\ $N_{m}=1$ \\ $\mu_{N_{v}-1}$} & \shortstack{nonlocal closure \\ $N_{m}=3$ \\ $\mu_{N_{v}-1}, \mu_{N_{v}-2}, \mu_{N_{v}-3}$} & \shortstack{filtering \\ $\chi/\Delta t$} \\
\hline
$4$ & 1.69 & 1.88 & $\xmark$ & $\xmark$ & -0.93& -1.31, 0, -0.30& 1.88\\
\hline
$6$ & 1.74 & 9.69 & 2.35 & $\xmark$ & -0.96 & -1.37, 0, -0.37 & 2.35 \\
\hline
$8$ & 1.72 & 17.65 & 4.81 & 2.74 & -0.97 & -1.40, 0, -0.40 & 2.75\\
\hline
$10$ & 1.68 & 6.46 & 13.31 & 11.19 & -0.97& -1.42, 0, -0.42& 3.13\\
\hline 
$12$ & 1.64 & 6.63 & 19.83 & 24.51 &-0.98 & -1.44, 0, -0.44 & 3.51
\end{tblr}
\label{tab:optimal_parameters}
\end{table}

\subsection{Linear Eigenvalue Analysis}\label{sec:eigenvalue-analysis}
The previous section focused on analyzing the behavior of the response function for the Vlasov equation alone. We now investigate the ability of the discrete models to replicate the dynamics of the linearized Vlasov-Poisson system, using the Landau damping problem as a test case. We set $\ell=4\pi$ such that wavenumbers $k$ are integer multiples of $1/2$. The dispersion relation $\omega(k)$ for linear Landau damping is given by $k^{2} = -R(\xi)$. We adopt a more straightforward approach~\cite{grant_1967_hermite, pezzi_2016_recurrence, canosa_1974_recurrence} by numerically calculating the damping rate through the real component of the least damped eigenvalue of the discretized linear Vlasov-Poisson equations:
\begin{equation*}
\frac{\mathrm{d}\bar{C}}{\mathrm{d} t} = \bar{\bar{Q}} \bar{C}, \qquad \mathrm{with} \qquad 
\bar{\bar{Q}}  \coloneqq -ik \begin{bmatrix}
\frac{\eta_{0}}{ik} & 1 & 0 & \ldots & & & \\
1 + \frac{1}{k^2} & \frac{\eta_{1}}{ik} & \sqrt{2} & 0 &\ldots && \\
 0 & \sqrt{2} & \frac{\eta_{2}}{ik} & \sqrt{3} & \\
 & & & \ddots & &   \\
 & \ldots & 0& \sqrt{N_{v}-2} & \frac{\eta_{N_{v}-2}}{ik}  & \sqrt{N_{v}-1} \\
 & & \ldots & & \sqrt{N_{v}-1} & \frac{\eta_{N_{v}-1}}{ik} - i \frac{k}{|k|} \mu_{N_{v}-1}
\end{bmatrix} \in \mathbb{C}^{N_{v} \times N_{v}},
\end{equation*}
where
\begin{equation*}
    \eta_{n} \coloneqq \begin{cases}
     -\nu \frac{n!(N_{v}-2\alpha)!}{(n-2\alpha+1)! (N_{v}-1)!} &\text{if artificial collisions}, \\
    -\frac{\chi}{\Delta t} \left(\frac{n}{N_{v} - 1}\right)^{36} &\text{if Hou-Li filtering},\\
    0 &\text{otherwise}.
\end{cases}
\end{equation*}

\begin{figure}
    \centering
    \begin{subfigure}{0.32\textwidth}
        \caption{Nonlocal closure with $N_{m}=1$ \newline \citet{smith_1997_closure}}
        \includegraphics[width=\textwidth]{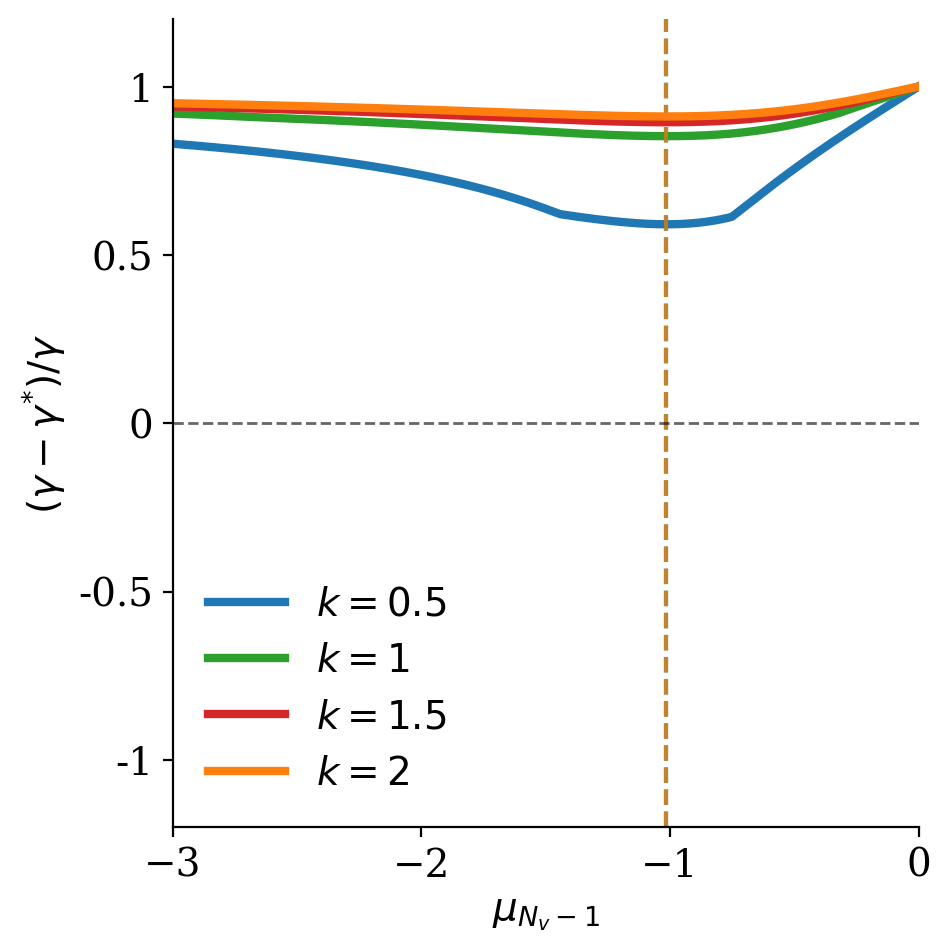}
    \end{subfigure}
    \begin{subfigure}{0.32\textwidth}
        \caption{Artificial collisions $\alpha =1$ \newline \citet{lenard_bernstein_1958_collisions} }
        \includegraphics[width=\textwidth]{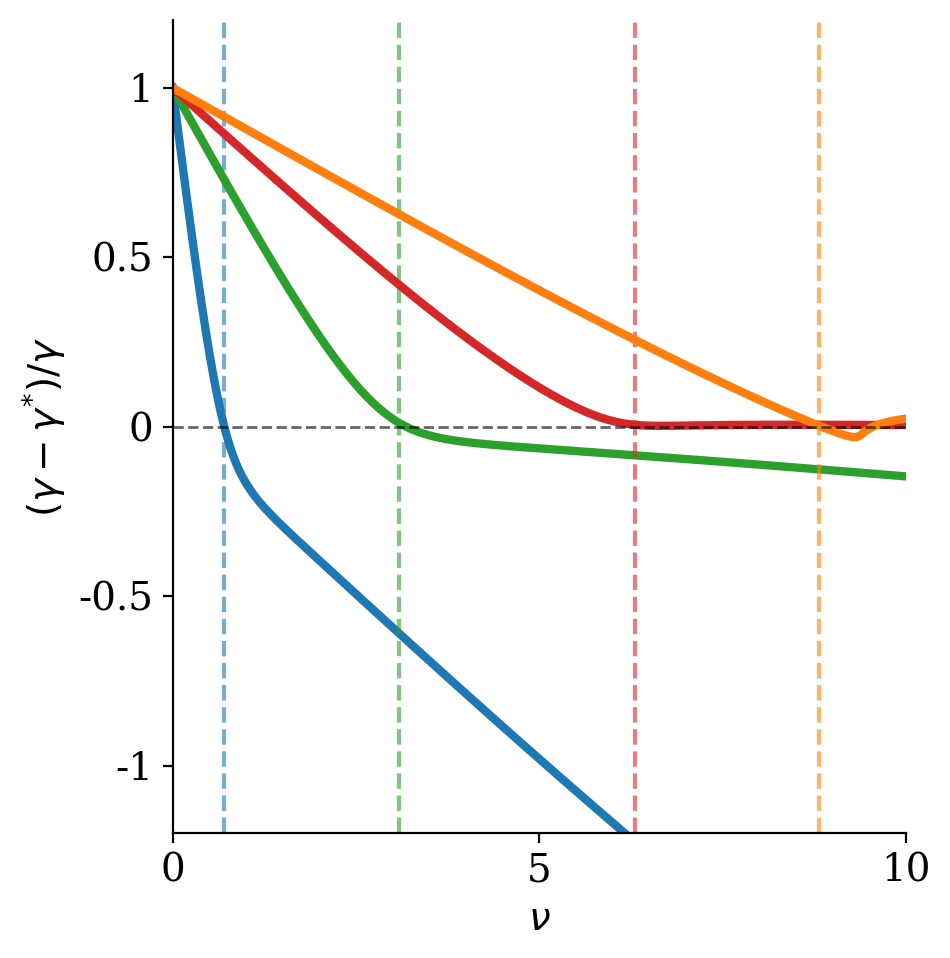}
    \end{subfigure}
    \begin{subfigure}{0.32\textwidth}
        \caption{Artificial collisions $\alpha =2$ \newline \citet{camporeale_2016}}
        \includegraphics[width=\textwidth]{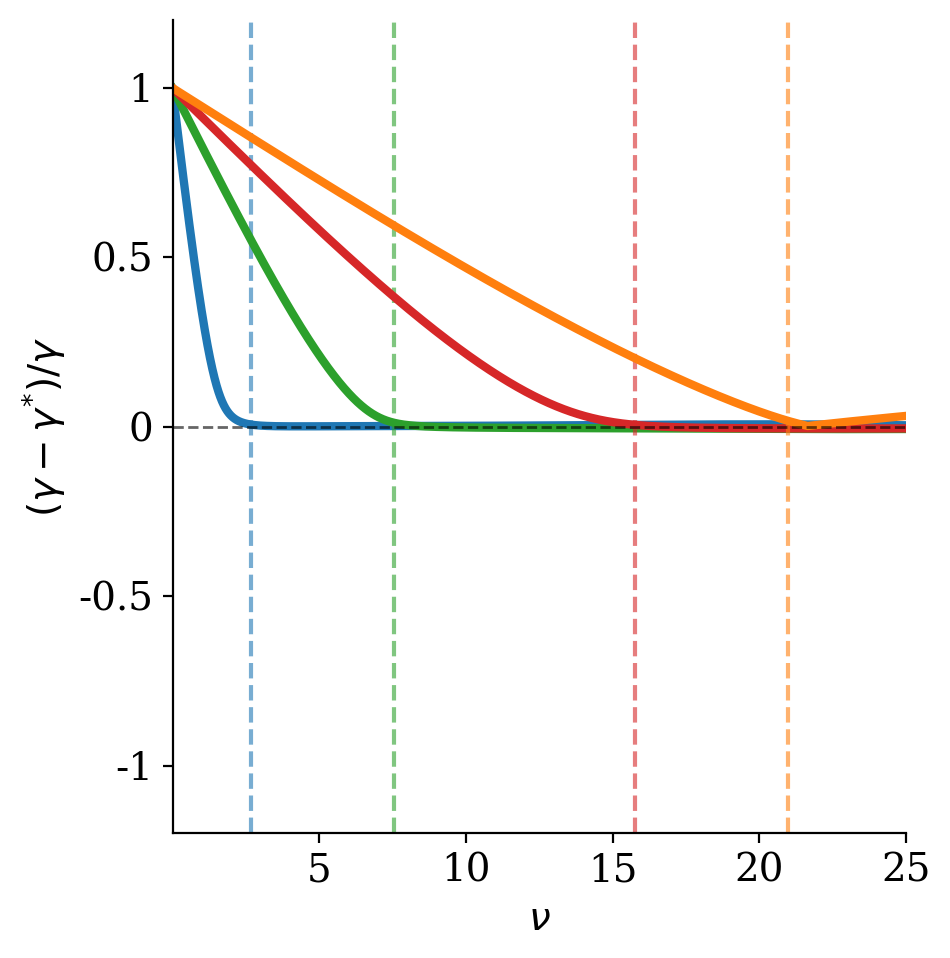}
    \end{subfigure}
    \begin{subfigure}{0.32\textwidth}
        \caption{Artificial collisions $\alpha =3$}
        \includegraphics[width=\textwidth]{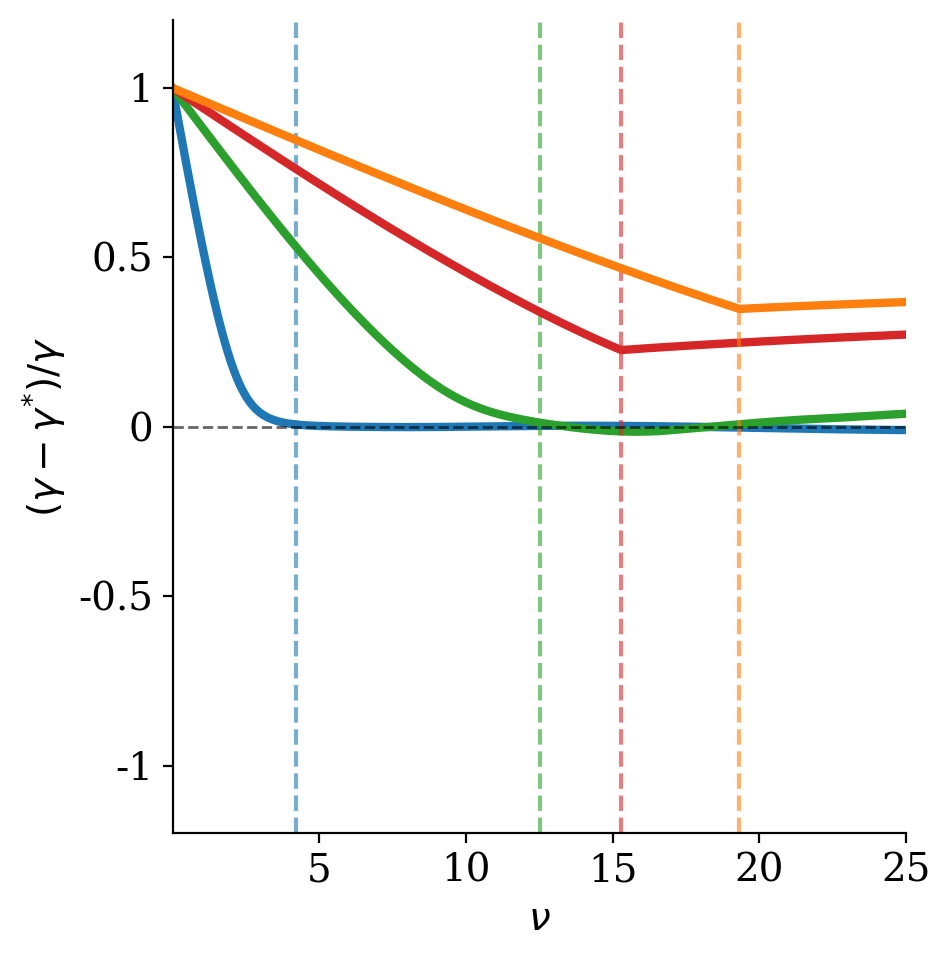}
    \end{subfigure}
    \begin{subfigure}{0.32\textwidth}
        \caption{Artificial collisions $\alpha =4$}
        \includegraphics[width=\textwidth]{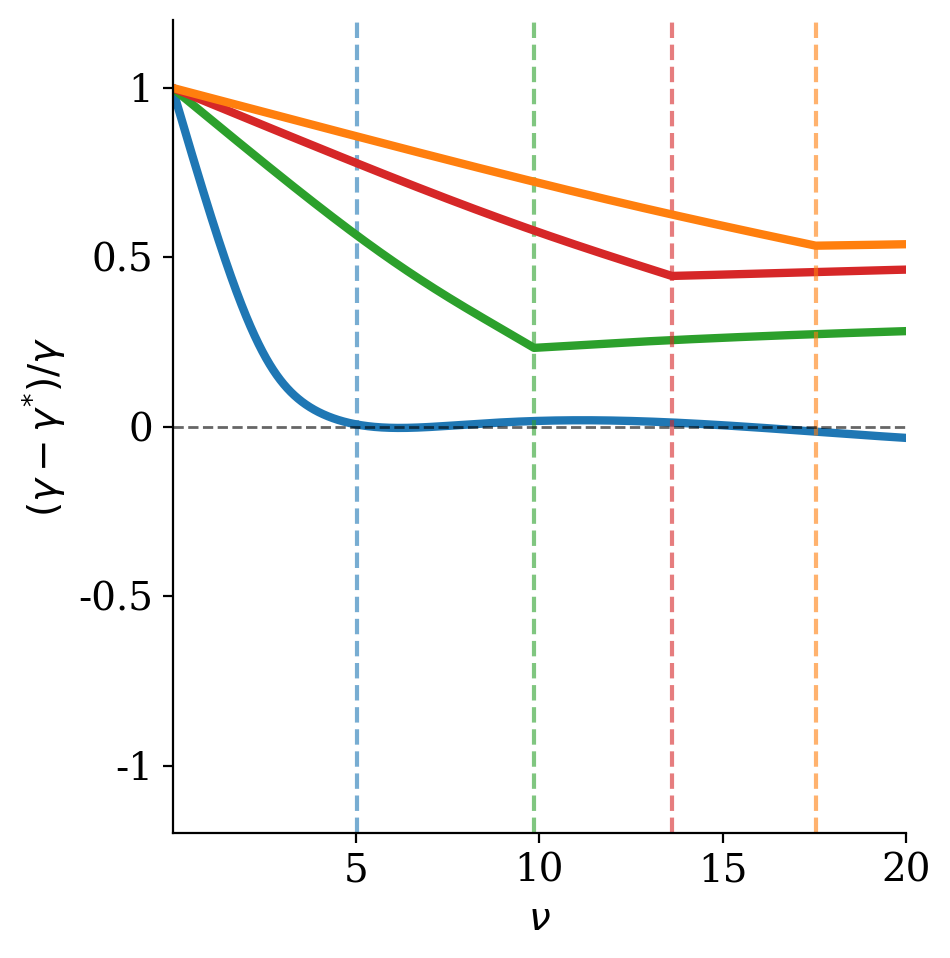}
    \end{subfigure}
        \begin{subfigure}{0.32\textwidth}
        \caption{\citet{hou_li_2007_filter} filter}
        \includegraphics[width=\textwidth]{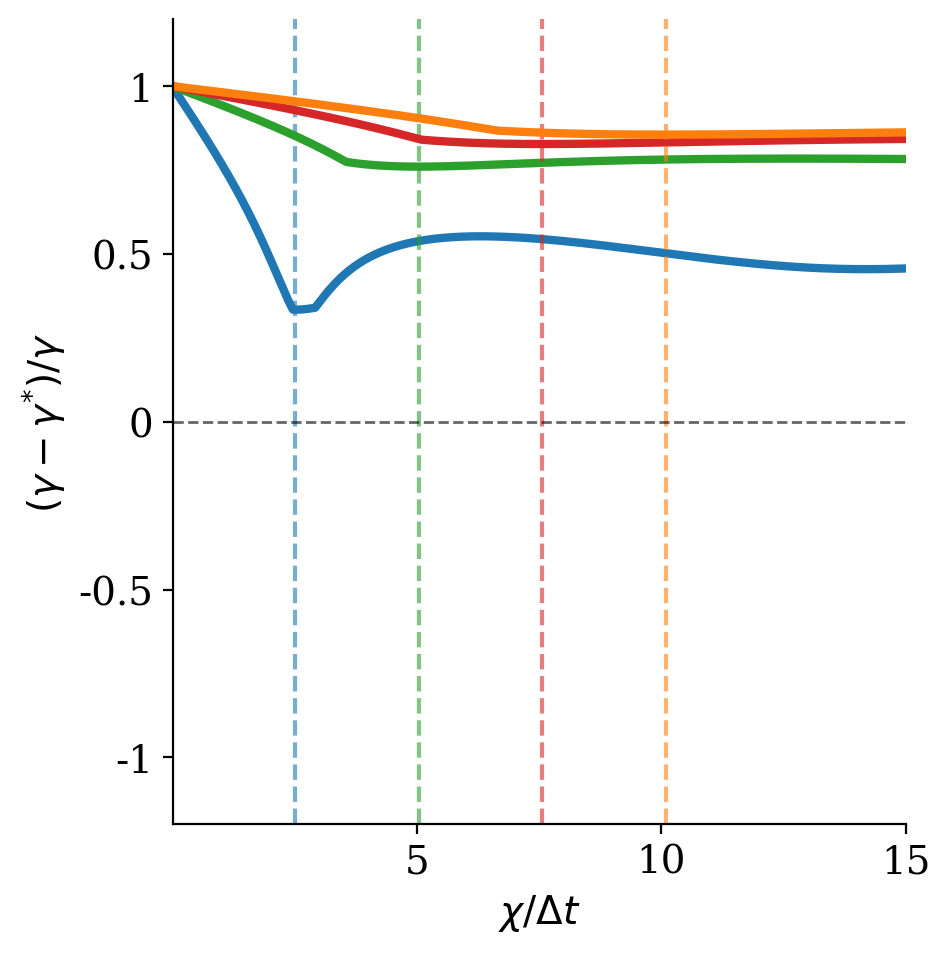}
    \end{subfigure}
    \caption{The relative error between the discrete system damping rate $\gamma^{*}$ and the analytic damping rate $\gamma$ for various spatial wavenumber $k \in \{0.5, 1, 1.5, 2\}$ with $N_{v}=20$. Positive values lead to underdamping of the modes, while negative values correspond to overdamping. When the free parameters of the methods are set to zero, we recover the collisionless with closure by truncation approximation which incorrectly predicts $\gamma^{*}=0$. Dashed vertical lines illustrate the optimal free parameter for each mode. The results show that only artificial collisions with $\alpha=2$ can recover the correct damping rate across a range of wavenumbers. }
    \label{fig:over-under-damping-error-20}
\end{figure}

Figure~\ref{fig:over-under-damping-error-20} shows the relative error of the Landau damping rate for various spatial wavenumbers as a function of the system's tunable parameter with $N_{v}=20$.
In this and the following sections, we restrict the nonlocal closure to $N_{m} = 1$ to avoid higher-dimensional parametric optimization and to constrain each method to a single free parameter.
The nonlocal closure underestimates the damping rate for all spatial wavenumbers, irrespective of the value of the tunable parameter, so that higher wavenumber modes are most severely underdamped. Furthermore, no value of the closure coefficient $\mu_{N_{v}-1}$ can spectrally alter the discrete system to accurately recover the correct damping rate. The same underdamping effect also applies to the filtering technique.
The LB operator, i.e. artificial collisions with $\alpha=1$, can replicate the correct damping for a single spatial mode, but not for a range of wavenumbers. As a result, if the system is initially excited with a range of modes, accurately capturing the damping rate of the highest wavenumber results in the LB operator overdamping all other modes, which is consistent with the results in~\citet{pezzi_2016_recurrence}. In contrast, setting $\alpha=2$ (as in \citet{camporeale_2016}) does not overdamp modes. As we increase $\alpha$ to $3$ and $4$, the higher wavenumber modes are underdamped, such that $\alpha=2$ provides the ideal balance.   

\begin{figure}
    \centering
    \begin{subfigure}{0.32\textwidth}
        \caption{Nonlocal closure with $N_{m}=1$ \newline \citet{smith_1997_closure}}
        \includegraphics[width=\textwidth]{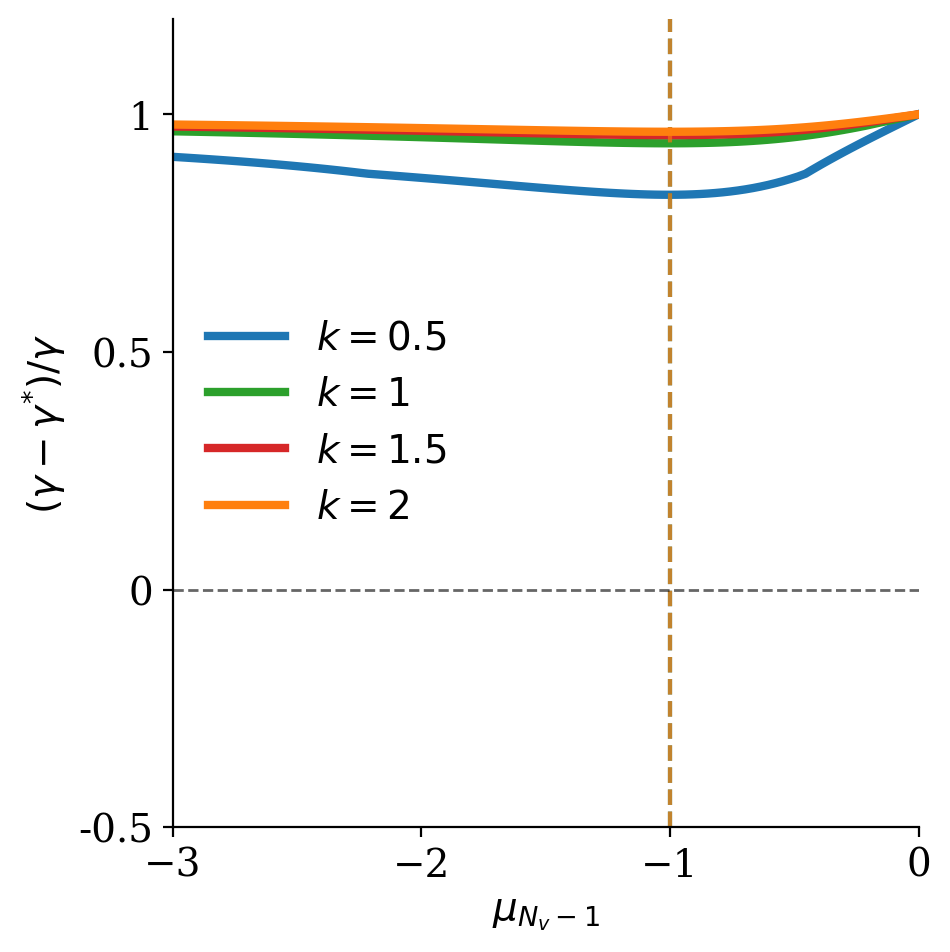}
    \end{subfigure}
    \begin{subfigure}{0.32\textwidth}
        \caption{Artificial collisions $\alpha =1$ \newline \citet{lenard_bernstein_1958_collisions} }
        \includegraphics[width=\textwidth]{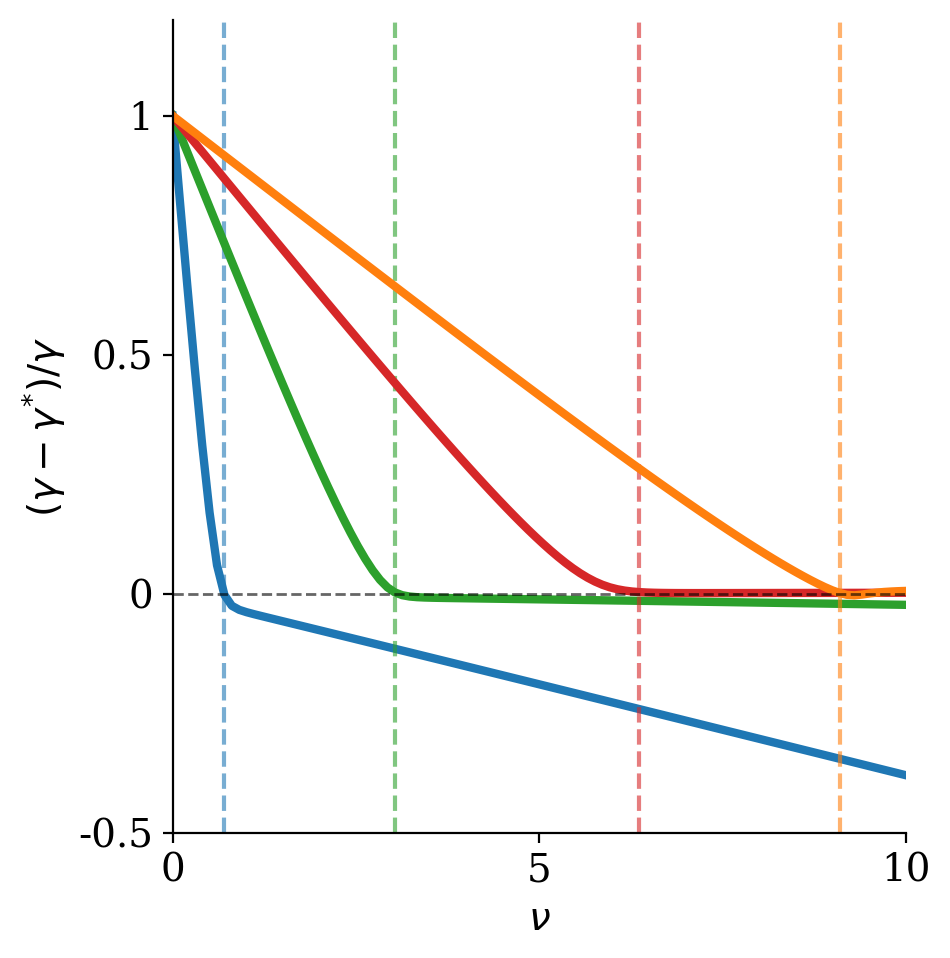}
    \end{subfigure}
    \begin{subfigure}{0.32\textwidth}
        \caption{Artificial collisions $\alpha =2$ \newline \citet{camporeale_2016}}
        \includegraphics[width=\textwidth]{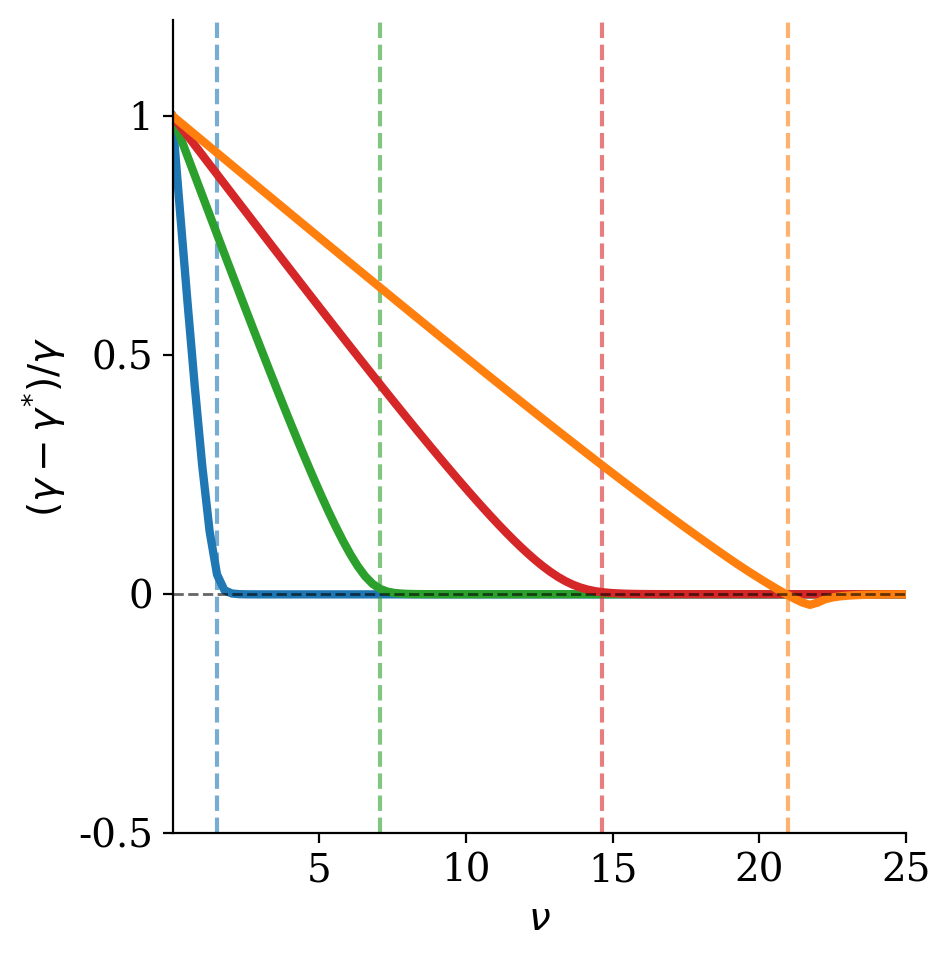}
    \end{subfigure}
    \begin{subfigure}{0.32\textwidth}
        \caption{Artificial collisions $\alpha =3$}
        \includegraphics[width=\textwidth]{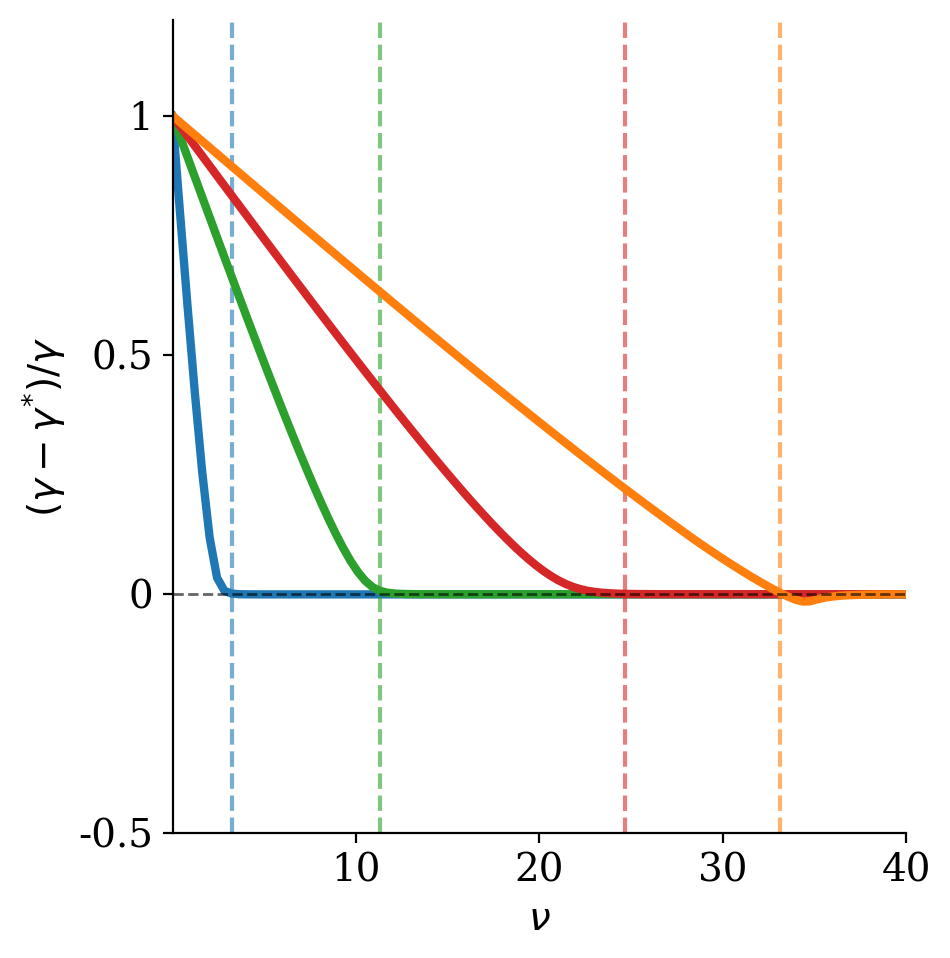}
    \end{subfigure}
    \begin{subfigure}{0.32\textwidth}
        \caption{Artificial collisions $\alpha =4$}
        \includegraphics[width=\textwidth]{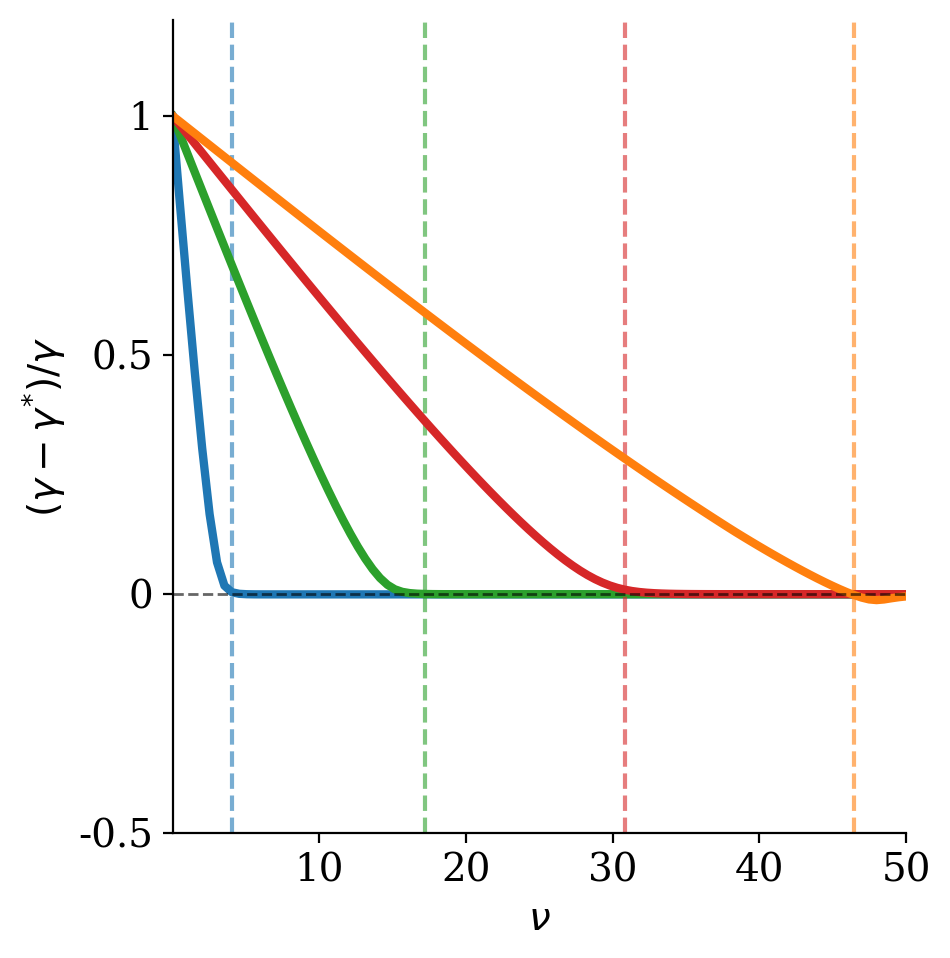}
    \end{subfigure}
        \begin{subfigure}{0.32\textwidth}
        \caption{\citet{hou_li_2007_filter} filter}
        \includegraphics[width=\textwidth]{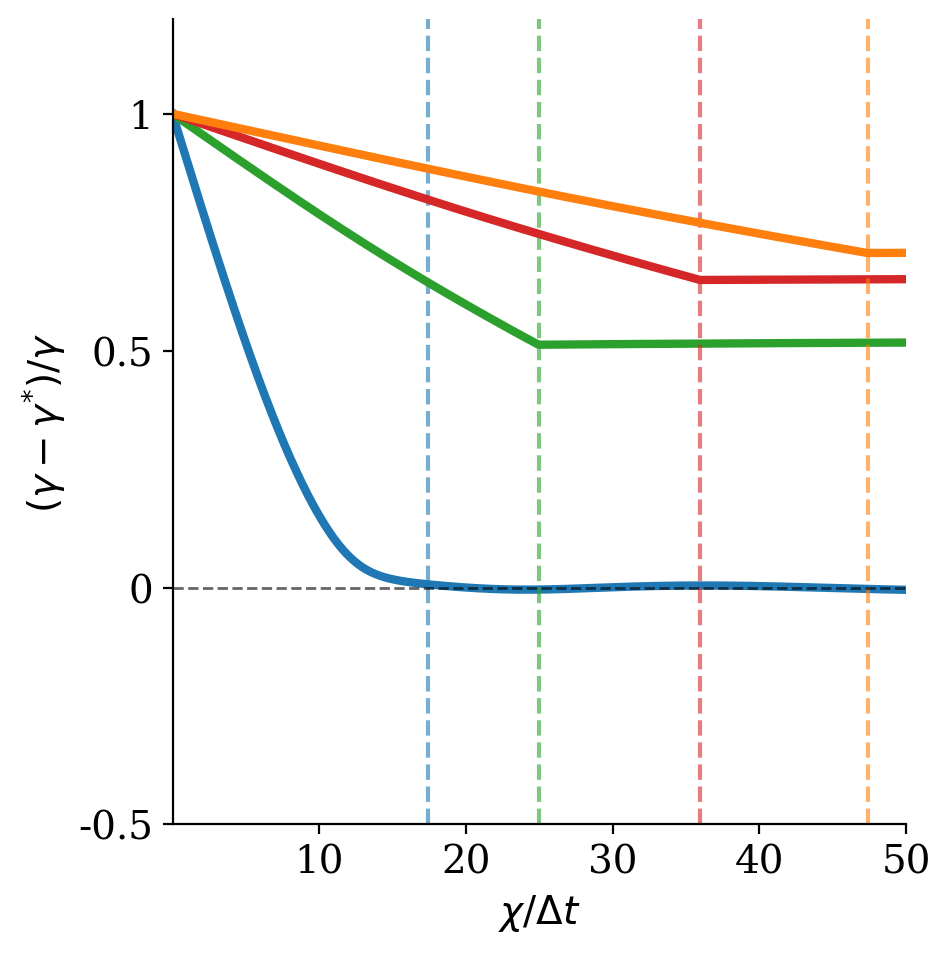}
    \end{subfigure}
    \caption{Same as Figure~\ref{fig:over-under-damping-error-20} with $N_{v}=100$. In comparison to $N_{v}=20$ in  Figure~\ref{fig:over-under-damping-error-20}, artificial collisions with $\alpha \in \{3, 4\}$ in subfigures~(d)--(e) no longer underdamp the higher wavenumber modes. Artificial collisions with $\alpha=1$ continue to overdamp smaller wavenumber modes.  Nonlocal closure and filtering approaches continue to underdamp most (or all) modes. }
    \label{fig:over-under-damping-error-100}
\end{figure}

We investigate whether the same behavior persists at higher velocity resolutions, specifically with $N_{v} = 100$, as shown in Figure~\ref{fig:over-under-damping-error-100}. The nonlocal closure continues to underdamp all modes. The LB operator still overdamps small wavenumber modes, while the filtering technique continues to underdamp larger wavenumber modes. However, both the filtering and the LB operator show improvement compared to Figure~\ref{fig:over-under-damping-error-20} as the velocity resolution increases. The artificial collision operator with $\alpha = \{3, 4\}$ no longer underdamps the larger wavenumber modes. Therefore, with sufficiently high-velocity resolution ($N_v > 100$), artificial collisions with $\alpha \in \{3, 4\}$ can accurately recover the correct Landau damping rate across a range of wavenumbers. However, in multidimensional simulations where computational limitations constrain velocity resolution, the artificial collision operator with $\alpha = 2$ remains the most effective choice, as demonstrated in Figure~\ref{fig:over-under-damping-error-20}.

\section{Numerical Results}\label{sec:section-4}
We use the AW Hermite discretization in velocity and Fourier discretization in space, as described in~\citet{camporeale_2016}, to simulate Landau damping with artificial collisions, filtering, and nonlocal closure approaches discussed in section~\ref{sec:section-3}. The linear and nonlinear Landau damping tests are initialized by perturbing the initial electron distribution function as follows
\begin{equation}\label{initial-condition}
    f(x, v, t=0) =  \frac{1 + \epsilon \sum_{k}\cos(k x)}{\sqrt{2\pi}} \exp\left(-\frac{v^2}{2}\right),
\end{equation}
where $\epsilon$ is the perturbation amplitude. We set the spatial length to $\ell = 4\pi$, such that wavenumbers $k$ are integer multiples of $1/2$. We use the second-order implicit midpoint integrator with the unpreconditioned Jacobian-Free-Newton-Krylov method with absolute and relative error set to $10^{-10}$~\cite{knoll_jfnk_2004}. The time step is $\Delta t = 10^{-2}$ and the number of Fourier spectral terms in space is $N_{x} = 10$ for linear Landau damping in section~\ref{sec:linear-landau-damping} and $N_{x}=100$ for nonlinear Landau damping in section~\ref{sec:nonlinear-landau-damping}. The optimal tuning parameter of each method for the linear and nonlinear Landau tests are listed in Table~\ref{tab:optimal_parameters_landau_damping}, where the optimal parameters minimize the damping error via the eigenvalue analysis in section~\ref{sec:eigenvalue-analysis}.

\noindent
\begin{table}
\SetTblrInner{rowsep=1pt}
\caption{The optimal tunable parameter of each method based on minimizing the damping rate error for the highest wavenumber excited. The three methods are artificial collisions in Eq.~\eqref{hypercollisions-normalized}, filtering in Eq.~\eqref{filtering-normalized}, and nonlocal closure in Eq.~\eqref{nonlocal-closure-form}. The linear Landau damping parameters in section~\ref{sec:linear-landau-damping} are optimized with $N_{v}=20$ and $k=1.5$ and the nonlinear Landau damping parameters in section~\ref{sec:nonlinear-landau-damping} are optimized with $N_{v}=300$ and $k=0.5$.}
\centering
\begin{tblr}{c |c |c | c| c| c}
& \shortstack{collisions \\ $\alpha=1$ \\ $\nu$ } & \shortstack{collisions\\ $\alpha=2$ \\ $\nu$}  & \shortstack{collisions \\ $\alpha=3$ \\ $\nu$}  & 
\shortstack{filtering \\ $\chi/\Delta t$} & 
\shortstack{nonlocal closure \\ $N_{m}=1$ \\ $\mu_{N_{v}-1}$}   \\
\hline
linear Landau damping in section~\ref{sec:linear-landau-damping} & 6.30 & 16.76 & 15.29&  7.56 & -1.01 \\
\hline
nonlinear Landau damping in section~\ref{sec:nonlinear-landau-damping} & 0.55 & 1.31 & 2.01 &12.23 & -1.00 \\
\end{tblr}
\label{tab:optimal_parameters_landau_damping}
\end{table}

\subsection{Linear Landau Damping} \label{sec:linear-landau-damping}
\begin{figure}
    \centering
    \begin{subfigure}{0.25\textwidth}
        \caption{Collisionless with \newline closure by truncation \newline $N_{v}=1,024$}
        \label{fig:recurrence-linear-landau-collisionless-1024}
        \includegraphics[width=\textwidth]{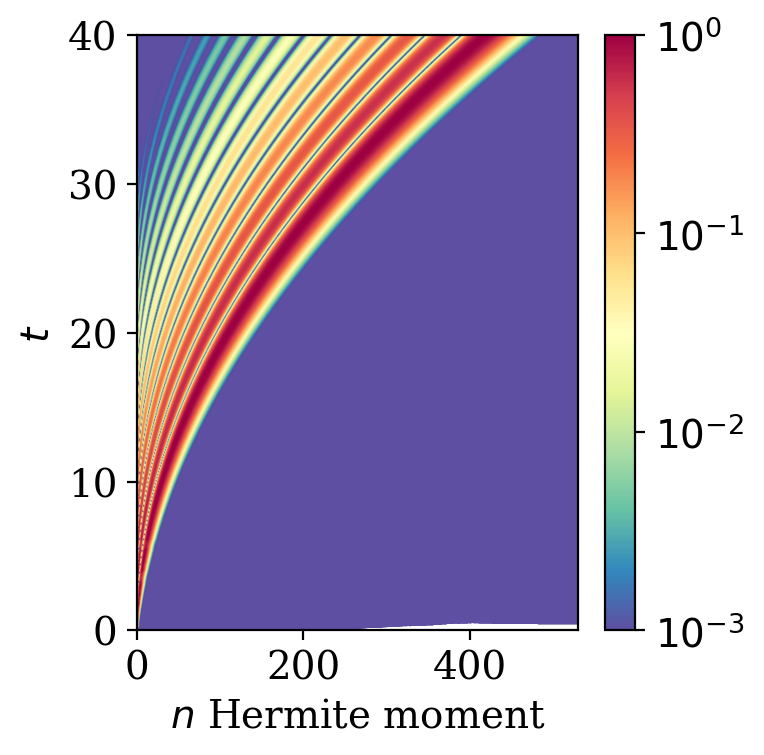}
    \end{subfigure}
    \hspace{-10pt}
    \begin{subfigure}{0.25\textwidth}
        \caption{Collisionless with \newline closure by truncation \newline $N_{v}=20$}
        \label{fig:recurrence-linear-landau-collisionless-20}
        \includegraphics[width=\textwidth]{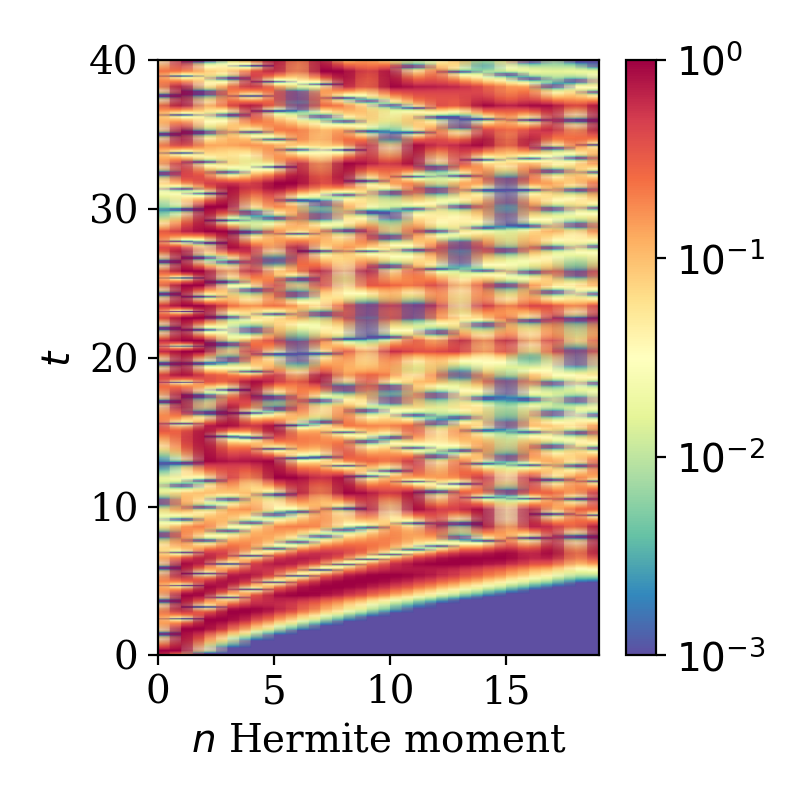}
    \end{subfigure}
    \hspace{-10pt}
    \begin{subfigure}{0.25\textwidth}
        \caption{Nonlocal closure with $N_{m}=1$ \newline \citet{smith_1997_closure}}
        \label{fig:recurrence-linear-landau-hp}
        \includegraphics[width=\textwidth]{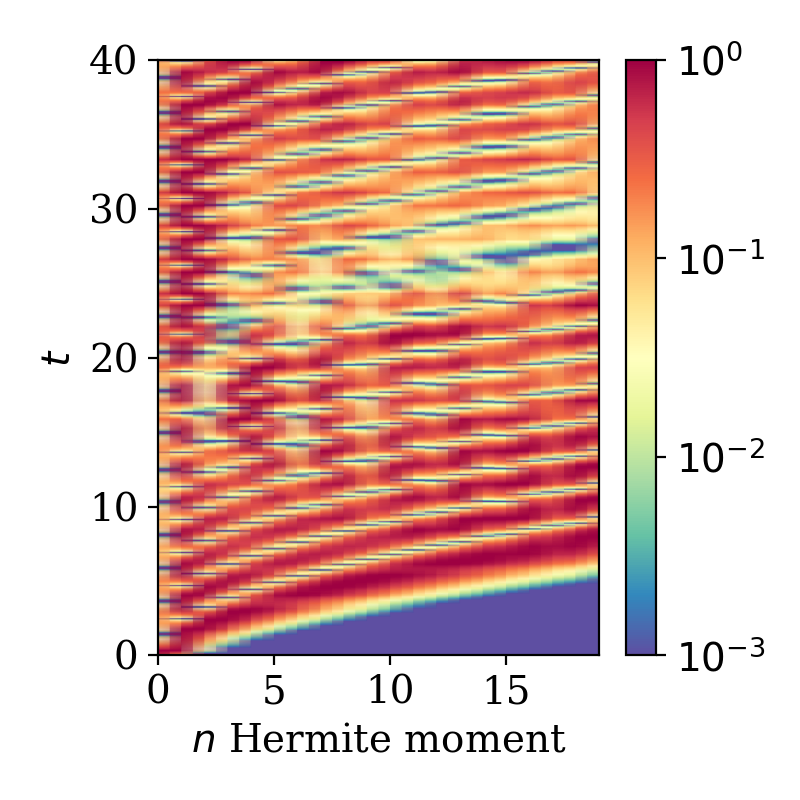}
    \end{subfigure}
    \hspace{-10pt}
    \begin{subfigure}{0.25\textwidth}
        \caption{Artificial collisions $\alpha=1$ \newline \citet{lenard_bernstein_1958_collisions}}
        \label{fig:recurrence-linear-landau-lb}
        \includegraphics[width=\textwidth]{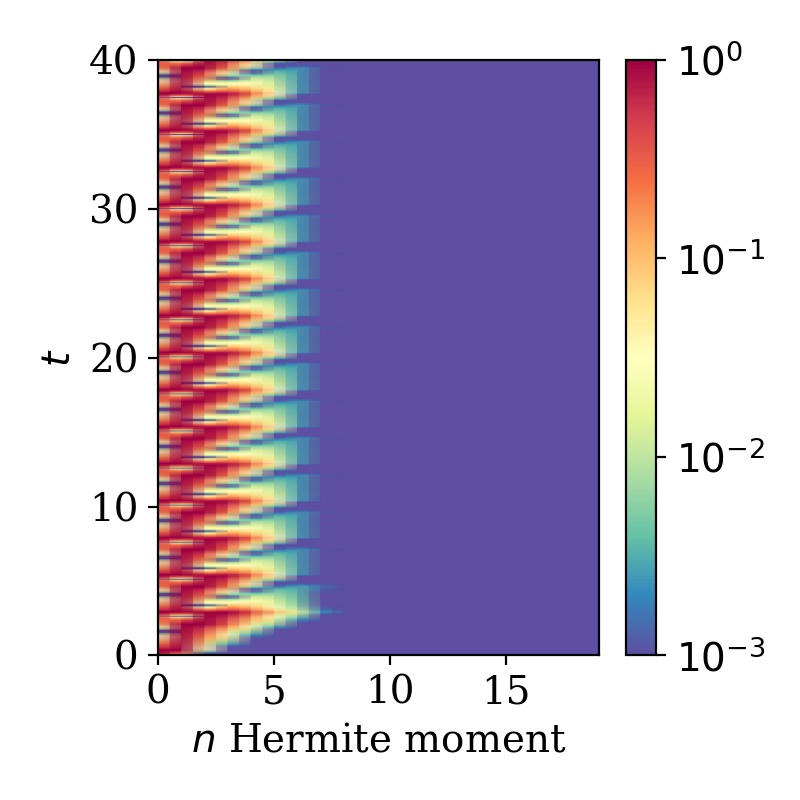}
    \end{subfigure}
    \hspace{-10pt}
    \begin{subfigure}{0.25\textwidth}
        \caption{Artificial collisions $\alpha=2$ \newline \citet{camporeale_2016}}
        \label{fig:recurrence-linear-landau-hyper2}
        \includegraphics[width=\textwidth]{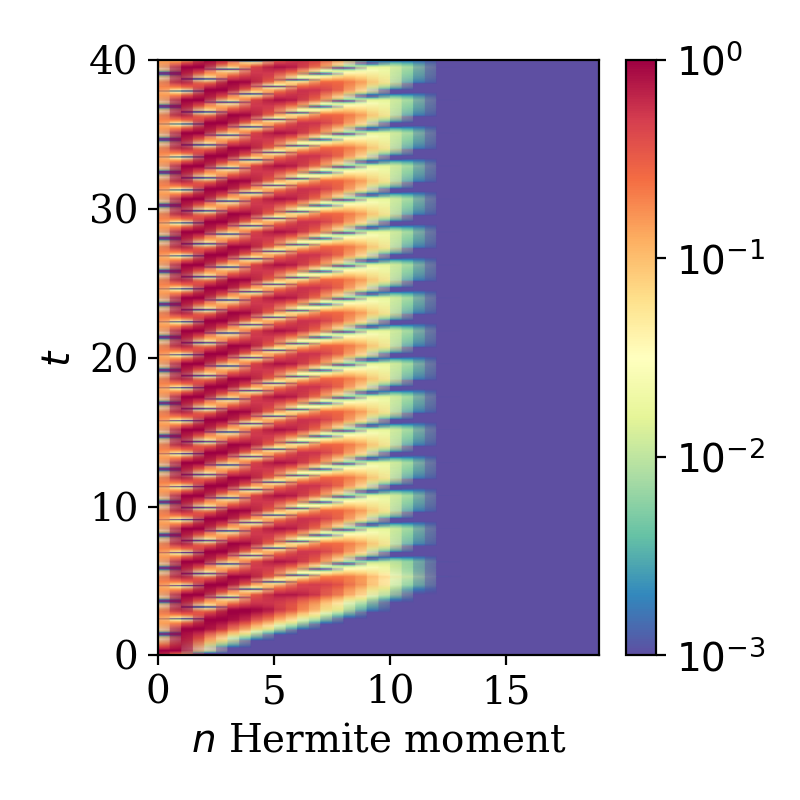}
    \end{subfigure}
    \hspace{-10pt}
    \begin{subfigure}{0.25\textwidth}
        \caption{Artificial collisions $\alpha=3$}
        \label{fig:recurrence-linear-landau-hyper3}
        \includegraphics[width=\textwidth]{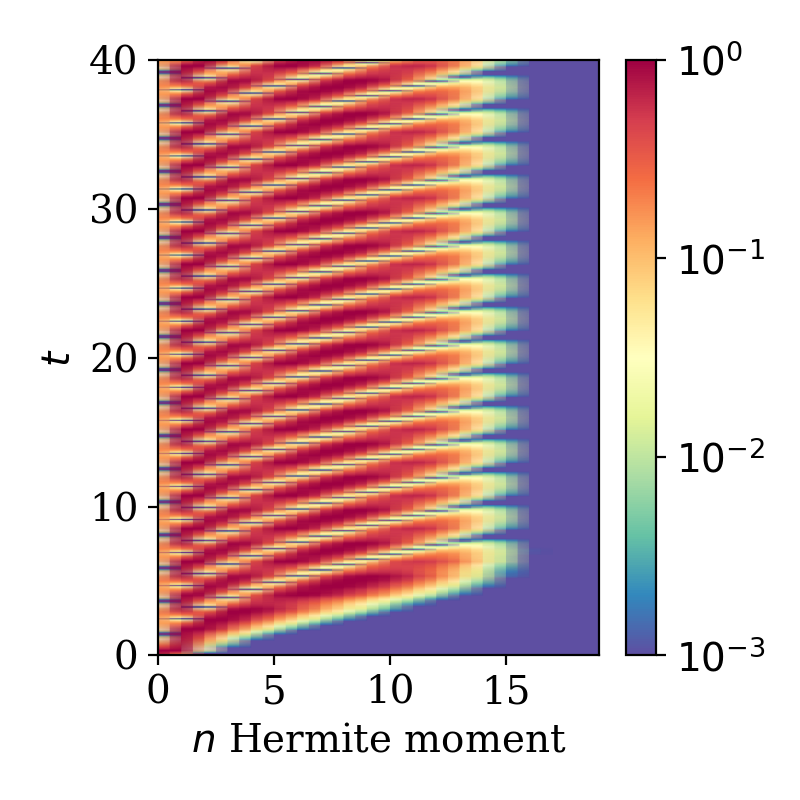}
    \end{subfigure}
    \hspace{-10pt}
    \begin{subfigure}{0.25\textwidth}
        \caption{\citet{hou_li_2007_filter} filter}
        \label{fig:recurrence-linear-landau-filter}
        \includegraphics[width=\textwidth]{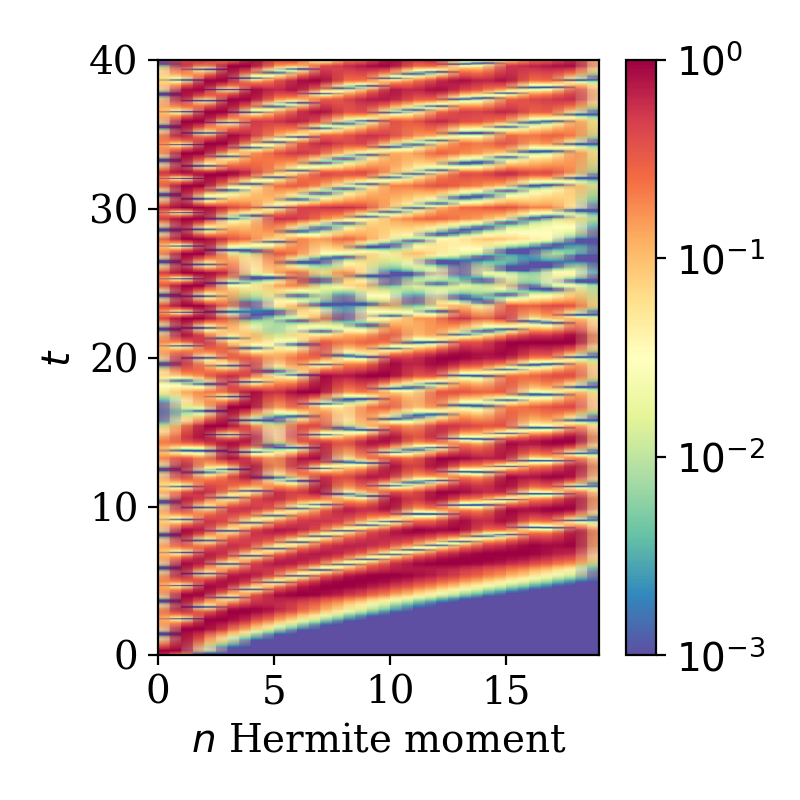}
    \end{subfigure}
    \caption{The squared normalized magnitude of the Hermite coefficients with $k=0.5$ for (a) $N_{v}=1,024$ and (b)-(g) $N_{v}=20$. The results indicate that artificial collisions in subfigures~(d)--(f) effectively mitigate recurrence caused by an artificial backpropagating Hermite cascade due to truncation, whereas nonlocal closure in subfigure~(c) and filtering in subfigure~(g) do not adequately suppress this backpropagation.}
    \label{fig:linear-landau-damping-recurrence}
\end{figure}

We investigate the impact of artificial collisions, filtering, and nonlocal closure approaches on the recurrence problem in linear Landau damping with $N_{v}=20$. Table~\ref{tab:optimal_parameters_landau_damping} lists the optimal tunable parameters for each method, which are derived through the eigenvalue analysis in section~\ref{sec:eigenvalue-analysis}. Initially, two modes with $k=0.5$ and $k=1.5$ are perturbed with amplitude $\epsilon=10^{-2}$ in Eq.~\eqref{initial-condition}. 
The velocity space Hermite moment cascade is shown in Figure~\ref{fig:linear-landau-damping-recurrence}. We plot the squared normalized magnitude of the Hermite coefficients with $k=0.5$, more specifically $|\hat{C}_{n}(k=0.5, t)|^{2}/\max_{n}(|\hat{C}_{n}(k=0.5, t)|^{2})$. The results show transport from the zeroth to the highest order Hermite mode, which for the collisionless, nonlocal closure, and filtering, gets reflected from the finite Hermite resolution and backpropagates. 
For the linear Landau damping problem, Hermite flux only propagates from the lower to higher order Hermite modes as shown in Figure~\ref{fig:recurrence-linear-landau-collisionless-1024} for the collisionless simulation with closure by truncation and $N_{v}=1,024$, and explained in greater detail in~\citet{parker_2015_hermite}. 
This backpropagation then artificially amplifies the zeroth order Hermite mode, resulting in a growth of the electric field. Artificial collisions suppress higher-order Hermite modes and prevent artificial backpropagation of Hermite flux.

\begin{figure}
    \centering
        \begin{subfigure}{0.325\textwidth}
        \caption{Collisionless with \newline closure by truncation}
        \label{fig:linear-landau-collisionless}
        \includegraphics[width=\textwidth]{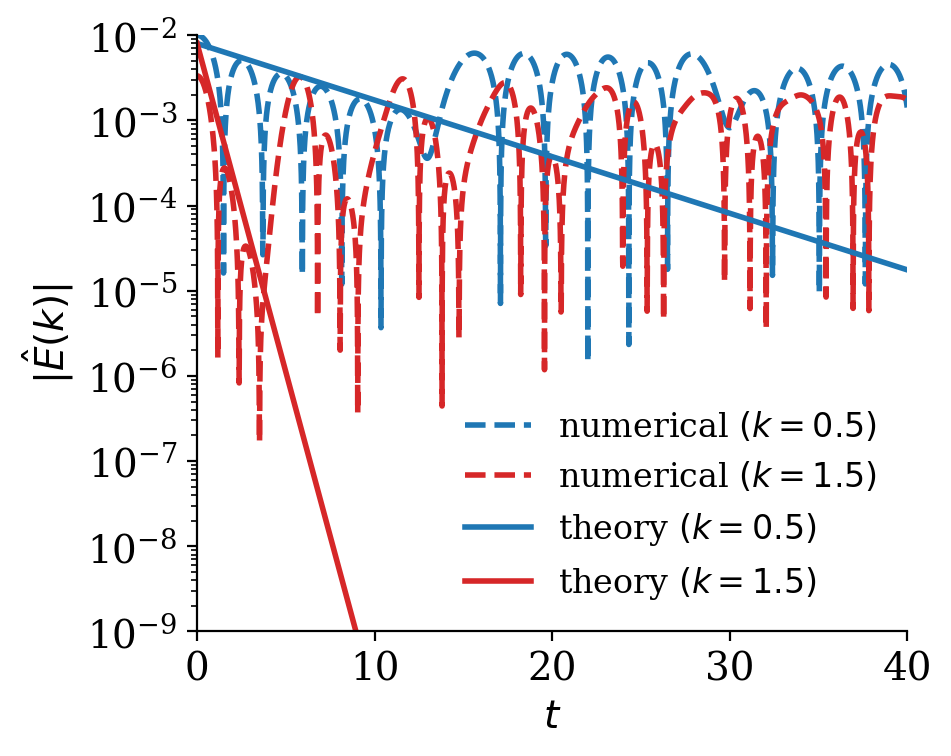}
    \end{subfigure}
    \begin{subfigure}{0.325\textwidth}
        \caption{Nonlocal closure with $N_{m}=1$ \newline \citet{smith_1997_closure}}
        \label{fig:linear-landau-hp}
        \includegraphics[width=\textwidth]{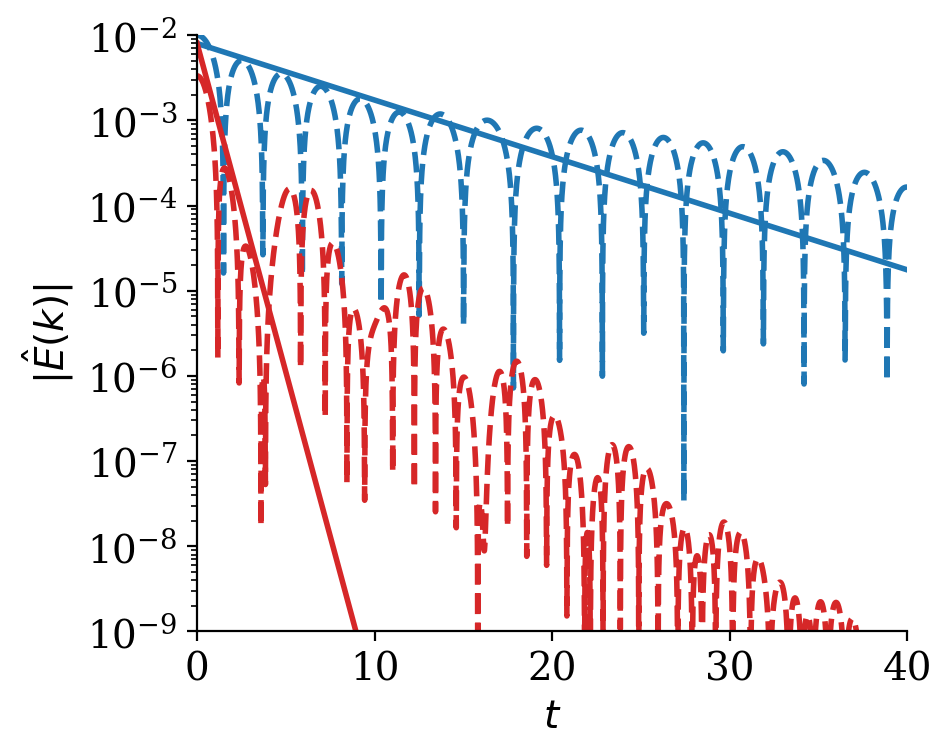}
    \end{subfigure}
    \begin{subfigure}{0.325\textwidth}
        \caption{Artificial collisions $\alpha=1$ \newline \citet{lenard_bernstein_1958_collisions}}
        \label{fig:linear-landau-lb}
        \includegraphics[width=\textwidth]{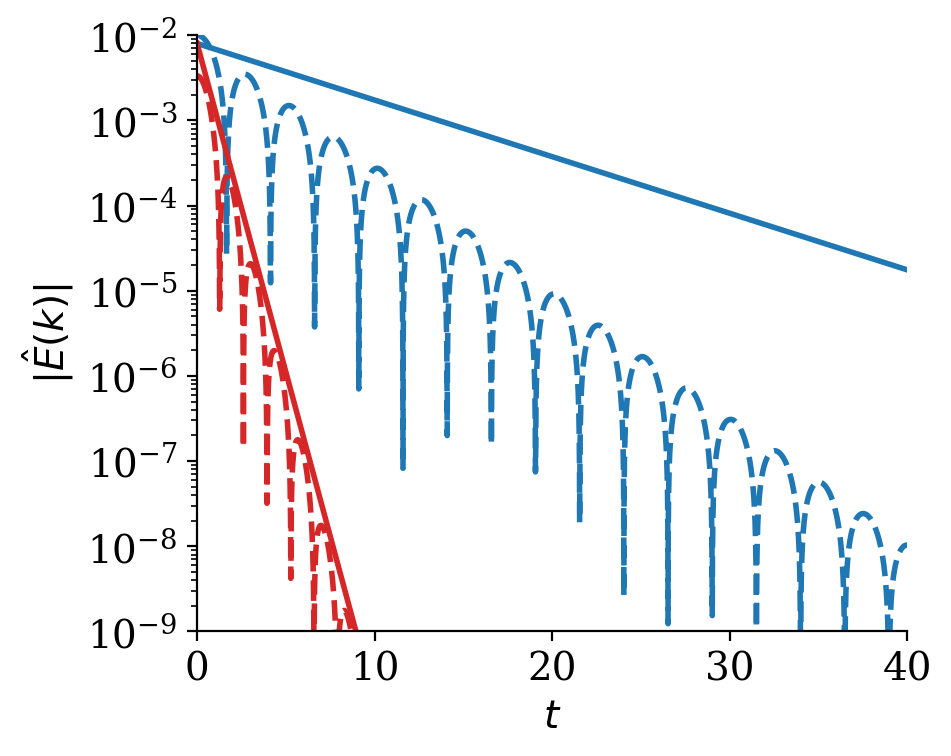}
    \end{subfigure}
    \begin{subfigure}{0.325\textwidth}
        \caption{Artificial collisions $\alpha=2$ \newline \citet{camporeale_2016}}
        \label{fig:linear-landau-hyper2}
        \includegraphics[width=\textwidth]{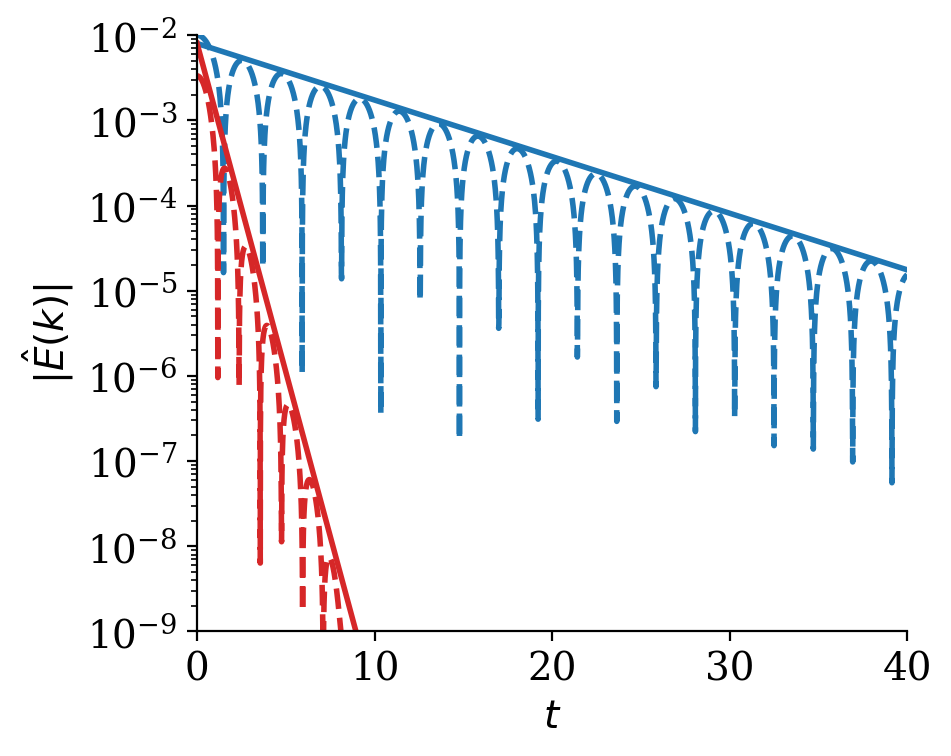}
    \end{subfigure}
    \begin{subfigure}{0.325\textwidth}
        \caption{Artificial collisions $\alpha=3$}
        \label{fig:linear-landau-hyper3}
        \includegraphics[width=\textwidth]{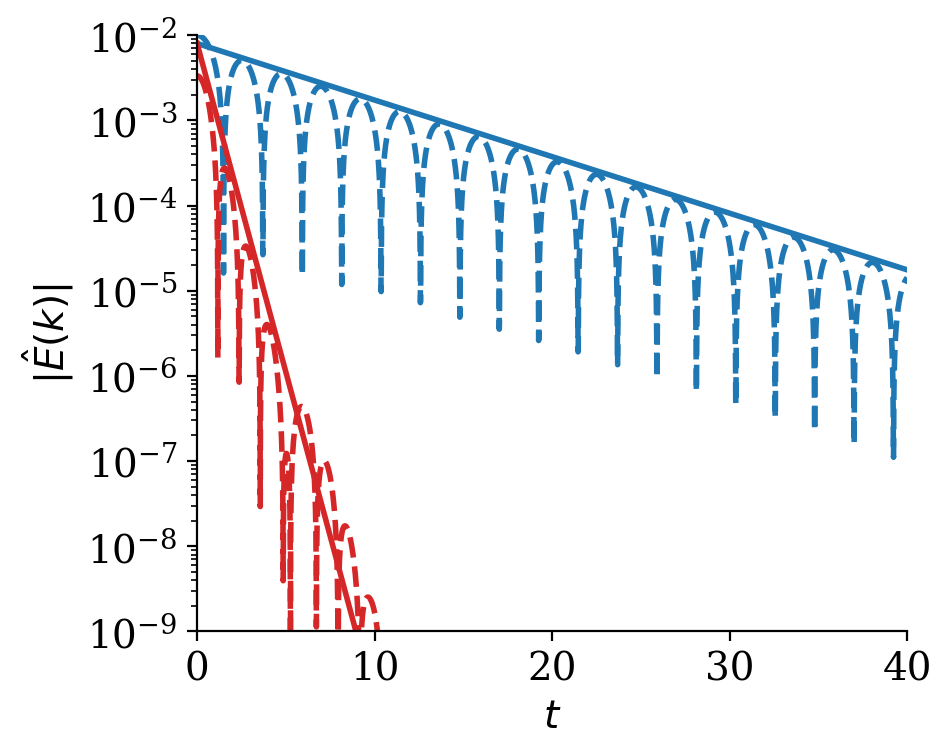}
    \end{subfigure}
    \begin{subfigure}{0.325\textwidth}
        \caption{\citet{hou_li_2007_filter} filter}
        \label{fig:linear-landau-filter}
        \includegraphics[width=\textwidth]{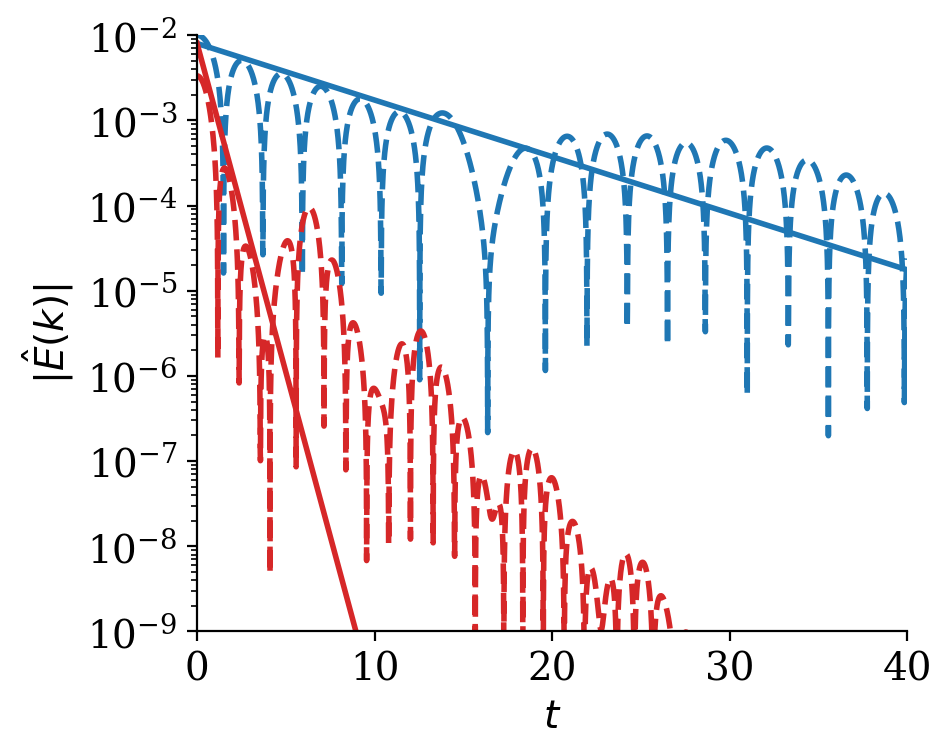}
    \end{subfigure}
    \caption{The linear Landau damping electric field Fourier coefficient magnitude of $k \in \{0.5, 1.5\}$ with $N_{v}=20$. The solid lines show the analytic damping rate obtained from linear theory, i.e. solving $k^2 = -R(\xi)$. Recurrence time is given by $T_{\mathrm{rec}}(k) \approx \sqrt{N_{v}}/k$, such that recurrence occurs at $T_{\mathrm{rec}}(k=1.5) \approx 4$ and $T_{\mathrm{rec}}(k=0.5) \approx 13$, as shown in subfigure~(a). The numerical results show that only artificial collisions with $\alpha=2$ in subfigure~(c) can mitigate recurrence without overdamping the smallest wavenumber perturbations or underdamping the largest wavenumber perturbations.}
    \label{fig:linear-landau-damping}
\end{figure}

Figure~\ref{fig:linear-landau-damping} shows the damping rate of the electric field. The closure by truncation without artificial collisions exhibits recurrence effects, with a recurrence period given by $T_{\mathrm{rec}}(k) \approx \sqrt{N_{v}} / k$, such that $T_{\mathrm{rec}}(k=1.5) \approx 4$ and $T_{\mathrm{rec}}(k=0.5) \approx 13$. The nonlocal closure with one tunable parameter and filtering results also show recurrence effects and underestimate the damping rate, especially for $k=1.5$. The recurrence period is slightly delayed in both approaches for $k=0.5$, compared to the collisionless case with closure by truncation. 
Additionally, although the LB operator (i.e. artificial collisions with $\alpha=1$) eliminates recurrence, it overdamps the $k=0.5$ wave. This is expected since we chose to match the optimal free parameter $\nu$ to recover the correct damping rate for $k=1.5$ wave; otherwise, if we chose the optimal $\nu$ to recover the correct damping rate for $k=0.5$ mode, we would underdamp the $k=1.5$ mode. 
Similarly, artificial collisions with $\alpha=3$ eliminate recurrence but underdamp the $k=1.5$ wave. These results confirm the findings from section~\ref{sec:eigenvalue-analysis}, with artificial collisions with $\alpha=2$ proving to be the most effective method for capturing the correct linear Landau damping phenomena for a range of wavenumbers in limited velocity resolution.  

\subsection{Nonlinear Landau Damping} \label{sec:nonlinear-landau-damping}
\begin{figure}
    \centering
        \begin{subfigure}{0.325\textwidth}
        \caption{Collisionless with \newline closure by truncation}
        \includegraphics[width=\textwidth]{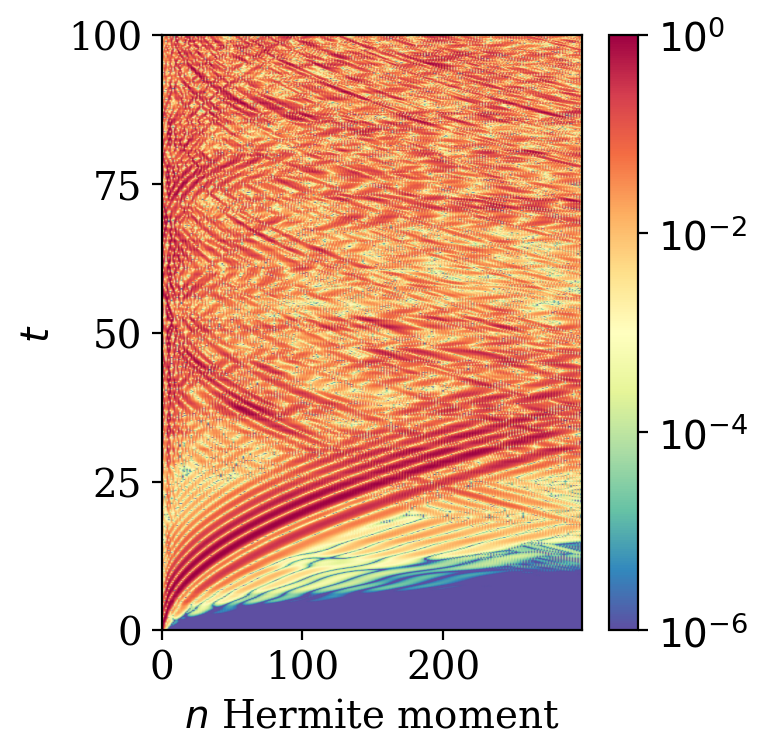}
    \end{subfigure}
    \begin{subfigure}{0.325\textwidth}
        \caption{Nonlocal closure with $N_{m}=1$ \newline \citet{smith_1997_closure}}
        \includegraphics[width=\textwidth]{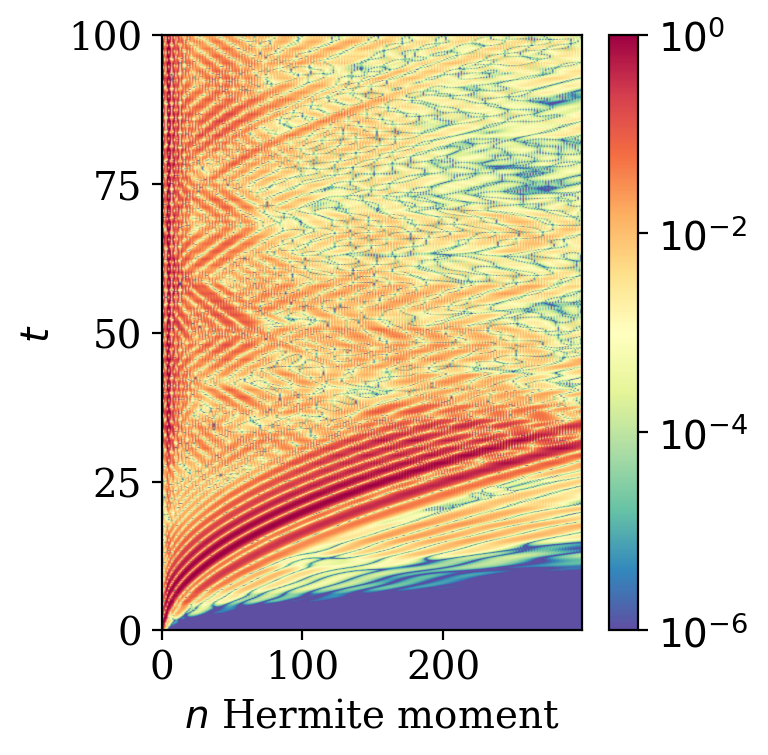}
    \end{subfigure}
    \begin{subfigure}{0.325\textwidth}
        \caption{Artificial collisions $\alpha=1$ \newline \citet{lenard_bernstein_1958_collisions}}
        \includegraphics[width=\textwidth]{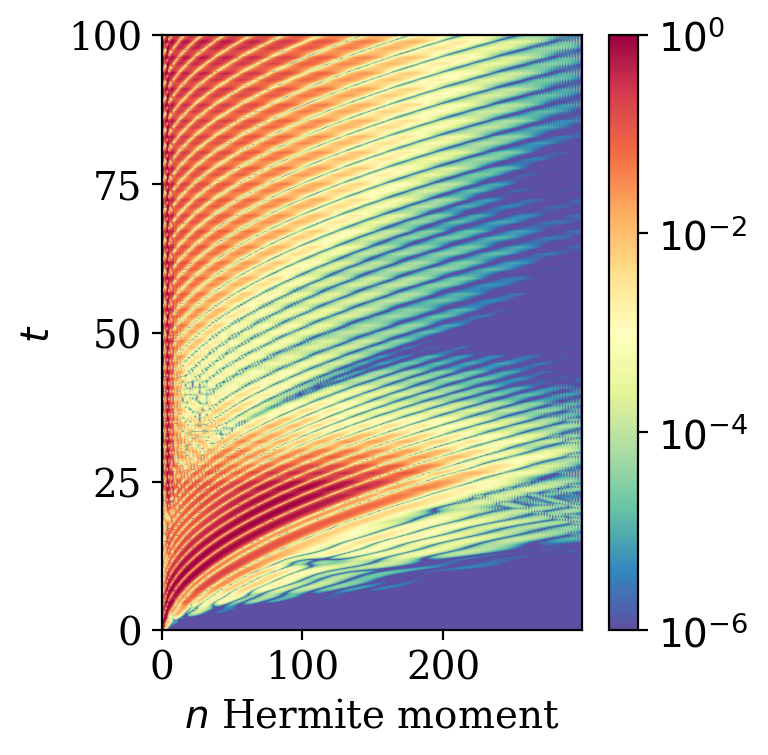}
    \end{subfigure}
    \begin{subfigure}{0.325\textwidth}
        \caption{Artificial collisions $\alpha=2$ \newline \citet{camporeale_2016}}
        \includegraphics[width=\textwidth]{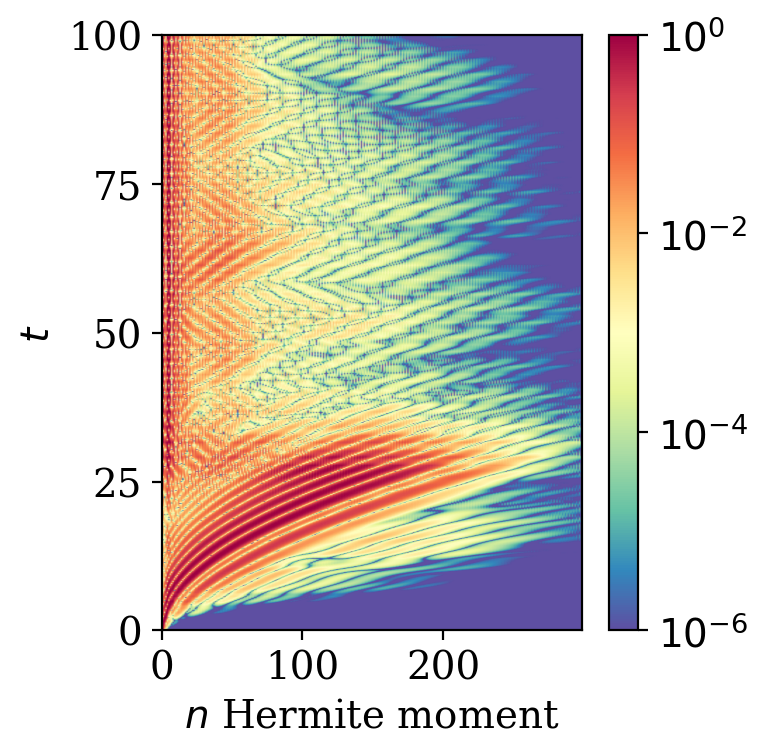}
    \end{subfigure}
    \begin{subfigure}{0.325\textwidth}
        \caption{Artificial collisions $\alpha=3$}
        \includegraphics[width=\textwidth]{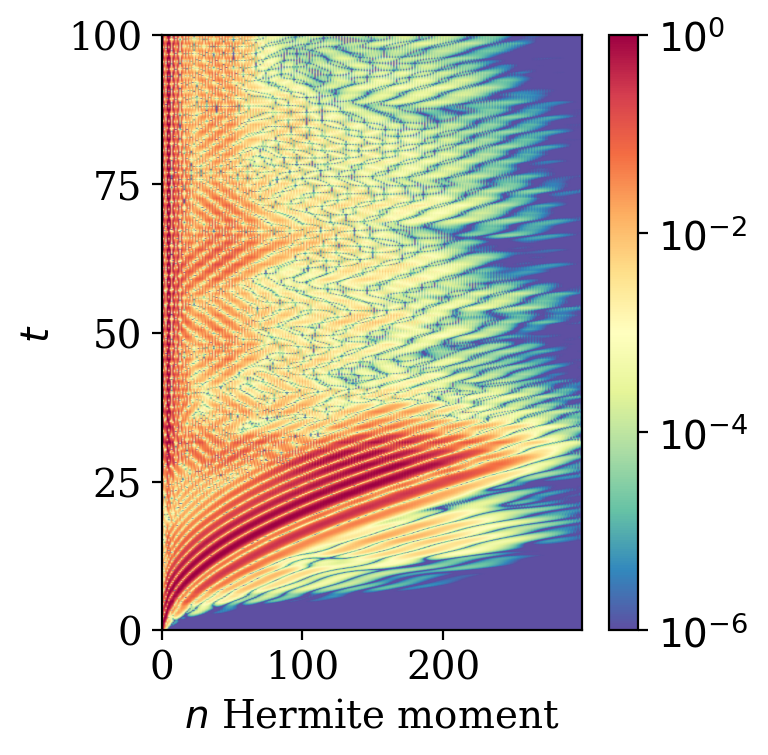}
    \end{subfigure}
    \begin{subfigure}{0.325\textwidth}
        \caption{\citet{hou_li_2007_filter} filter}
        \includegraphics[width=\textwidth]{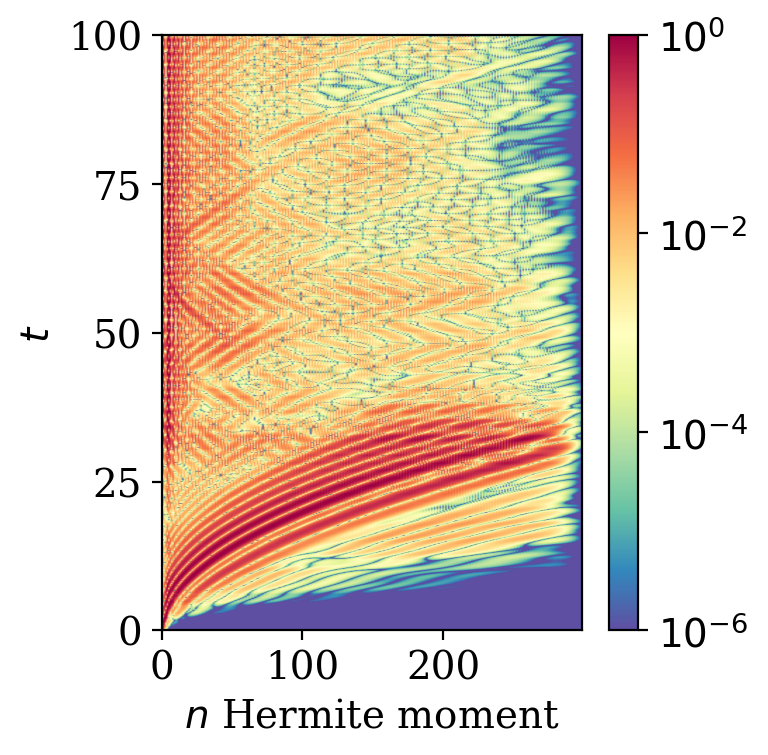}
    \end{subfigure}
    \caption{Same as Figure~\ref{fig:linear-landau-damping-recurrence} for nonlinear Landau damping with $N_{v}=300$. The nonlinear dynamics lead to the backward propagation of Hermite flux from higher to lower moments. Moreover, the finite dimensionality in velocity space introduces an artificial backward propagation of Hermite flux. The results from the nonlocal closure in subfigure~(b) show that the method mitigates some of this backward propagation compared to subfigure~(a). Artificial collisions in subfigures~(c)--(e) and filtering in subfigure~(f) significantly damp the highest modes.}
    \label{fig:nonlinear-landau-damping-recurrence}
\end{figure}

For nonlinear Landau damping, we set the initial perturbation with wavenumber~$k=0.5$ and amplitude~$\epsilon=0.5$ in Eq.~\eqref{initial-condition}. The tunable parameters for each method are listed in Table~\ref{tab:optimal_parameters_landau_damping} and are determined through the eigenvalue analysis in section~\ref{sec:eigenvalue-analysis} with~$N_{v}=300$, which aims to minimize errors in the linear theory damping rate. 
\citet{parker_2015_hermite} describe the nonlinear Landau damping dynamics in terms of Hermite flux, where the initial linear decay phase of the electric field corresponds to forward-propagating Hermite modes. The subsequent growth phase is associated with backward-propagating Hermite modes. Eventually, a balance is reached between the forward and backward Hermite fluxes, leading to the nonlinear saturation phase. Therefore, the method's suppression of recurrence is more subtle in the nonlinear case since the goal is to suppress artificial backpropagation stemming from the finite resolution while permitting physical backpropagation caused by the nonlinear dynamics. 
Figure~\ref{fig:nonlinear-landau-damping-recurrence} shows the squared normalized magnitude of the Hermite coefficients with $k=0.5$ (similar to Figure~\ref{fig:linear-landau-damping-recurrence} for linear Landau damping). The results indicate that LB collisions, i.e. artificial collisions with $\alpha=1$, prevent the backpropagation of Hermite flux from higher to lower Hermite coefficients. Hypercollisions and filtering permit backpropagation while significantly damping the highest Hermite modes. Conversely, the nonlocal closure allows the distribution function to spread across all velocity scales including the highest Hermite mode. 

\begin{figure}
    \centering
    \begin{subfigure}{0.325\textwidth}
        \caption{Collisionless with \newline closure by truncation}
        \includegraphics[width=\textwidth]{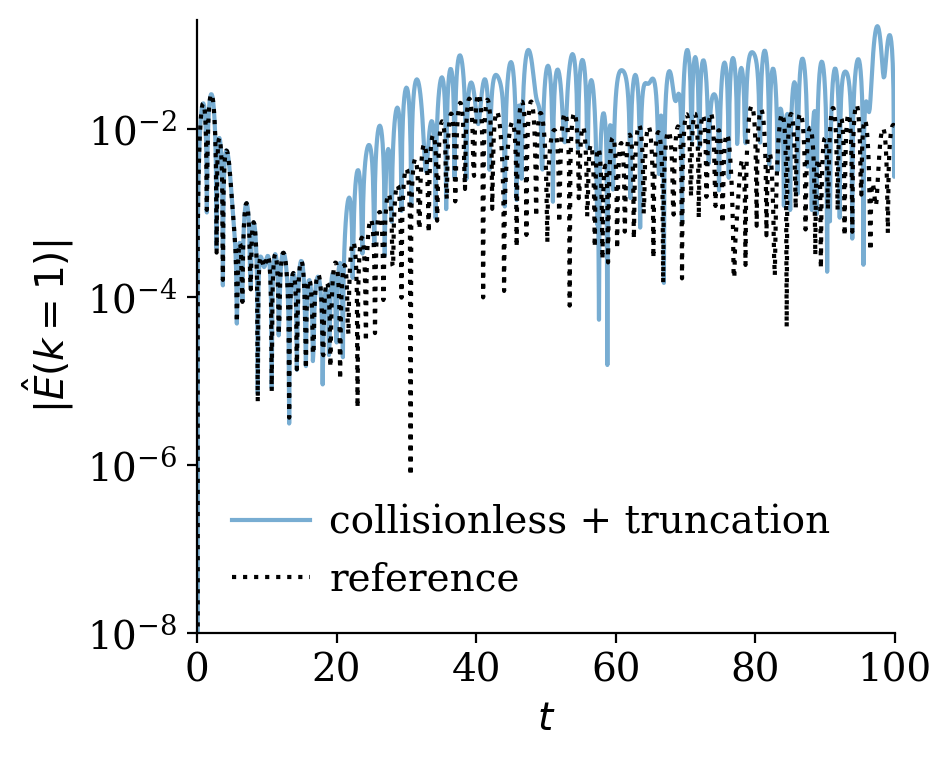}
    \end{subfigure}
    \begin{subfigure}{0.325\textwidth}
        \caption{Nonlocal closure with $N_{m}=1$ \newline \citet{smith_1997_closure}}
        \includegraphics[width=\textwidth]{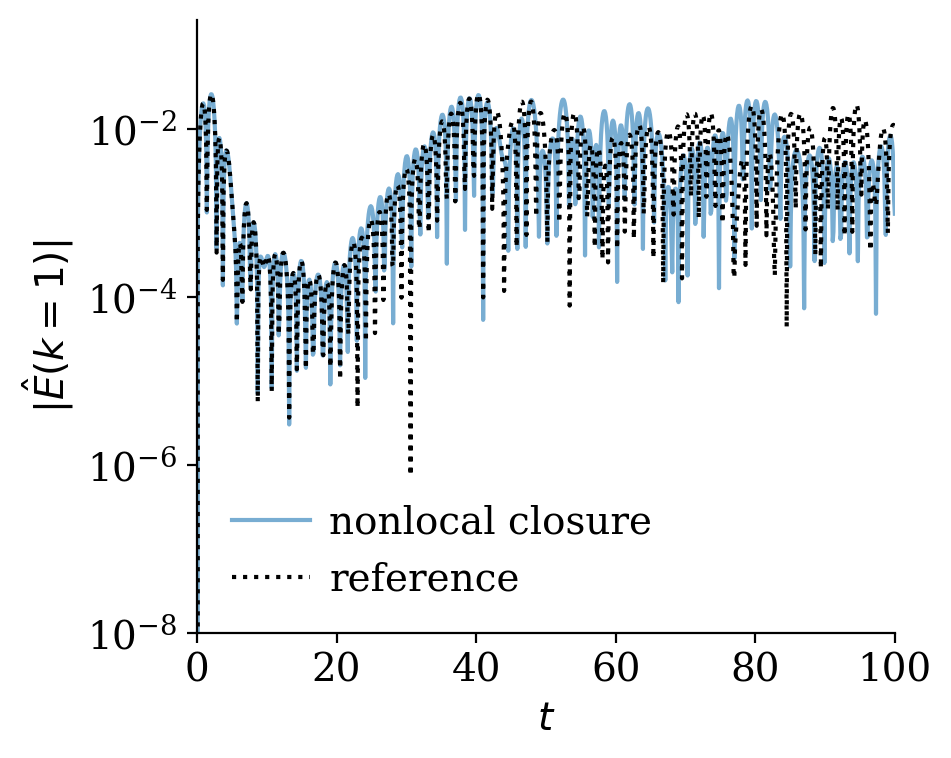}
    \end{subfigure}
    \begin{subfigure}{0.325\textwidth}
        \caption{Artificial collisions $\alpha=1$ \newline \citet{lenard_bernstein_1958_collisions}}
        \includegraphics[width=\textwidth]{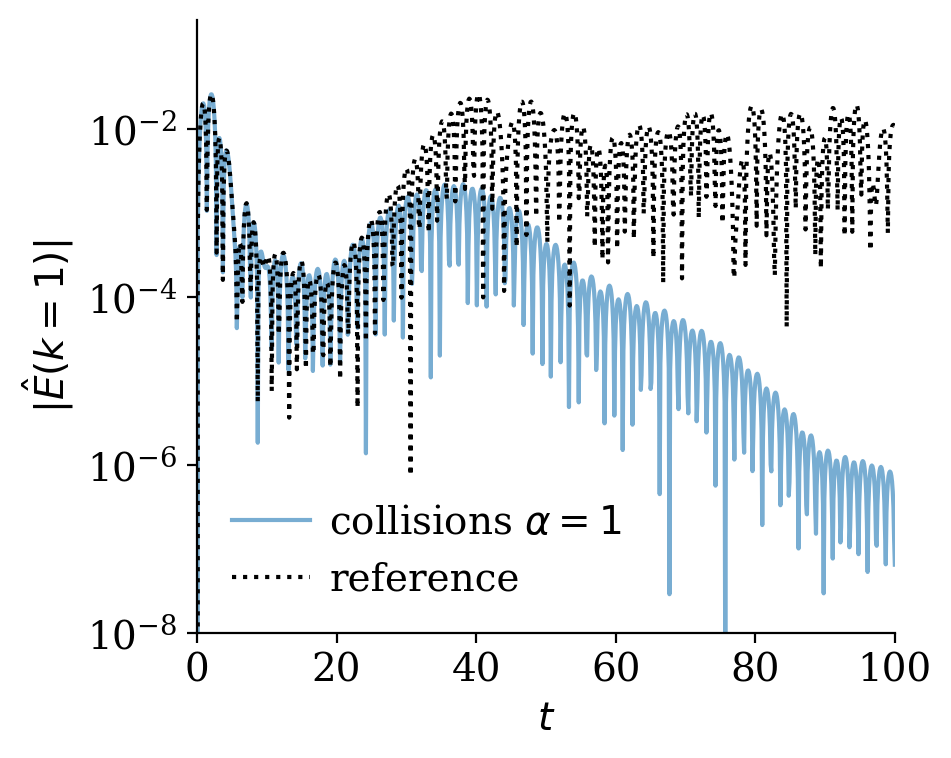}
    \end{subfigure}
    \begin{subfigure}{0.325\textwidth}
        \caption{Artificial collisions $\alpha=2$ \newline \citet{camporeale_2016}}
        \includegraphics[width=\textwidth]{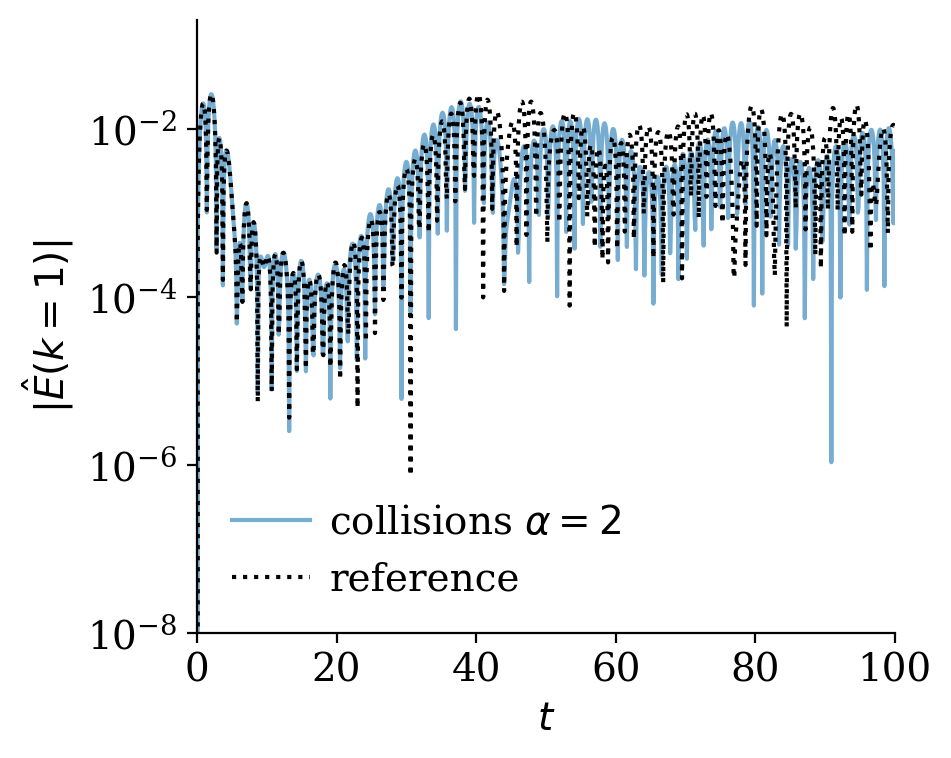}
    \end{subfigure}
    \begin{subfigure}{0.325\textwidth}
        \caption{Artificial collisions $\alpha=3$}
        \includegraphics[width=\textwidth]{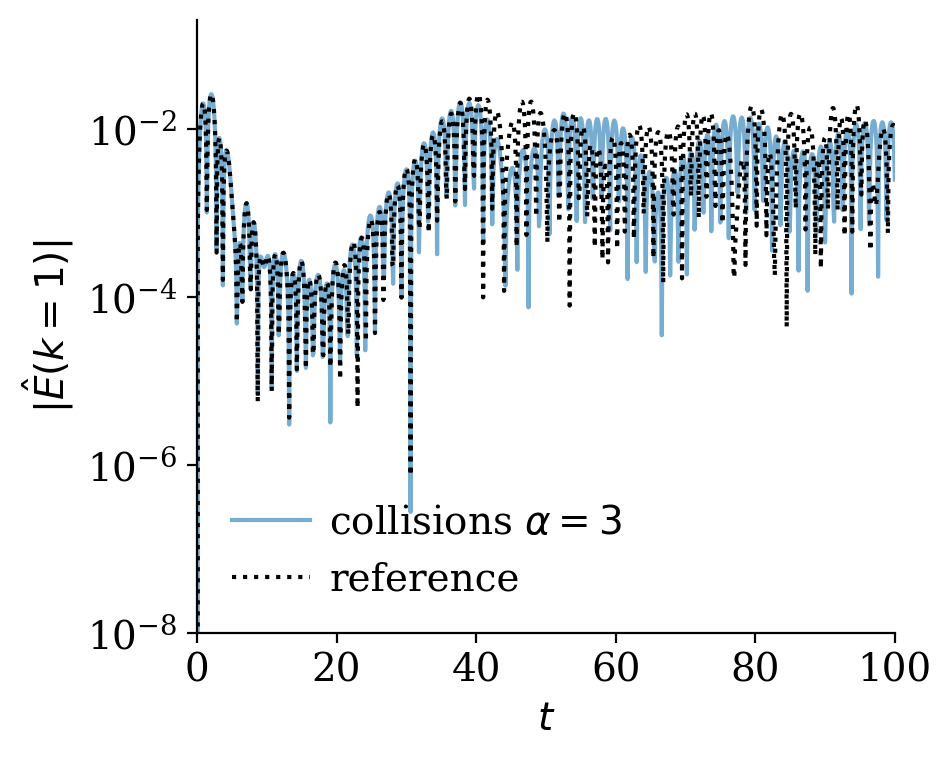}
    \end{subfigure}
    \begin{subfigure}{0.325\textwidth}
        \caption{\citet{hou_li_2007_filter} filter}
        \includegraphics[width=\textwidth]{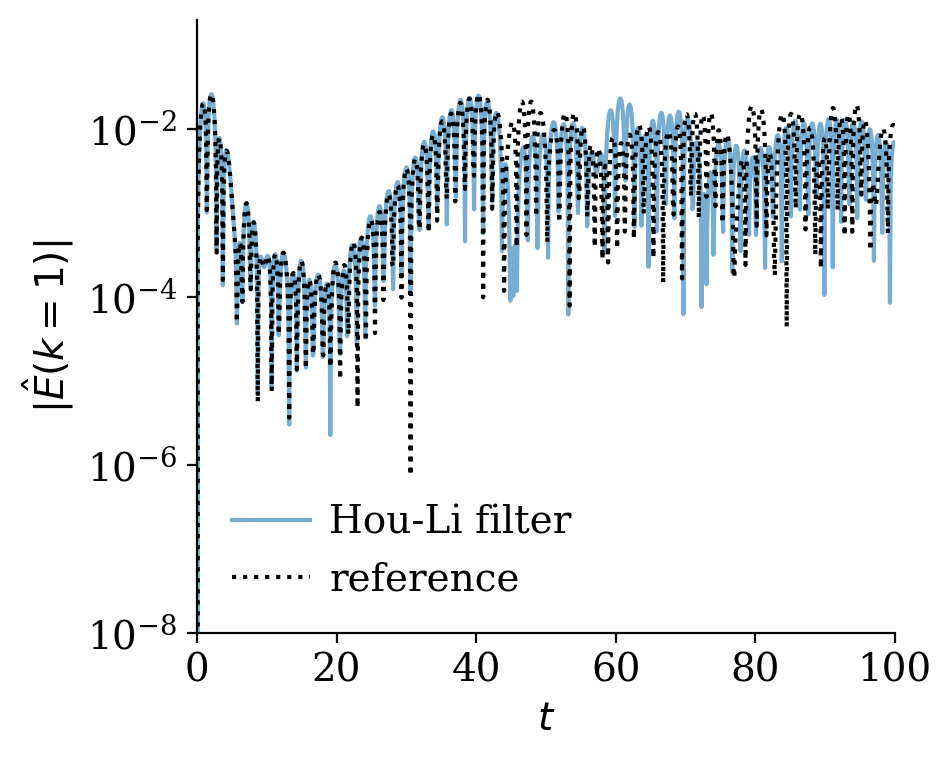}
    \end{subfigure}
    \caption{A comparison of the methods estimating the magnitude of the second Fourier mode of the electric field (i.e., $k=1$) in nonlinear Landau damping for $N_v=300$ against a high-resolution reference solution. In subfigure (c), the LB operator significantly alters the system's dynamics, driving the plasma toward thermodynamic equilibrium. In contrast, all other methods in subfigures~(b) and (d)--(f) qualitatively capture the collisionless dynamics during the saturation phase.}
    \label{fig:nonlinear-landau-electric-field}
\end{figure}

Figure~\ref{fig:nonlinear-landau-electric-field} compares the magnitude of the electric field second Fourier mode (i.e. $k=1$) of the different methods against the reference solution obtained by a high-resolution dissipationless second-order central finite difference scheme presented in~\citet{shiroto_2019_fd} with grid resolution $N_{x}=N_{v}=16,384$.
The electric field dynamics for the nonlocal closure, filtering, and higher-order artificial collisions ($\alpha \geq 2$) are similar, as both permit the backpropagation of Hermite flux, leading to the correct qualitative behavior of nonlinear saturation.
In contrast, the LB operator, i.e. artificial collisions with $\alpha=1$, entirely suppresses the backpropagation of Hermite flux, which overdamps the dynamics and incorrectly drives the system to thermal equilibrium. The LB results shown here are consistent with the previous analysis by~\citet{pezzi_2016_recurrence}. Moreover, as expected, the collisionless case with closure by truncation exhibits recurrence, leading to incorrect growth in the electric field.

\begin{figure}
    \centering
    \begin{subfigure}{0.245\textwidth}
        \caption{Collisionless with  \newline closure by truncation}
        \includegraphics[width=\textwidth]{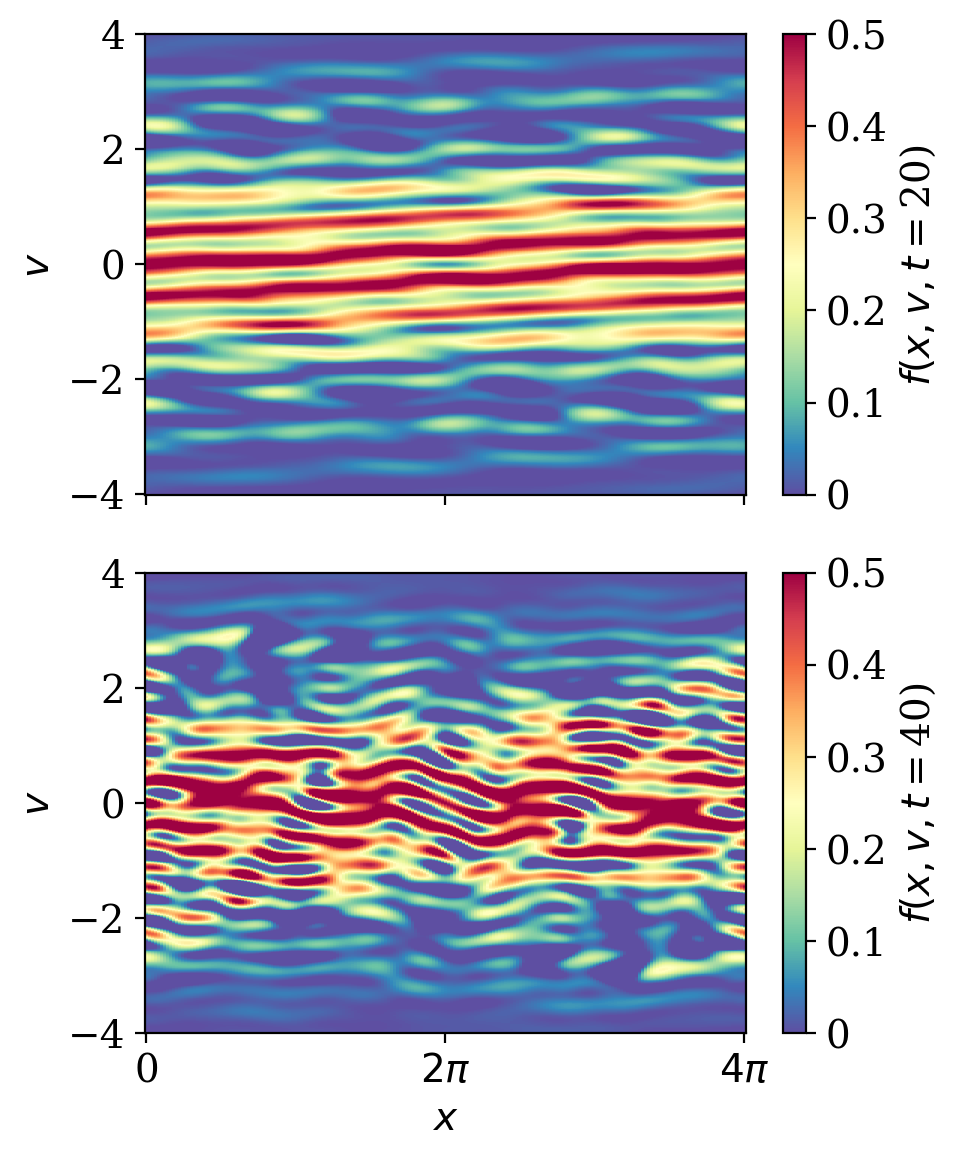}
    \end{subfigure}
    \begin{subfigure}{0.245\textwidth}
        \caption{Nonlocal closure \newline with $N_{m}=1$ \cite{smith_1997_closure}}
        \includegraphics[width=\textwidth]{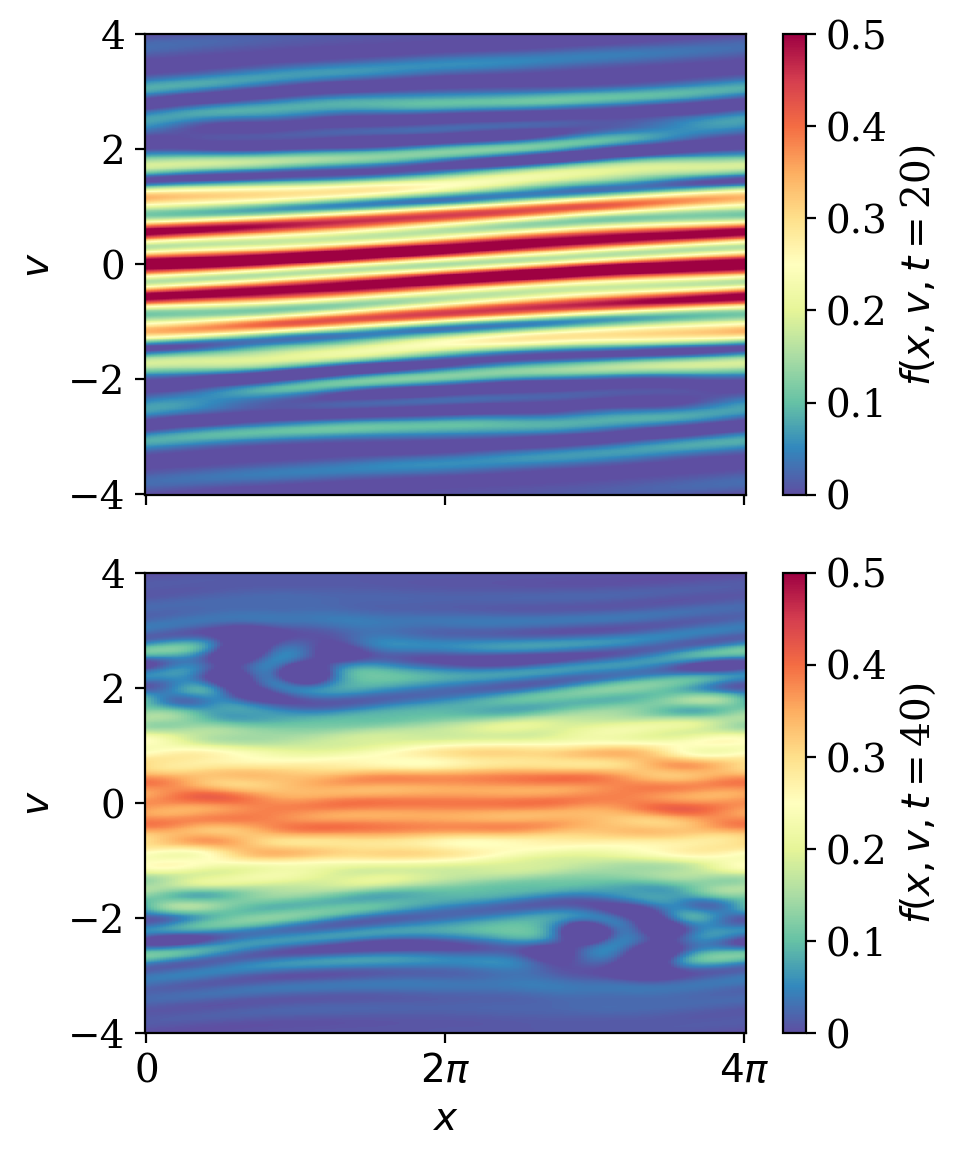}
    \end{subfigure}
    \begin{subfigure}{0.245\textwidth}
        \caption{Artificial collisions $\alpha=1$ \newline \citet{lenard_bernstein_1958_collisions}}
        \includegraphics[width=\textwidth]{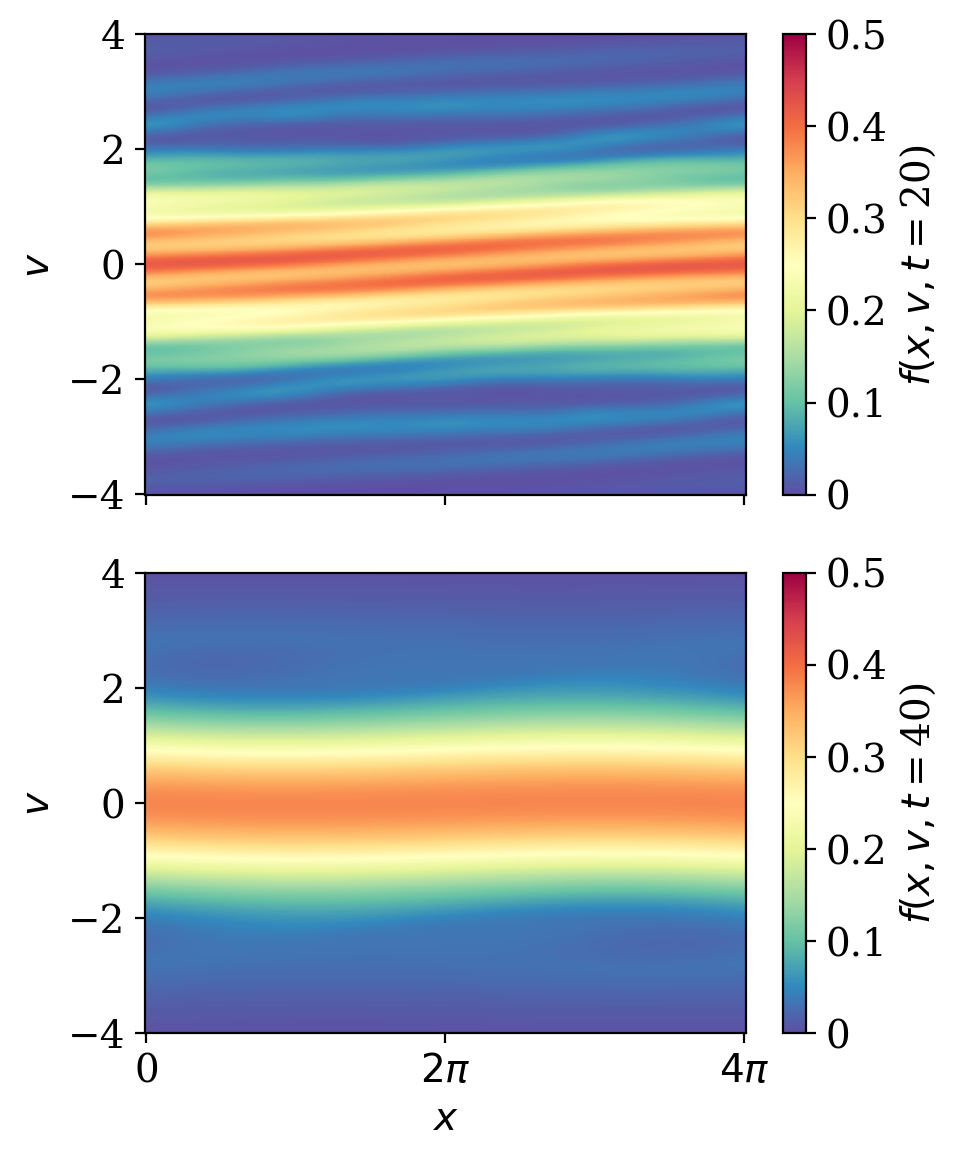}
    \end{subfigure}
    \begin{subfigure}{0.245\textwidth}
        \caption{Artificial collisions $\alpha=2$ \newline \citet{camporeale_2016}}
        \includegraphics[width=\textwidth]{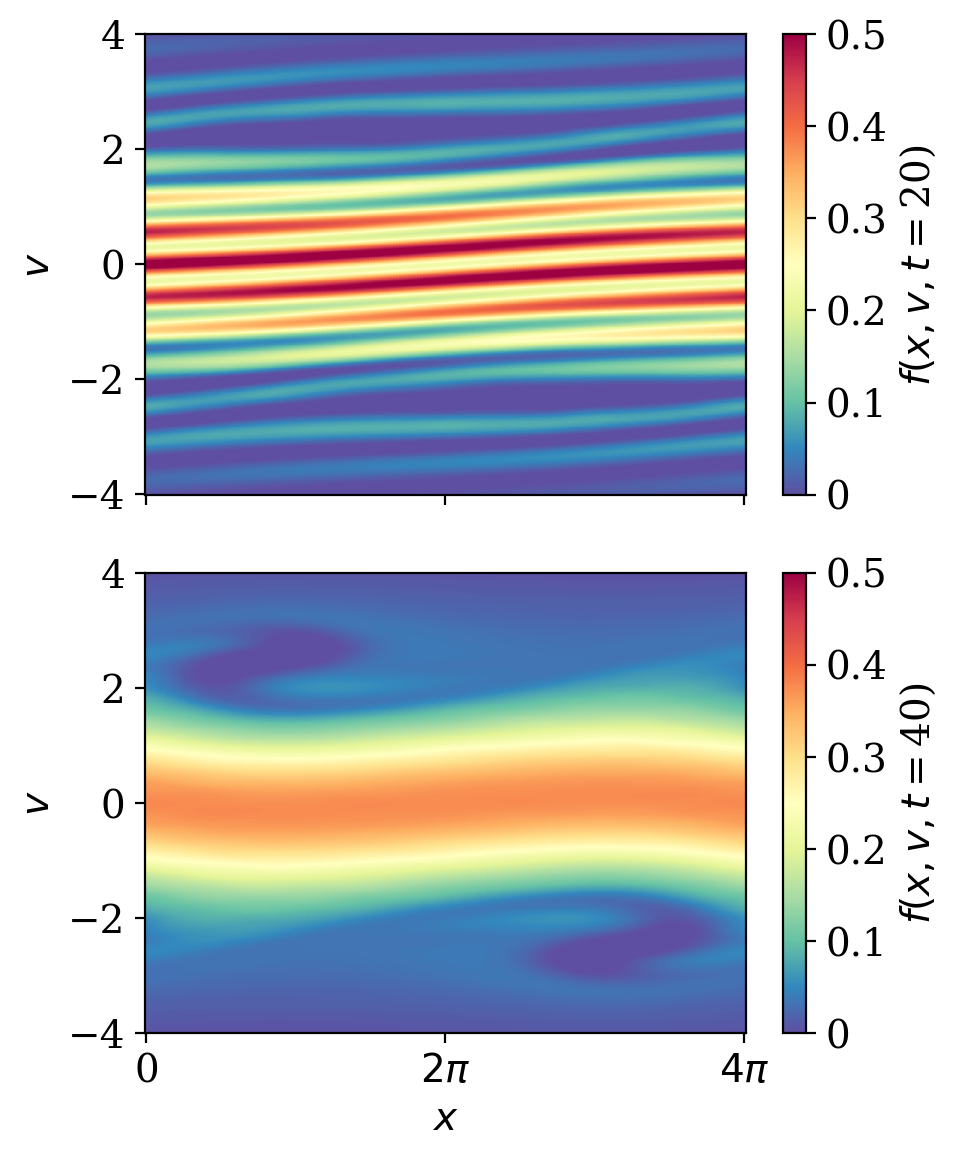}
    \end{subfigure}
    \begin{subfigure}{0.245\textwidth}
        \caption{Artificial collisions $\alpha=3$}
        \includegraphics[width=\textwidth]{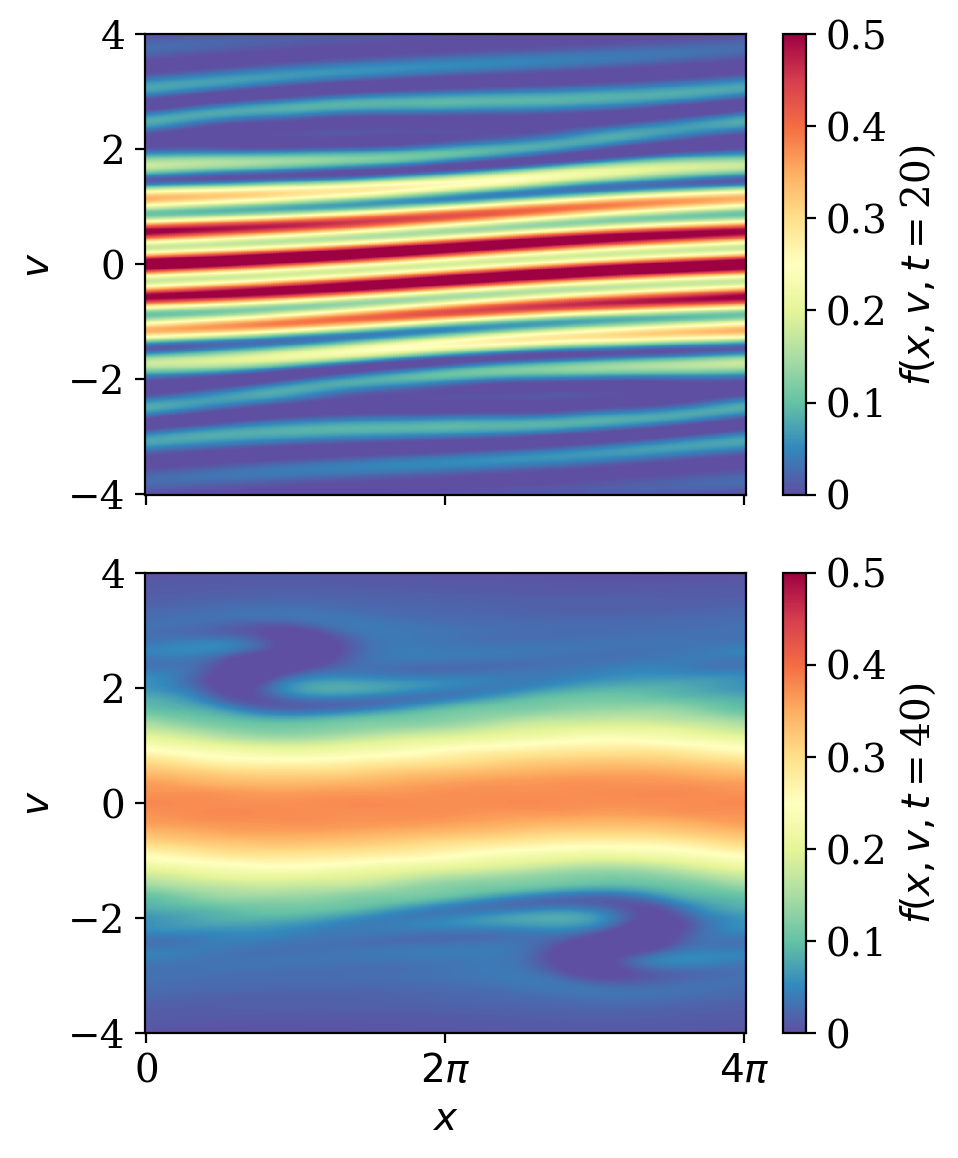}
    \end{subfigure}
    \begin{subfigure}{0.245\textwidth}
        \caption{\citet{hou_li_2007_filter} filter}
        \includegraphics[width=\textwidth]{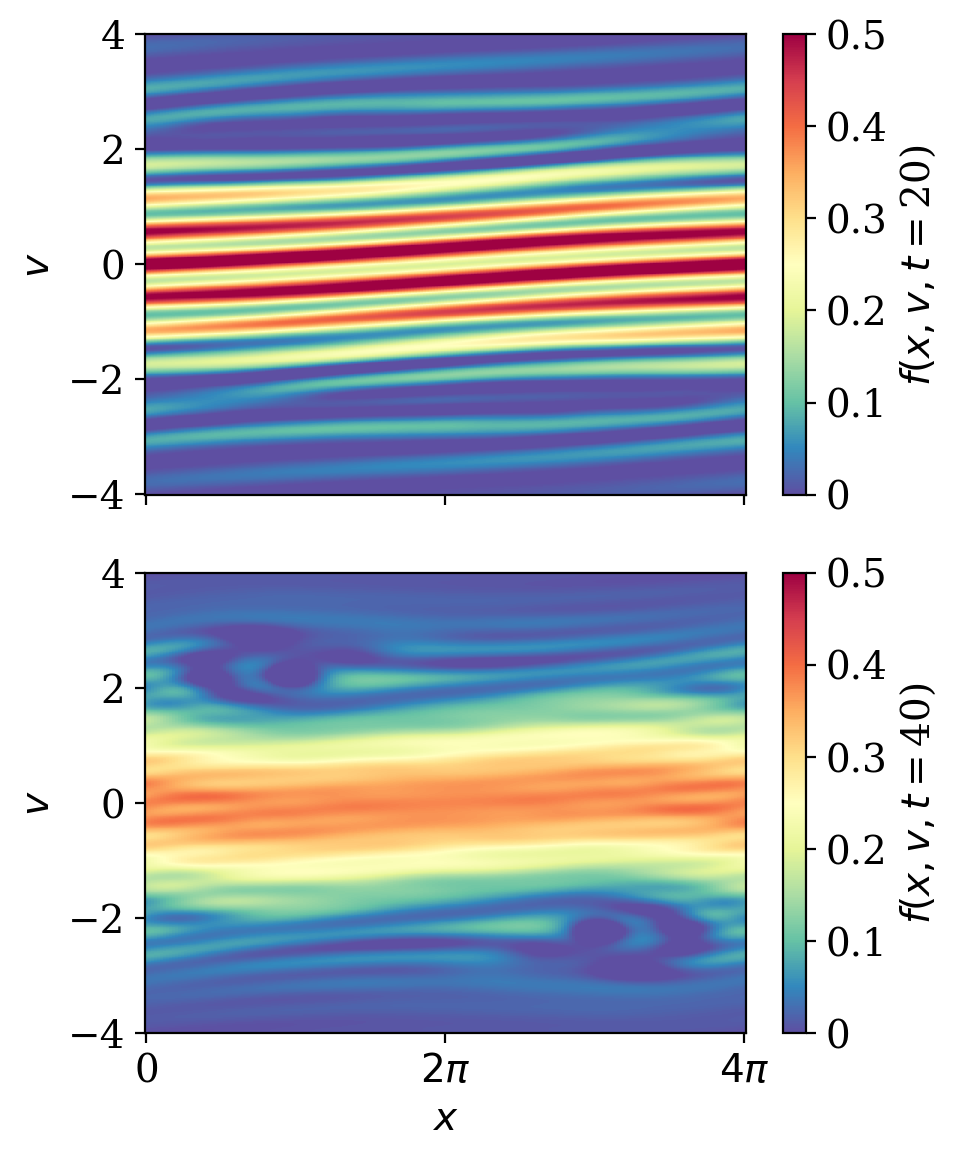}
    \end{subfigure}
        \begin{subfigure}{0.245\textwidth}
        \caption{Reference solution}
        \label{fig:nonlinear-landau-phase-space-reference}
        \includegraphics[width=\textwidth]{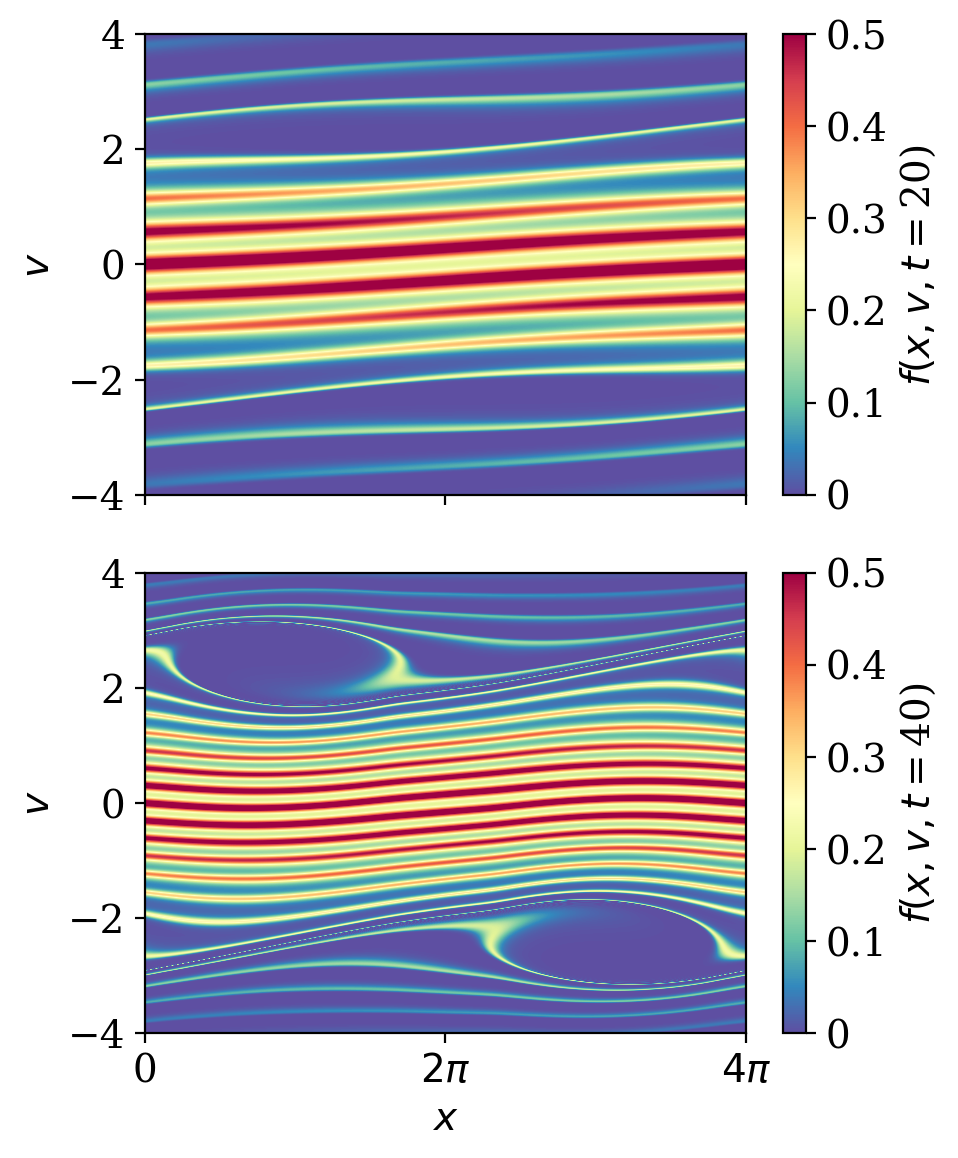}
    \end{subfigure}
    \caption{Nonlinear Landau damping electron distribution function $f(x, v, t)$ at time instances $t = 20$ and $t = 40$ with $N_{v} = 300$. Each method is compared to the reference solution in subfigure~(g), which is obtained using a high-resolution finite difference scheme. While the velocity resolution of $N_{v}=300$ is insufficient to fully resolve the filamentation at $t = 40$, all simulations besides the collisionless with closure by truncation in subfigure~(a) and LB collisions in subfigure~(c) capture the correct particle trapping vortex around the phase speed $\omega_{r}/k = \pm 2.8$.}
    \label{fig:nonlinear-landau-phase-space}
\end{figure}

\begin{figure}
    \centering
    \begin{subfigure}{0.45\textwidth}
        \caption{Comparison of $f(x=3\pi, v, t=40)$}
        \includegraphics[width=\textwidth]{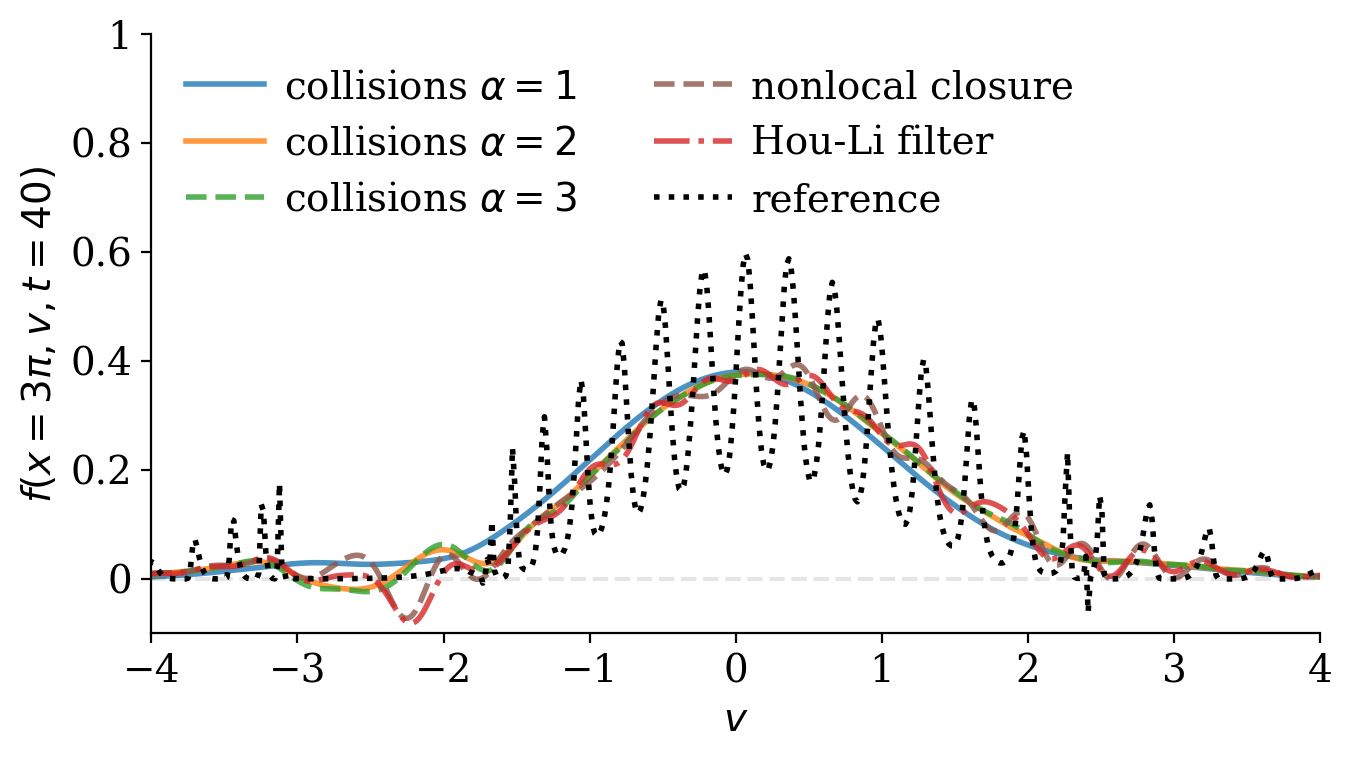}
    \end{subfigure}
        \begin{subfigure}{0.45\textwidth}
        \caption{Zoomed in version of subfigure~(a)}
        \includegraphics[width=\textwidth]{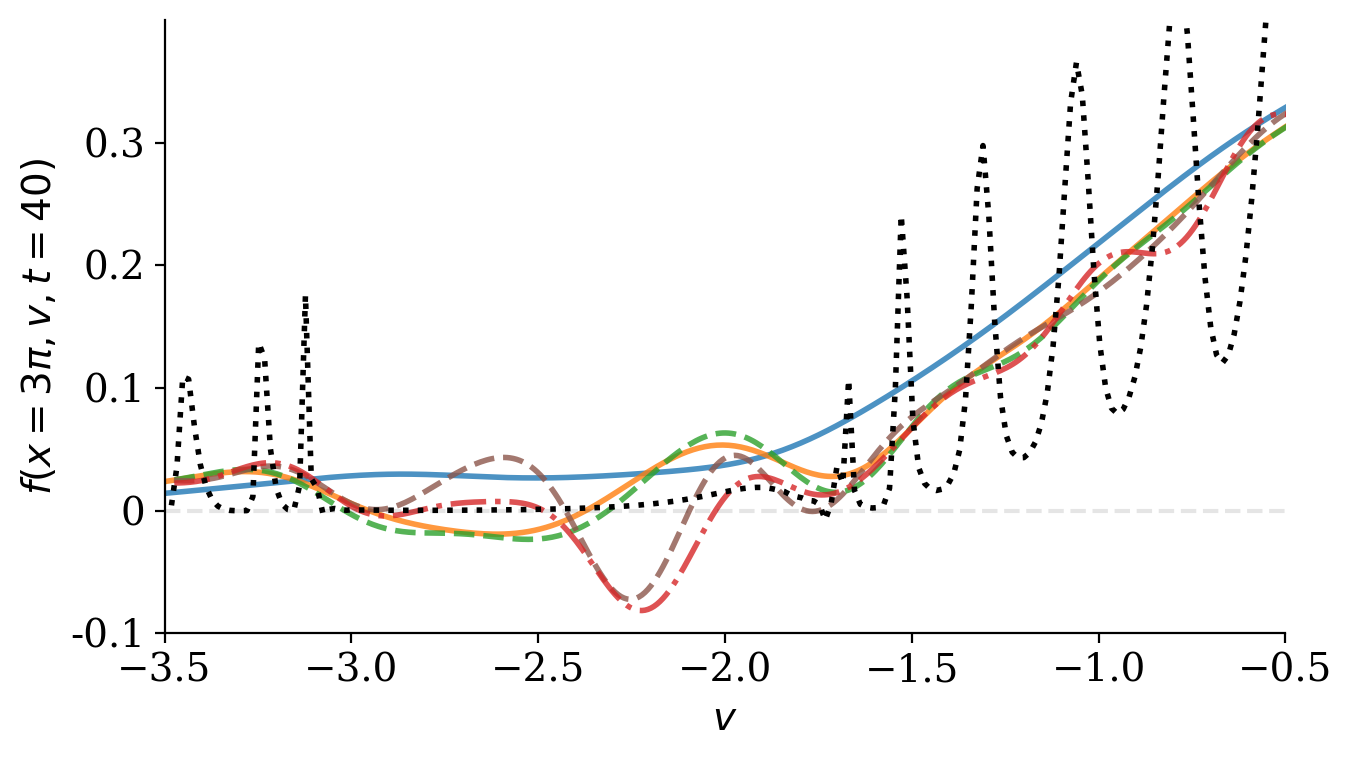}
    \end{subfigure}
    \caption{Cross-section of the nonlinear Landau damping electron distribution function $f(x=3\pi, v, t=40)$ with $N_v = 300$, shown for (a) $v \in [-4, 4]$ and (b) a zoomed-in view with $v \in [-3.5, -0.5]$. All methods, except for artificial collisions with $\alpha=1$, produce a smoothed approximation of the reference solution. However, in the region $-3 \leq v \leq -2$, these methods result in negative values for the distribution function.}
    \label{fig:f-cross-section-40}
\end{figure}

The electron distribution function in phase space is shown in Figure~\ref{fig:nonlinear-landau-phase-space}, which becomes highly oscillatory due to filamentation in velocity space. The oscillations then roll up to form a vortex centered at the phase speed $\omega_{r}/k = \pm 2.8$. 
The velocity resolution of $N_{v}=300$ does not resolve the filamentation microstructures at $t=40$ by comparison to the reference solution in Figure~\ref{fig:nonlinear-landau-phase-space-reference}, yet all methods besides LB collisions capture the correct particle trapping vortex around the phase speed. 
At $t = 40$, the shortest velocity wavelength of the filaments in the reference solution is $\lambda_{v} \approx 0.01$. According to Eq.~\eqref{scale-hermite}, resolving such velocity scales with the Hermite discretization requires $N_{v} \approx (2\pi/\lambda_{v})^2 = 394,784$.
The phase space dynamics are distorted by numerical filamentation artifacts in the collisionless case with closure by truncation.  
The filamentation artifacts shown at $t=40$ for the nonlocal closure and filtering approaches are a result of limited velocity resolution. The hypercollisions simulation with $\alpha\geq 2$ smooths out any strong deformations of the particle distribution function while maintaining the correct plasma dynamics characteristics. 
Additionally, Figure~\ref{fig:f-cross-section-40} shows a cross-section of the electron distribution function at $t=40$ and $x=3\pi$, highlighting that artificial collisions are more effective in preserving the positivity of the electron distribution function compared to nonlocal closure and filtering techniques. 
In the spectral solver, the particle distribution function is not guaranteed to remain positive due to the orthogonal nature of the velocity basis functions, potentially leading to unphysical solutions. To enforce positivity, a square-root formulation can be used; see the work by~\citet{issan_2024_antisymmetry}, which employs this transformation for the symmetrically weighted Hermite discretization.
\citet{pagliantini_2023_adaptive} showed that positivity can be improved by dynamically adjusting the Hermite tunable parameters over time or increasing the artificial collisional frequency $\nu$ in Eq.~\eqref{hypercollisions-normalized} with $\alpha=2$. This implies that the loss of positivity arises from insufficient spectral resolution. It is worth noting that even the reference solution exhibits negativity near $v\approx 2.5$, as the finite difference scheme does not enforce positivity. 

\section{Conclusions}\label{sec:section-5}
This study evaluated the effectiveness of artificial collisions, filtering, and nonlocal closure methods in mitigating filamentation and recurrence artifacts in Hermite-based Vlasov-Poisson simulations. Our analysis, including comparisons of linear kinetic response function, dispersion relation, and nonlinear simulation, indicates that higher-power artificial collisions (hypercollisions) provide the most robust solution for approximating correct Landau damping rates across a broad range of wavenumbers, particularly in multidimensional simulations constrained by limited velocity resolution. Filtering and nonlocal closures underestimate the damping rates, particularly at high wavenumber modes. In the nonlinear simulations, both hypercollisions, nonlocal closures, and Hou-Li filtering approximately distinguish between backward-propagating Hermite flux stemming from the nonlinear term and the finite truncation and mainly damp the latter. The LB collisional operators, on the other hand, completely prevent backward flux propagation from higher to lower Hermite modes.

The analysis highlights hypercollisions as a practical approach for enhancing accuracy in kinetic plasma simulations, as they effectively suppress artificial recurrence without compromising the accurate representation of collisionless dynamics. Future research could explore extending these methods to electromagnetic Vlasov-Maxwell simulations, including deriving nonlocal closures in Hermite space that capture electromagnetic wave-particle interaction such as cyclotron resonance similar to the four-moment fluid model in~\citet{jikei_2021} but generalized to higher-moment models in Hermite space. Additionally, it would be interesting to examine a hybrid formulation of the different methods, e.g. nonlocal closure method coupled with filtering or artificial collisions. Lastly, future work could explore the response function approximation for non-Maxwellian equilibrium distributions~\cite{fan_2022_nonmaxwellian}, such as Kappa or Cauchy distributions with suprathermal tails~\cite{pierrard_2010_kappa}. For these non-Maxwellian distributions, the finite AW Hermite expansion would approximate both the equilibrium and perturbation components, whereas the zeroth term in the AW Hermite expansion represents exactly a Maxwellian equilibrium. Additionally, other spectral discretizations, such as the Legendre basis, may be well-suited for non-Maxwellian distributions, which similarly to the AW Hermite basis also exhibit fluid-kinetic coupling properties~\cite{manzini_2016_legendre}.

\appendix
\section{Klimas Filter} \label{sec:Appendix-A}
\citet{klimas_1987_filter} proposed to solve for an exponentially filtered distribution function 
\begin{equation*}
    \bar{f}(x, v, t) \coloneqq  \int_{\mathbb{R}} \Theta(v-v') f(x, v', t)\mathrm{d} v', \qquad \mathrm{where}\qquad  \Theta(v) \coloneqq \frac{1}{\sqrt{2\pi}v_{0}}\exp\left( -\frac{1}{2} \left(\frac{v}{v_{0}}\right)^{2}\right).
\end{equation*}
The exponentially filtered Vlasov-Poisson equations~\eqref{vlasov-continuum}--\eqref{poisson-continuum} become
\begin{align*}
    \left(\frac{\partial}{\partial t} + v \frac{\partial}{\partial x}+ \frac{\partial \phi(x, t)}{\partial x}\frac{\partial}{\partial v} \right) \bar{f}(x, v, t)  &= -v_{0}^2 \frac{\partial^{2}}{\partial x\partial v}\bar{f}(x, v, t), \\
    -\frac{\partial^{2} \phi(x, t)}{\partial x^{2}} &= 1 - \int_{\mathbb{R}} \bar{f}(x,v, t) \mathrm{d}v.
\end{align*}
The filter modifies only the right-hand side of the Vlasov equation, introducing a second derivative that combines velocity and spatial derivatives. Additionally, the charge density and current density are invariant under filtering, i.e.~$\int_{\mathbb{R}} f(x, v, t) \mathrm{d}v = \int_{\mathbb{R}} \bar{f}(x, v, t) \mathrm{d} v$ and $\int_{\mathbb{R}} vf(x, v, t) \mathrm{d}v = \int_{\mathbb{R}} v\bar{f}(x, v, t) \mathrm{d} v$, such that the evolution of the electric field remains unchanged for any value of $v_{0} \in \mathbb{R}$~\cite{klimas_1987_filter}.
By applying the AW Hermite discretization in velocity, Fourier discretization in space, and using the recursive property in Eq.~\eqref{recursive-AW-1}, we obtain 
\begin{equation*}
     - v_{0}^{2} \frac{\partial^2}{\partial x \partial v} \bar{f} \qquad \Rightarrow \qquad  ik v_{0}^2  \sqrt{n} \hat{C}_{n-1}.
\end{equation*}
Because the Klimas filter does not correspond to a true dissipation, the Landau root is not represented by a discrete mode as in the cases considered in the main text. We have therefore tested it with dynamical simulations with different values of $v_0$ and found that recurrence effects are not eliminated when the Klimas filter is applied to the AW Hermite discretization with closure by truncation.  Indeed, ~\citet{klimas_2018_recurrence} show that the Klimas filter can suppress recurrence in doubly Fourier-transformed solvers with imposed proper boundary conditions. Note that it is important to maintain $v_{0} \lessapprox 1$ since the convolution of two Gaussian distributions is $\mathcal{G}\left(\mu_{1}, \sigma_{1}\right) \ast \mathcal{G}\left(\mu_{2}, \sigma_{2}\right) = \mathcal{G}\left(\mu_{1}+\mu_{2}, \sqrt{\sigma_{1} + \sigma_{2}}\right)$, such that setting the parameter $v_{0}$ much greater than the electron thermal velocity results in oversmoothing of the solution and the same eigenvalue analysis performed in section~3.5 predicts discrete unstable modes.

\section*{Code Availability}
The public repository~\url{https://github.com/opaliss/Vlasov_Hermite_recurrence_study.git} contains a collection of Jupyter notebooks and modules in Python~3.9 with the code used in this study.

\section*{Acknowledgment}
O.I. appreciates the useful and informative discussions with Federico Halpern and Chris Holland.
O.I. was partially supported by the Los Alamos National Laboratory (LANL) Student Fellowship sponsored by the Center for Space
and Earth Science (CSES). CSES is funded by LANL's Laboratory Directed Research and Development (LDRD) program under project
number 20210528CR.
The LANL LDRD Program supported O.I., G.L.D., O.K., and O.C. under project number 20220104DR. LANL is operated by Triad National Security, LLC, for the National Nuclear Security Administration of the US Department of Energy (Contract No. 89233218CNA000001). 

\bibliography{references}
\end{document}